\documentclass[11pt]{article}

\usepackage{amsmath,amssymb,amsthm,mathtools,fullpage,hyperref}
\usepackage{enumitem} 
\usepackage{booktabs,multirow,array,makecell}
\usepackage{thmtools}
\usepackage{thm-restate}
\usepackage{xcolor}
\usepackage{amsmath,amssymb}
\usepackage{tikz}
\usetikzlibrary{decorations.pathreplacing,calc,arrows.meta,positioning,fit,backgrounds}
\definecolor{iso}{HTML}{2E9E5B}
\definecolor{col}{HTML}{D1495B}
\definecolor{cmp}{HTML}{9AA0A6}
\definecolor{ink}{HTML}{222222}
\colorlet{isoL}{iso!18}
\colorlet{colL}{col!20}

\makeatletter
\renewcommand\thmt@autorefsetup{%
  \@xa\def\csname\thmt@envname autorefname\@xa\endcsname\@xa{\thmt@thmname}%
}
\makeatother

\newcolumntype{C}[1]{>{\centering\arraybackslash}m{#1}}

\newtheorem{theorem}{Theorem}[section]
\newtheorem{lemma}[theorem]{Lemma}

\theoremstyle{definition}
\newtheorem{definition}[theorem]{Definition}
\theoremstyle{remark}
\newtheorem{remark}[theorem]{Remark}
\newtheorem{fact}[theorem]{Fact}
\newtheorem{claim}[theorem]{Claim}

\newcommand{\R}{\mathbb{R}}
\newcommand{\Z}{\mathbb{Z}}
\newcommand{\E}{\mathbb{E}}
\newcommand{\Prb}{\mathbb{P}}
\newcommand{\1}{\mathbf{1}}

\newcommand{\tr}{\operatorname{tr}}
\newcommand{\col}{\operatorname{col}}
\newcommand{\diag}{\operatorname{diag}}
\newcommand{\vol}{\operatorname{vol}}
\newcommand{\Id}{\mathrm{I}}
\newcommand{\supp}{\operatorname{supp}}
\newcommand{\smin}{\sigma_{\min}}

\newcommand{\eps}{\varepsilon}
\newcommand{\set}[1]{\left\{#1\right\}}
\newcommand{\paren}[1]{\left(#1\right)}
\newcommand{\bracket}[1]{\left[#1\right]}
\newcommand{\abs}[1]{\left|#1\right|}
\newcommand{\norm}[1]{\left\|#1\right\|}
\newcommand{\ip}[2]{\left\langle #1, #2 \right\rangle}

\title{Optimal Injectivity of the SparseStack}
\author{
  Lorenzo Beretta\\
  IBM Cambridge
  \and
  Simone Bombari\\
  NYU \& Flatiron Institute
  \and
  Margalit Glasgow\\
  Massachusetts Institute of Technology
  \and
  Raphael Arkady Meyer\\
  UC Berkeley, ICSI, \& LBNL
  \and
  Kijun Shin\\
  Carnegie Mellon University
  \and
  David P. Woodruff\\
  Google \& Carnegie Mellon University
}
\date{}

\begin{document}

\maketitle
\pagenumbering{gobble}

\begin{abstract}
An Oblivious Subspace Injection (OSI) is a random sketching matrix that reliably preserves the minimum singular value of any fixed subspace with high constant probability, requiring only a one-sided spectral lower bound without controlling the upper singular values. While classical Oblivious Subspace Embeddings (OSEs) fundamentally require row sparsity $\Omega(\log r)$ at linear sketch dimension $k = O(r)$, we show that one-sided injectivity permits strictly sparser sketches. Specifically, we resolve the optimal sparsity threshold for linear-dimension OSIs, establishing matching (up to $\log\log r$ factors) upper and lower bounds of $\tilde{\Theta}(\sqrt{\log r})$.

First, we prove that the SparseStack sketch, formed by horizontally concatenating $\zeta$ independent CountSketch blocks, achieves constant injectivity at linear embedding dimension $k = \zeta b = \Theta(r)$ with row sparsity only $\zeta = O(\sqrt{\log r})$. This operates well below the $\Theta(\log r)$ barrier where standard operator-norm concentration fails due to coupon-collector bucket collisions. Second, we establish a matching lower bound: any isotropic random sketching matrix satisfying a mild row-overlap condition requires row sparsity $\zeta = \Omega(\sqrt{\log r / \log \log r})$ to achieve linear-dimension injectivity, confirming that the $\sqrt{\log r}$ scale is fundamental.

To bypass the lack of an operator-norm upper bound in the $\zeta = O(\sqrt{\log r})$ regime, we decouple the sketch into two independent halves and introduce three main techniques:
(i) a uniform lower bound for \emph{flat} vectors (vectors with large tail norm) via a randomized rounding argument, where discretization error is controlled using solely the Frobenius norm rather than the operator norm;
(ii) a Gaussian truncation and minimax duality argument showing that all non-flat vectors escaping injectivity in the first half must concentrate their mass on a single, universal coordinate subset $I \subseteq [n]$ of size $O(r/\zeta)$, completely bypassing union bounds over supports; and
(iii) a two-tier coordinate isolation analysis in the independent second half that suppresses the mutual collision probability to $e^{-\Omega(\zeta^2)} \ll 1/r$, allowing a union bound to reliably isolate and preserve these heavy coordinates.
\end{abstract}

\newpage
\pagenumbering{arabic}
\setcounter{page}{1}

\tableofcontents

\paragraph{AI Disclosure Statement.} The authors utilized the colosseum harness \cite{LWDMZ26} on top of Gemini 4 throughout the mathematical development of this work. Specifically, the vast majority of the technical proofs and mathematical derivations in this manuscript were generated and drafted by the model. The human authors independently established several key structural lemmas, improved the exposition, and carefully verified all mathematical steps. Furthermore, this work is accompanied by a fully functional, machine-checked formal proof in \textsc{Lean}~4. The human authors audited the formalization to confirm that all definitions and theorem statements faithfully and accurately encode the intended mathematical results. The machine-checked formal proof in \textsc{Lean}~4 is publicly available at \href{https://github.com/sharaelong/optimal-injectivity-sparsestack}{GitHub repository}. The authors take full responsibility for the accuracy, rigor, and conclusions of this work. 
\newpage

\section{Introduction}\label{sec:introduction}

Sketching is the workhorse of randomized linear algebra: we replace a
large matrix by a small random image of it and do the expensive work on
the image \cite{Sarlos06,HMT11,Woodruff14,MT20}.
To sketch a matrix $A\in\R^{n \times d}$, we draw a tall random matrix $\Omega\in\R^{n \times k}$ and form the product $\Omega^\top A$.
To measure the cost of a sketching method, we need to consider two aspects.
First is the embedding dimension \(k\), which controls the size of (and therefore cost of working on) the image;
when the problem is
governed by an $r$-dimensional subspace, the natural goal is $k=O(r)$.
Second, we need to account for the cost of computing \(\Omega^\top A\).
Indeed, if \(\Omega\) is a dense Gaussian matrix, then computing \(\Omega^\top A\) can easily become the computation bottleneck.

To resolve this, a very popular approach is to use \emph{sparse} random matrices~\cite{CW13,MM13,KN14,MDMEM23,CNRSP26}.
If \(\Omega\) has at most \(\zeta\) nonzeros per row, then we can form the sketch \(\Omega^\top A\) in \(O(\zeta \operatorname{nnz}(A))\) time.
We call \(\zeta\) the \emph{row sparsity} of \(\Omega\).
This paper studies a single question: how sparse can the test matrix be when $k=O(r)$ while still being reliable for numerical algorithms?


\subsection{Embeddings and Injections}\label{subsec:embvsinj}

Classically, randomized numerical linear algebra (RNLA) methods have used the \emph{Oblivious Subspace Embedding (OSE)} condition to characterize when a sketching matrix is reliable \cite{Sarlos06,Woodruff14,Tropp15,MT20,CEMT25}.
We say that \(\Omega\in\R^{n \times k}\) is an OSE if, for every matrix \(Q\in\R^{n \times r}\) with orthonormal columns, we have
\[
    (1 - \eps)\Id_r \preceq Q^\top \Omega \Omega^\top Q \preceq (1 + \eps)\Id_r \qquad \text{with high probability.}
\]
Equivalently, the OSE guarantee ensures that vectors in the range of \(Q\) are neither stretched nor squished by large amount.
A long line of work has shown that sparse sketching matrices with embedding dimension \(k = O(r)\) and constant dilation \(\eps=1/2\) must have row sparsity \(\zeta = \Omega(\log r)\), and that this rate is achieved by the SparseStack matrix (see below) \cite{Cohen16,NN13,CDDR24,CDMD24,CDMD26,Heidary26}.

While the OSE defintion is natural, and is guaranteed to hold for many sketching matrices, it is often also overkill.
Several works in RNLA show that many downstream algorithms, like methods for low-rank approximation~\cite{HMT11,Woodruff14,MW17,MT20,CEMT25} and least-squares regression~\cite{RT08,DMMS11,CEMT25}---do not actually require a two-sided guarantee.
Instead, it suffices for the sketch to satisfy a \emph{one-sided} lower bound:
\begin{equation}\label{eq:one-sided-def}
  Q^\top \Omega \Omega^\top Q \succeq \alpha \Id_r, \qquad \text{or equivalently} \qquad \bigl\|\Omega^\top Q x\bigr\|^2 \ge \alpha \norm{x}^2 \quad \text{for all } x \in \R^r,
\end{equation}
where $\alpha > 0$ is called the \emph{injectivity parameter}, all subject the normalization criteria \(\E[\Omega\Omega^\top]=I\).
While this weaker criteria was known or folklore in specific contexts, its systematic study was initiated by Cama\~no, Epperly, Meyer, and Tropp~\cite{CEMT25}, who introduced the term \emph{Oblivious Subspace Injection (OSI)}.
It will be helpful to think of the OSI condition in terms of spectra.
Indeed, equation \ref{eq:one-sided-def} holds if and only if
\[
  \smin(\Omega^\top Q)^2 \ge \alpha \qquad \text{with high constant probability}.
\]
The central question we want to understand is, at linear sketch dimension $k = \Theta(r)$, how small can we make the row sparsity \(\zeta\) while maintaining constant injectivity \(\alpha\).
Most specifically, can a sparse matrix be an OSI with sparsity \(\zeta = o(\log r)\), proving that the OSEs formalism is unable to fully justify when sparse matrices are reliable numerical tools?
We answer this in the affirmative.

\paragraph{The SparseStack sketch.}
To minimize both sketching time and storage, a popular family of structured random matrices in RNLA is CountSketch~\cite{CW13,CCF02,TZ04} and its multi-block generalization, \emph{SparseStack}~\cite{WDLSA09,NN13,MM13,PP13,Cohen16,CEMT25}.
Throughout this paper, a SparseStack sketch matrix $\Omega_{\zeta, b} \in \R^{n \times k}$ has $n \ge r$ rows, row sparsity $\zeta$, and total embedding dimension $k = \zeta b = \Theta(r) \ll n$.

We define these matrices directly in terms of independent Rademacher signs and sparse coordinate selections. Let $e_1, \dots, e_b \in \R^b$ denote the standard basis vectors. A CountSketch matrix $\Phi \in \R^{n \times b}$ has rows given by
\begin{equation}\label{eq:countsketch-def}
  \Phi = \begin{bmatrix}
    \varrho_1 e_{s_1}^\top \\
    \vdots \\
    \varrho_n e_{s_n}^\top
  \end{bmatrix} \in \R^{n \times b},
  \qquad \text{where} \qquad
  \begin{aligned}
    \varrho_i &\sim \textsc{Rademacher}\ \text{i.i.d.}; \\
    s_i &\sim \textsc{Uniform}\{1, \dots, b\}\ \text{i.i.d.}
  \end{aligned}
\end{equation}
That is, each row of $\Phi$ contains a single non-zero entry $\varrho_i \in \{-1, 1\}$ placed at an independent, uniformly chosen column index $s_i \in [b]$.

A SparseStack matrix $\Omega_{\zeta, b} \in \R^{n \times k}$ with row sparsity $\zeta$, block size $b$, and dimension $k = \zeta b$ horizontally concatenates $\zeta$ independent CountSketch matrices $\Phi_1, \dots, \Phi_\zeta \sim \Phi$, scaled by $\zeta^{-1/2}$:
\begin{equation}\label{eq:sparsestack-def}
  \Omega_{\zeta, b} := \frac{1}{\sqrt{\zeta}} \begin{bmatrix} \Phi_1 & \cdots & \Phi_\zeta \end{bmatrix}
  = \frac{1}{\sqrt{\zeta}} \begin{bmatrix}
    \varrho_{1,1} e_{s_{1,1}}^\top & \cdots & \varrho_{1,\zeta} e_{s_{1,\zeta}}^\top \\
    \varrho_{2,1} e_{s_{2,1}}^\top & \cdots & \varrho_{2,\zeta} e_{s_{2,\zeta}}^\top \\
    \vdots & \ddots & \vdots \\
    \varrho_{n,1} e_{s_{n,1}}^\top & \cdots & \varrho_{n,\zeta} e_{s_{n,\zeta}}^\top
  \end{bmatrix} \in \R^{n \times k},
\end{equation}
where $\{\varrho_{it}\}$ are i.i.d. Rademacher random variables and $\{s_{it}\}$ are i.i.d. uniform random indices on $[b]$.
By construction, every row of $\Omega_{\zeta, b}$ has exactly $\zeta$ non-zero entries of magnitude $\zeta^{-1/2}$ and unit Euclidean norm. Since distinct rows have mean-zero inner products, $\Omega_{\zeta, b}$ is isotropic in expectation.

\paragraph{Proof methods for OSEs and the $\Theta(\log r)$ barrier.}
In the two-sided $(1 \pm \eps)$-OSE setting, SparseStack is the block-independent model within the OSNAP framework of Nelson and Nguy\~{\^e}n~\cite{NN13} and Cohen~\cite{Cohen16}. Recent work by Heidary~\cite{Heidary26} established that dimension $k = \Theta(r/\eps^2)$ and row sparsity $\zeta = \Theta((\log r)/\eps)$ are both sufficient and optimal for a two-sided $(1 \pm \eps)$-embedding.

Existing methods establish injectivity by viewing the SparseStack as an average of independent CountSketch blocks, giving
\begin{equation}\label{eq:gram-iid}
  A^\top A = Q^\top \Omega_{\zeta, b} \Omega_{\zeta, b}^\top Q = \frac{1}{\zeta} \sum_{t=1}^\zeta Q^\top \Phi_t \Phi_t^\top Q.
\end{equation}
In particular, using matrix comparison theorems, Tropp~\cite{Tropp25} proved that row sparsity $\zeta = \Theta(\log r)$ ensures $\smin(A)^2 \ge \Omega(1)$ at linear sketch dimension $k = \Theta(r)$. The question of whether constant injectivity is achievable with row sparsity strictly below $\Theta(\log r)$ was posed by Cama\~no et~al.~\cite{CEMT25} and has remained open.

Techniques that analyze $A^\top A$ directly as a sum of independent blocks hit a conceptual barrier at row sparsity $\Theta(\log r)$. Below this threshold, the largest singular value $\sigma_{\max}(A)$ is no longer bounded by $O(1)$ with high probability. It is easy to see this via a coupon-collector argument: hashing $r$ basis coordinates into $b = \Theta(r/\zeta)$ buckets in a block inevitably creates overloaded buckets where $\Theta(\log(r/\zeta)/\log\log(r/\zeta^2)) = \omega(\zeta)$ coordinates collide, and a unit vector aligned with those colliding coordinates has its entries add up to give $\|\Omega_{\zeta, b}^\top Qx\| = \omega(1)$. 

Nonetheless, it has been experimentally shown that SparseStacks, as well as other sparse sketching matrices, are OSIs with favorable values of \(\alpha\) even when \(\zeta = 4\) and \(k=2r\) \cite{MDMEM23, CEMT25}.
Motivated by these questions, Cama\~no et al.~\cite{CEMT25} conjectured that constant row sparsity $\zeta = O(1)$ might suffice for an OSI at linear dimension $k = \Theta(r)$. However, Huang, Rudelson, and Tikhomirov~\cite{HRT26} recently refuted this conjecture, showing that constant row sparsity fails for any OSI distribution. This leaves open the central question: \emph{What is the true optimal sparsity for an Oblivious Subspace Injection?} Does injectivity inherently require $\Theta(\log r)$ sparsity as in the two-sided OSE case, or does the relaxation to one-sided injectivity permit a strictly sparser sketch?

\subsection{Main Results}\label{subsec:main-results}

In this paper, we resolve this question by establishing matching (up to $\log \log r$ factors) upper and lower bounds on the row sparsity of OSI at linear embedding dimension, showing that the optimal threshold is $\tilde{\Theta}(\sqrt{\log r})$.

Our first main result is an upper bound proving that SparseStack achieves constant injectivity at linear embedding dimension with row sparsity only $O(\sqrt{\log r})$, well below the $\Theta(\log r)$ barrier of two-sided embeddings.

\begin{restatable}{theorem}{thmmain}\label{thm:main}
There exist universal constants $C, K > 0$ and $\alpha \in (0, 1)$ such that the following holds. For every $n \ge r \ge 2$, every fixed isometry $Q \in \R^{n \times r}$ satisfying $Q^\top Q = \Id_r$, and integer sketch parameters satisfying
\[
  \zeta = O(\sqrt{\log r}) \text{ with } \zeta \ge C\sqrt{\log(er)}, \qquad b = \Theta\!\paren{\frac{r}{\zeta}}, \qquad k = \zeta b \ge Kr,
\]
the sketched matrix $A = \Omega_{\zeta, b}^\top Q \in \R^{k \times r}$ formed from the $\zeta$-block SparseStack sketch $\Omega_{\zeta, b} \in \R^{n \times k}$ satisfies
\[
  \Prb\bracket{\smin(A)^2 \ge \alpha} \ge \frac{19}{20}.
\]
Moreover, $\E\bracket{\Omega_{\zeta, b}\Omega_{\zeta, b}^\top} = \Id_n$ and the total sketch dimension is linear in $r$, namely $k = \zeta b = \Theta(r)$.
\end{restatable}

Our second main result is a general sparsity lower bound showing that this $\sqrt{\log r}$ scale is essentially unavoidable. Recently, Huang, Rudelson, and Tikhomirov~\cite{HRT26}, denoted by HRT, proved that constant row sparsity $\zeta = O(1)$ cannot achieve an OSI by introducing three structural assumptions on the sketching matrix: uniform entry-magnitude bounds of order $O(\zeta^{-1/2})$ on all non-zero entries, deterministic row-norm and support-size regularity, and a bound on pairwise row-support overlaps. In this work, we eliminate the first two assumptions and retain only the \emph{overlap condition} to establish the $\tilde{\Omega}(\sqrt{\log r})$ sparsity lower bound:

\begin{definition}[Overlap condition]\label{def:condition-h}
A matrix $S \in \R^{n \times k}$ with rows $s_1, \dots, s_n$ satisfies the \emph{overlap condition} if the number of pairs of distinct rows $\{u, v\} \subset [n]$ sharing two or more non-zero coordinates satisfies $\sum_{1 \le u < v \le n} \mathbf{1}_{\{|\supp(s_u) \cap \supp(s_v)| \ge 2\}} \le \eta \frac{n^2}{k}$ for an \emph{overlap parameter} $\eta = o(1)$. Specifically, we need $\eta = O((\log \log r / \log r)^3)$.
\end{definition}

Here the $o(1)$ bound on $\eta$ is a function of the subspace dimension $r$: as we will show in the proof by requiring $\eta \le (C_H \zeta^6)^{-1}$ for a constant $C_H > 0$, under the target sparsity scale $\zeta = \Theta(\sqrt{\log r / \log \log r})$ our analysis ultimately sets $\eta$ to be on the order of $\Theta((\log \log r / \log r)^3)$. Because this requirement is only a logarithmically decaying function of $r$, it is far more permissive than the polynomial decay exhibited by standard sparse sketching matrices. Specifically, in ensembles such as SparseStack and OSNAP~\cite{NN13,Cohen16} at linear dimension $k = \Theta(r)$, any two rows share a given column with probability $\zeta/k$ and thus share two or more columns with probability at most $\binom{\zeta}{2}(\zeta/k)^2 = O(\zeta^4/k^2)$; summing over all $\binom{n}{2}$ row pairs yields an actual overlap parameter of order $\eta = O(\zeta^4/k) = \tilde{O}(1/r) \ll (\log \log r / \log r)^3$, showing a polynomial-versus-polylogarithmic gap and confirming that the overlap condition encompasses all standard sparse sketching constructions.

\begin{restatable}{theorem}{thmlowerbound}\label{thm:intro-lower-bound}
For sufficiently large $r$, let $n \ge \Omega(r \log^3 r)$, $k \le Kr$ for a constant $K \ge 1$, and $\alpha \in (0, 1)$. If an isotropic random sketching matrix $\Omega \in \R^{n \times k}$ with row sparsity $\zeta$ satisfies the \hyperref[def:condition-h]{overlap condition} with  the overlap parameter $\eta = O((\log \log r / \log r)^3)$ with high probability, and is an $(r, \alpha)$-OSI, then the row sparsity must satisfy
\[
  \zeta = \Omega\left( \sqrt{\frac{\log r}{\log \log r}} \right).
\]
\end{restatable}

\paragraph{Separation of OSIs from OSEs.}
Together, Theorems~\ref{thm:main} and~\ref{thm:intro-lower-bound} separate OSI from OSE. Nelson and Nguy\~{\^e}n~\cite{NN13} proved that for any $n \ge 100r^2$, any $(r, 1 \pm \eps)$-OSE at linear dimension $k = O(r)$ requires row sparsity $\Omega(\log r / \log \log r)$. Theorem~\ref{thm:main} demonstrates that SparseStack is an $(r, \alpha)$-OSI with row sparsity $O(\sqrt{\log r})$, showing that one-sided injectivity requires strictly less sparsity than two-sided preservation:
\[
  \zeta_{\mathrm{OSI}} = \Theta(\sqrt{\log r}) \ll \Theta(\log r) = \zeta_{\mathrm{OSE}}.
\]

\paragraph{Algorithmic consequences.}
Injectivity alone supports several algorithmic primitives in randomized numerical linear algebra. Following the reductions of Cama\~no et al.~\cite{CEMT25}:
\begin{enumerate}[leftmargin=*,itemsep=2pt]
  \item \textbf{Sketch-and-solve least-squares regression:} Given an overdetermined system $\min_{x \in \R^r} \|Bx - u\|_2$ with $B \in \R^{n \times r}$ of full column rank and $u \in \R^n$, an $(r, \alpha)$-OSI $\Omega \in \R^{n \times k}$ produces an estimator $\hat{x} \in \operatorname{argmin}_x \|\Omega^\top(Bx - u)\|_2$ that satisfies
  \[
    \|B\hat{x} - u\|_2^2 \le \frac{C}{\alpha} \min_{x \in \R^r} \|Bx - u\|_2^2
  \]
  with probability at least $9/10$. By Theorem~\ref{thm:main}, computing the sketch $\Omega^\top B$ requires only $O(\mathrm{nnz}(B)\sqrt{\log r})$ operations, improving upon the $\Omega(\mathrm{nnz}(B)\log r)$ cost required by two-sided OSEs.
  \item \textbf{Low-rank approximation and randomized SVD:} For a matrix $M$, computing randomized SVD and Nystr\"om approximations~\cite{MW17} via an $(r, \alpha)$-OSI achieves constant-factor accuracy with probability at least $4/5$~\cite[Theorem~1.3]{CEMT25}. Theorem~\ref{thm:main} reduces the sketching cost to $O(\mathrm{nnz}(M)\sqrt{\log r})$.
\end{enumerate}

\subsection{Union Bounds and Flat and Spiky Vectors.}\label{subsec:why_sqrtlog}
\textbf{Union Bound Approaches}
An alternative approach matrix concentration theorems is to establish point-wise lower bounds on $\| Ax\|$ for each fixed $x\in S^{r-1}$, and then obtain the desired bound on $\sigma_{\min}(A)$ through a union bound over $S^{r-1}$. Naive net $\eps$-net arguments obviously fail here.  To see this concretely, consider a 2-sparse unit vector such as $Qx = (e_1 + e_2)/\sqrt{2}$. In any given CountSketch block $t$, its image $\Phi_t^\top Qx = \frac{1}{\sqrt{2}}(\varrho_{1t} e_{s_{1t}} + \varrho_{2t} e_{s_{2t}})$ is identically zero precisely when coordinates $1$ and $2$ select the same bucket ($s_{1t} = s_{2t}$, which occurs with probability $1/b$) and receive opposite signs ($\varrho_{1t} = -\varrho_{2t}$, which occurs with probability $1/2$). This collision cancellation annihilates the vector in block $t$ with probability $1/(2b)$. Since the $\zeta$ blocks are independent, the vector is completely canceled across the entire sketch with probability
\begin{equation}\label{eq:spiky-annihilation-prob}
  \Prb\bracket{\Omega_{\zeta, b}^\top Qx = 0} = (2b)^{-\zeta} = e^{-\zeta \log(2b)}.
\end{equation}
Since a naive $\eps$-net over the unit sphere $S^{r-1}$ contains $e^{\Theta(r)}$ vectors, which would demand $\zeta \log(2b) = \Omega(r)$ and force an impractical near-linear sparsity $\zeta = \tilde{\Omega}(r)$. Furthermore, even for vectors whose coordinates are spread out so that pointwise concentration holds with failure probability $e^{-\Omega(r)}$, extending a pointwise bound over an $\eps$-net requires controlling the discretization error via the operator norm:
\begin{equation}\label{eq:net-discretization-error}
  \norm{\Omega_{\zeta, b}^\top Q(x - x')} \le \norm{\Omega_{\zeta, b}^\top Q} \cdot \norm{x - x'}.
\end{equation}
This step crucially relies on an $O(1)$ upper bound on $\|\Omega_{\zeta, b}^\top Q\|$, which as discussed above, due to the coupon-collector analysis, ceases to hold when $\zeta \ll \log r$.

\paragraph{Combinatorial Random Matrix Approaches} The combinatorial random matrix literature has developed a more intricate toolbox for taking union bounds over $S^{r-1}$. The core of this approach is distinguishing between \em compressible vectors, \em which have mass concentrated on a small set of entries, and \em incompressible vectors, \em which have mass spread more evenly among the components. The key observation is that there are fewer compressible vectors: for example, in the extreme compressible case of $1$-sparse vectors on $S^{r-1}$, there are only $r$ such vectors. Thus even if the corresponding ``bad'' events have larger probabilities for the compressible vectors (eg. that event that $x$ gets compressed by a random matrix), the union bound over such compressible $x$ is smaller. For incompressible vectors, typically the ``bad'' events have much smaller probabilities. 

The more levels of compression that are considered when breaking up the union bound, the stronger the results are, and these tools have been successful applied to proving bounds on the singular values of combinatorial random matrices~\cite{rudelson3956smallest}, sharp bounds on their singularity probabilities~\cite{costello2006random, tikhomirov2020singularity, jain2020sharp}, and for random matrices below the $O(\log(r)/r)$ sparsity threshold, sharp bounds on their rank~\cite{costello2010rank, demichele2024rank, glasgow2025exact}.

In our proof, we will take a coarse approach that distinguishes \em spiky \em vectors from \em flat \em vectors (see Definition~\ref{def:flat}), which are similar to the notions of compressible and incompressible vectors in this literature. Ultimately though, one strength of our proof is that we do \em not \em require taking intricate union bounds at many levels of compressibility.


\subsection{Comparison with Prior Work}\label{subsec:related-work}

\paragraph{Subspace embeddings (OSEs).}
The study of subspace embeddings was initiated by Sarl\'os~\cite{Sarlos06}. Clarkson and Woodruff~\cite{CW13} showed that CountSketch yields an input-sparsity OSE with dimension $k = O(r^2)$. To achieve linear dimension $k = O(r)$, Nelson and Nguy\~{\^e}n~\cite{NN13} introduced the OSNAP framework and conjectured that row sparsity $\zeta = O(\log r)$ suffices. Intermediate bounds were obtained by Cohen~\cite{Cohen16} ($k = O(r \log r)$, $\zeta = O(\log r)$) and Chenakkod et al.~\cite{CDDR24} ($k = O(r)$, $\zeta = O(\log^4 r)$). Recently, Heidary~\cite{Heidary26} settled the conjecture, proving that SparseStack is an $(r, 1 \pm \eps)$-OSE with dimension $k = O(r/\eps^2)$ and sparsity $\zeta = O(\eps^{-1}\log(r/\delta))$. Nelson and Nguy\~{\^e}n~\cite{NN13} also showed that any linear-dimension OSE requires row sparsity $\Omega(\log r / \log \log r)$, confirming that $\Theta(\log r)$ is essentially optimal for two-sided embeddings.

\begin{table}[t]
\centering
\small
\begin{tabular}{@{}llcc@{}}
\toprule
\textbf{Guarantee} & \textbf{Source} & \textbf{Dimension $k$} & \textbf{Row Sparsity $\zeta$} \\ \midrule
OSE & CountSketch~\cite{CW13,NN13} & $O(r^2)$ & $1$ \\
OSE & Cohen~\cite{Cohen16} & $O(r \log r)$ & $O(\log r)$ \\
OSE & Chenakkod et al.~\cite{CDDR24} & $O(r)$ & $O(\log^4 r)$ \\
OSE & Heidary~\cite{Heidary26} & $O(r)$ & $O(\log r)$ \\ \midrule
OSI & Tropp~\cite{Tropp25}; Cama\~no et al.~\cite{CEMT25} & $O(r)$ & $O(\log r)$ \\
OSI & \textbf{Theorem~\ref{thm:main}} (Upper Bound) & $O(r)$ & $\mathbf{O(\sqrt{\log r})}$ \\
OSI & \textbf{Theorem~\ref{thm:intro-lower-bound}} (Lower Bound) & $O(r)$ & $\mathbf{\Omega\left(\sqrt{\frac{\log r}{\log \log r}}\right)}$ \\
\bottomrule
\end{tabular}
\caption{Comparison of dimension and row sparsity parameters for SparseStack and general sketching distributions at fixed accuracy and constant failure probability. The OSE bounds require two-sided distortion $1 \pm \eps$, whereas the OSI bounds guarantee one-sided constant injectivity $\smin^2 \ge \alpha$.}
\label{tab:comparison}
\end{table}

\paragraph{Oblivious Subspace Injections (OSIs).}
Cama\~no, Epperly, Meyer, and Tropp~\cite{CEMT25} introduced OSIs and showed that one-sided injectivity suffices for core RNLA tasks. Using comparison theorems for the minimum eigenvalue of random positive-semidefinite matrices~\cite{Tropp25}, they showed that SparseStack achieves injectivity with $k = O(r)$ and $\zeta = O(\log r)$. Subsequent Gaussian universality results~\cite{Tropp25} sharpened these guarantees, but existing analyses inherently rely on operator-norm concentration around the mean, which restricts them to $\zeta = \Omega(\log r)$. In contrast, our Theorem~\ref{thm:main} achieves injectivity at $\zeta = O(\sqrt{\log r})$ by decoupling the sketch and avoiding operator-norm concentration entirely.

\paragraph{Sparsity lower bounds.}
For two-sided embeddings, Nelson and Nguy\~{\^e}n~\cite{NN13} proved that any linear-dimension OSE requires row sparsity $\Omega(\log r / \log \log r)$. For one-sided injections, Cama\~no et al.~\cite{CEMT25} conjectured that constant sparsity $\zeta = O(1)$ might suffice (Problem~5.5 in the Simons Institute open problem collection). Huang, Rudelson, and Tikhomirov~\cite{HRT26} refuted this by proving that constant row sparsity fails for any OSI, using matching-rooted trees to construct cancellation vectors. Theorem~\ref{thm:intro-lower-bound} establishes that any isotropic distribution over row-sparse matrices requires row sparsity $\zeta = \Omega(\sqrt{\log r / \log \log r})$, showing that the $\sqrt{\log r}$ scale is fundamental to all OSIs.

\subsection{Conventions and Notation}\label{subsec:conventions-organization}

All vector spaces considered in this paper are real and finite-dimensional. For $d \ge 1$, we write $S^{d-1} = \{x \in \R^d : \|x\| = 1\}$ for the Euclidean unit sphere and $B_2^d = \{x \in \R^d : \|x\| \le 1\}$ for the Euclidean unit ball in $\R^d$. Vector norms $\|\cdot\|$ are Euclidean, matrix norms $\|\cdot\|$ without subscripts denote the spectral (operator) norm, and $\|\cdot\|_F$ denotes the Frobenius norm. For a vector $z \in \R^n$ and a coordinate subset $I \subseteq [n]$ (with complement $I^c = [n] \setminus I$), we write $z_I \in \R^{|I|}$ for the subvector of coordinates indexed by $I$. Similarly, for a matrix $V \in \R^{n \times d}$, we write $V_I \in \R^{|I| \times d}$ for the row submatrix indexed by $I$.

For a $\zeta$-block SparseStack matrix $\Omega_{\zeta, b} = \zeta^{-1/2}[\Phi_1 \ \cdots \ \Phi_\zeta] \in \R^{n \times k}$ ($k = \zeta b$) as in~\eqref{eq:sparsestack-def}, the normalization factor $\zeta^{-1/2}$ is always determined by the number of blocks $\zeta$ in the subscript. When viewing the transpose $\Omega_{\zeta, b}^\top \in \R^{k \times n}$ (or an individual block $\Phi_t^\top \in \R^{b \times n}$) as the sketching operator acting on data vectors in $\R^n$, its $i$-th column vector is the 1-sparse bucket vector $\phi_{it} = \varrho_{it} e_{s_{it}} \in \R^b$ corresponding to the $i$-th coordinate of the input data. $\Omega_{\zeta, b, I} \in \R^{|I| \times k}$ denotes the row submatrix of $\Omega_{\zeta, b}$ indexed by $I$, so that $\Omega_{\zeta, b}^\top z = \Omega_{\zeta, b, I}^\top z_I + \Omega_{\zeta, b, I^c}^\top z_{I^c} \in \R^k$ decomposes the sketch of $z \in \R^n$.

\section{High-Level Overview}\label{sec:overview}

\subsection{Upper Bound Overview}\label{subsec:upper-overview}

In this subsection, we provide an intuitive, high-level overview of the proof of Theorem~\ref{thm:main}, explaining why standard sketching arguments break down in the $\zeta = O(\sqrt{\log r})$ sparsity regime and how our proof overcomes these barriers. Because our proof splits the SparseStack into two equal halves, from this section onward we consider for notational convenience a $2\zeta$-block SparseStack $\Omega_{2\zeta, b} \in \R^{n \times 2k}$ with row sparsity $2\zeta = O(\sqrt{\log r})$ and total embedding dimension $2k = 2\zeta b = \Theta(r)$, so that each half has $\zeta$ blocks and dimension $k = \zeta b$.

\paragraph{Why standard arguments fail: flat versus spiky vectors.}
Our goal is to show that for a fixed $r$-dimensional subspace, the column space of an isometry $Q \in \R^{n \times r}$, the SparseStack sketch matrix $\Omega_{2\zeta, b} \in \R^{n \times 2k}$ with total embedding dimension $2k = 2\zeta b = \Theta(r)$ and row sparsity $2\zeta = O(\sqrt{\log r})$ does not \emph{break injectivity} on any unit vector $y = Qx$ for $x \in S^{r-1} = \{x \in \R^r : \|x\| = 1\}$; meaning that no unit vector in the subspace has its sketched norm $\|\Omega_{2\zeta, b}^\top y\|$ shrink below a constant threshold $\sqrt{\alpha}$. To understand the main difficulty, we distinguish between spiky vectors, that have a large portion of their mass concentrated on $\Theta(r/\zeta)$ coordinates, and flat vectors, which have norm well-distributed.

For a vector $y \in \R^n$, and an integer threshold $m$, define
\begin{equation}\label{eq:dm-def}
  d_m(y) := \min_{\substack{I \subseteq [n] \\ |I| \le m}} \norm{y_{I^c}},
\end{equation}
where $I^c := [n] \setminus I$.

\begin{definition}[Flat and Spiky Vectors]\label{def:flat}
We say a vector $y \in \R^n$ is $(m, \eps)$-\em flat\em, if $d_m(y) \geq \eps$. We say $y$ is $(m, \eps)$-\em spiky \em if $d_m(y) < \eps$. 
\end{definition}
When $m$ and $\eps$ are clear from context, we will just call these flat and spiky vectors respectively.  In our proofs, we will take $m = \Theta(r/\zeta)$.


As seen in~\eqref{eq:spiky-annihilation-prob} and~\eqref{eq:net-discretization-error}, these two regimes cannot be analyzed directly using a union bound in the $2\zeta = O(\sqrt{\log r})$ sparsity regime. Here, our strategy is to split the $2\zeta$-block stack into two independent halves of $\zeta = O(\sqrt{\log r})$ blocks each, inducing a first half $X = (\Omega_{\zeta, b}^{(1)})^\top Q \in \R^{k \times r}$ and an independent second half $Y = (\Omega_{\zeta, b}^{(2)})^\top Q \in \R^{k \times r}$ ($k = \zeta b$), and to analyze flat vectors and spiky vectors through a coordinated two-stage process.

\subsubsection{Preserving flat vectors norm.}
We first show that the first half $X = (\Omega_{\zeta, b}^{(1)})^\top Q$ simultaneously preserves the norm of all flat unit vectors $y = Qx$ with high probability (Theorem~\ref{thm:localization}). For a fixed $x \in S^{r-1}$ with $d_m(y) \ge \eps$, we decompose $y = h + az$ into its top-$m$ coordinates $h$ and its tail $az$, where $a = d_m(y) \ge \eps$ and $\|z\| = 1$. Since each coordinate of $az$ is at most $1/\sqrt{m}$, dividing by $a \ge \eps$ gives $\|z\|_\infty^2 \le \frac{1}{m\eps^2}$. Because every coordinate of $z$ satisfies $|z_i| \le 1/\sqrt{m\eps^2}$ and its random signs are independent of $h$, its norm does not cancel across buckets, yielding in Lemma~\ref{lem:small-ball} for any fixed $x$, 
\[
  \Prb\bracket{\norm{(\Omega_{\zeta, b}^{(1)})^\top Qx} < \frac{\eps}{10}} \le 2\exp\!\paren{-\Omega\paren{\zeta \cdot \min\set{b,\, m\eps^2}}}.
\]
Choosing $m = \Theta(r/\zeta)$ and $k = \zeta b = \Theta(r)$ makes this failure probability for a fixed $x$ at most $e^{-\Omega(r)}$.

To extend this bound from a fixed $x$ to all flat unit vectors $y = Qx$ ($x \in S^{r-1}$) \emph{without} bounding the operator norm $\|X\|$, we show that for any $\eta \in (0, 1)$, one can find a finite point set $\mathcal{N}_\eta$ of size $|\mathcal{N}_\eta| \le (16/\eta)^r$ such that for every $x \in S^{r-1}$, there exists $x' \in \mathcal{N}_\eta$ satisfying
\[
  \norm{x - x'} \le \eta \qquad \text{and} \qquad \norm{X(x - x')}^2 \le \frac{\eta^2}{4r} \norm{X}_F^2
\]
(Lemma~\ref{lem:unbiased-lattice}). Since $\E[\|X\|_F^2] = r$, Markov's inequality ensures $\|X\|_F^2 = O(r)$ with high constant probability, which bounds the approximation error $\|X(x - x')\|$ by $O(\eta)$ without using $\|X\|$. Taking a union bound over all $|\mathcal{N}_\eta| \le (16/\eta)^r$ points where each fixed point of $\mathcal{N}_\eta$ has failure probability $e^{-\Omega(r)}$ yields the following theorem:

\begin{restatable}{theorem}{thmlocalization}\label{thm:localization}
Fix any constant $\eps \in (0, 1)$ and failure parameter $\delta_{\mathrm{loc}} \in (0, 1)$. There exists a constant $\lambda = \Omega(\eps)$ depending only on $\eps$ and $\delta_{\mathrm{loc}}$ such that for every fixed isometry $Q \in \R^{n \times r}$ with $r$ sufficiently large, every $1 \le \zeta \le r$, every $k = \zeta b = \Omega(r)$, and $m = \Theta(r/\zeta)$ with sufficiently large implied constants, the sketch $X = (\Omega_{\zeta, b}^{(1)})^\top Q \in \R^{k \times r}$ satisfies
\[
  \Prb\bracket{\exists x \in S^{r-1} : d_m(Qx) \ge \eps,\ \norm{Xx} < \lambda} \le \delta_{\mathrm{loc}}.
\]
\end{restatable}

\subsubsection{Finding a common support for problematic spiky vectors.}
By Theorem~\ref{thm:localization}, we know that all flat vectors are already well-preserved by $\Omega_{\zeta, b}^{(1)}$, so we only need to consider spiky unit vectors $y = Qx$ satisfying $d_m(y) < \eps$. To preserve the norm of all spiky vectors in $\col(Q)$ simultaneously, it is not enough that each individual spiky vector concentrates on \emph{some} set of $m$ coordinates, since we cannot afford a union bound over all $\binom{n}{m}$ possible $m$-coordinate subsets. Instead, we show that all spiky unit vectors $y = Qx \in \col(Q)$ satisfying $\|Xx\| < \lambda$ must in fact share a \emph{single, common} set $I \subseteq [n]$ of $O(m)$ coordinates where their norm concentrates, avoiding any union bound over candidate supports.


The basic idea is as follows. By maneuvering By Theorem~\ref{thm:localization}, we are able to show that there exists some $H = \eps^2 \Id_r + \lambda^{-2} X^\top X$, such that for all $x$, we have 
\begin{equation}\label{eq:dmH}
    (d_m(Qx))^2 \leq x^{\top}Hx.
\end{equation}
In other words, for each $x$, there exists a set $I$ with $|I| \leq m$ such that $\|Q_{I^c}x\|^2 - x^{\top}Hx \leq 0$. 

We would like to leverage a minimax-style argument to show that there exists some $I$, such that for all $x$, we have $\|Q_{I^c}x\|^2 - x^{\top}Hx \leq 0$. Unfortunately there are two obstacles. 
\begin{enumerate}
    \item The first obstacle is the minimization over $I$ is not over a convex space.
    \item The second obstacle is the the objective function $\|Q_{I^c}x\|^2 - x^{\top}Hx$ is not concave in $x$. 
\end{enumerate}
To resolve the second obstacle, we show that up to the loss of some constant factors, we can attain the guarantee $\|Q_{I^c}x\|^2 - x^{\top}Hx \leq 0$ \em simultaneously with the same $I$ \em for a whole \em distribution \em of $x \sim \mathcal{N}(0, R)$. Formally, using Lemma~\ref{lem:gaussian-trunc}, we attain that 
\begin{equation}
   \min_{|I| \leq 4m} \mathbb{E}_{x \sim R} x^{\top}Q_{I^c}x  - \frac{4}{3}\tr(RH) \leq 0.
\end{equation}
Notice that the objective above in linear in $R$. Further, by relaxing from minimization over $I$ to minimization over the convex set of fractional weight vector $w \in [0, 1]^n$ with $\sum w_i \leq 4m$.

We can now use minimax duality over $\mathcal{R}$ and $w$, and ultimately round back that resulting $w^*$ to find a single $I$ with the desired property.

Formally we prove the following result.

\begin{restatable}{theorem}{propenergycert}\label{prop:energy-cert}
Let $Q \in \R^{n \times r}$ be an isometry. Suppose a matrix $X \in \R^{k \times r}$ satisfies
\[
  d_m(Qx) \ge \eps, \quad \norm{x} = 1 \quad \Longrightarrow \quad \norm{Xx} \ge \lambda
\]
for some $\eps > 0$ and $\lambda > 0$. Then there exists a coordinate set $I \subseteq [n]$ with $|I| \le 8m$ such that
\[
  Q_{I^c}^\top Q_{I^c} \preceq \frac{8}{3}\eps^2 \Id_r + \frac{8}{3\lambda^2} X^\top X.
\]
\end{restatable}

This theorem implies the following: since $\|(Qx)_I\|^2 + \|(Qx)_{I^c}\|^2 = \|Qx\|^2 = 1$ for every unit vector $x \in S^{r-1}$, we have
\[
  \norm{(Qx)_I}^2 = 1 - \norm{(Qx)_{I^c}}^2 \ge 1 - \frac{8}{3}\eps^2 - \frac{8}{3\lambda^2}\norm{Xx}^2.
\]
In particular, choosing $\eps = 1/3$ shows that for every unit vector $x$ whose sketch norm $\|Xx\|$ is small, approximately at least $\frac{19}{27}$ of the squared norm of $Qx$ is guaranteed to reside on the single coordinate set $I$ of size $|I| \le 8m = O(r/\zeta)$.

\subsubsection{Preserving the heavy coordinates in the second half.}
Combining Theorem~\ref{thm:localization} and Theorem~\ref{prop:energy-cert}, the first half $(\Omega_{\zeta, b}^{(1)})^\top $ with high probability preserves all flat vectors $y = Qx$ and localizes the norm of all spiky vectors $y = Qx$ on which injectivity fails under $(\Omega_{\zeta, b}^{(1)})^\top$ (i.e., with small $\|(\Omega_{\zeta, b}^{(1)})^\top y\|$) onto a single small coordinate set $I \subseteq [n]$ of size $N := |I| \le 8m = O(r/\zeta)$. Therefore, the remaining task is to show that whenever $\|(\Omega_{\zeta, b}^{(1)})^\top y\|$ is small, the independent second half $(\Omega_{\zeta, b}^{(2)})^\top$ preserves the norm of $y$, ensuring that injectivity is not broken on $y$ under the full sketch. 

Why is it easier to prove that $\Omega_{\zeta, b}^{(2)}$ handles these spiky vectors even though $\Omega_{\zeta, b}^{(1)}$ and $\Omega_{\zeta, b}^{(2)}$ share the same distribution? The reason is that we have now restricted the spiky coordinates from all $n$ coordinates down to a single known set $I$ of size $N = O(m)$. Using the shorthand $\Omega := \Omega_{\zeta, b}^{(2)}$, we can expand the squared sketch norm $\|\Omega_{\zeta, b}^{(2)}y\|^2$ as follows: 

\begin{equation}\label{eq:foil}
\begin{aligned}
    \|(\Omega_{\zeta, b}^{(2)})^{\top}y\|^2 &=  \norm{\Omega_{I}^{\top}y_I + \Omega_{I^c}^{\top}Q_{I^c}x}^2\\
    &= \norm{\Omega_{I}^{\top}y_I}^2 + 2y_I^\top (\Omega_{I} \Omega_{I^c}^{\top} Q_{I^c})x + \norm{\Omega_{I^c}^{\top}Q_{I^c}x}^2. 
\end{aligned}
\end{equation}

\paragraph{Controlling the cross term of collisions between $I$ and $I^c$.} In Section~\ref{sec:coupling}, we will show that [a small modification of] this cross term in \eqref{eq:foil} is smaller than a sufficiently small constant. To do this, we will bound the operator norm of $\Omega_{I} \Omega_{ I^c}^{\top} Q_{I^c}$ using Lemma~\ref{lem:cross-gram} below. The intuition behind Lemma~\ref{lem:cross-gram} is that the matrix $\Omega_{ I}\Omega_{I^c}^{\top} \in \R^{|I| \times |I^c|}$ has entries which are are non-zero with probability roughly $\Theta(\zeta^2/r)$. Indeed, $\left(\Omega_{I} \Omega_{I^c}^{\top}\right)_{ij}$ is a dot-product between two independent, $\zeta$-sparse vectors in $\R^k$, so they have a non-zero dot product with probability roughly $\Theta(\zeta^2/k) = \Theta(\zeta^2/r)$. If $\zeta = \Omega(\sqrt{\log(r)})$, and thus the non-zero probability is $\Omega(\log(r)/r)$, we can expect universal behavior from $\Omega_{I} \Omega_{I^c}^{\top}$. Leveraging a result of \cite{Heidary26} for OSEs with entry non-zero probability $\Omega(\log(r)/r)$, we show that [a slight modification of] the cross-term matrix $\Omega_{I} \Omega_{I^c}^{\top}Q_{I^c}$ has constant operator norm in expectation. 

Of course, because the entries of $\Omega_{I} \Omega_{I^c}^{\top}$ are not independent, proving this requires a careful coupling argument to a suitable matrix with i.i.d. entries. We leave the details of this argument to Section~\ref{sec:coupling}. We state the main result of the section here.

\begin{restatable}{lemma}{crossgram}\label{lem:cross-gram}
  There exists a universal constant $C_{\mathrm{cg}} > 0$ such that the following holds. Let $\Omega \in \R^{n \times k}$ ($k = \zeta b$) be a SparseStack sketch with $\zeta \ge 1$ blocks and $b \ge 2\zeta$ buckets per block. Let $I, J \subseteq [n]$ be disjoint coordinate sets with $|I| = N \ge 1$, and let $W \in \R^{|J| \times d}$ ($d \ge 1$) be a fixed isometry. 
Let $T \subseteq [k]$ be the set of rows in which $\Omega_{I}^\top$ has exactly one non-zero entry, and let $\widetilde{\Omega}^{\top}$ denote the restrictions of $\Omega^{\top}$ to rows in $T$.

For every integer $p \ge \max\{1, \log(2d)\}$, setting $D_p = d + 2p + 1$, we have conditionally on $\Omega_{I}$:
\[
  \E\bracket{\norm{\widetilde{\Omega}_I\widetilde{\Omega}_{J}^{\top}W} \;\Big|\; \Omega_{I}} \le C_{\mathrm{cg}} \paren{\sqrt{\frac{N + D_p}{k} + \frac{p+1}{\zeta^2}} + \sqrt{\frac{N d}{\zeta^2 b^3} \binom{\zeta}{3}}}.
\]
If $N = 0$ or $d = 0$, then the expression is trivially $0$.
\end{restatable}

\paragraph{Avoiding collisions among heavy coordinates in $I$} In Section~\ref{sec:repair}, we show that the term $\norm{\Omega_{I}^{\top}y_I}^2$ is large. Here we leverage the fact that $N = |I| = O(r/\zeta)$ to show that the columns of $\Omega_{\zeta}^{\top}$ do not have many collisions, and thus most entries in $y_I$ are directly copied to $\zeta$ entries in $\Omega_{I}^{\top}y_I$ (up to a sign change, and a factor of $1/\sqrt{\zeta}$).

Indeed, we can show that any coordinate $i \in I$ lands in a bucket containing no other element of $I$ with probability at least $1 - N/b \ge 15/16$. When a coordinate is isolated in its bucket, no other coordinate of $I$ can cancel its contribution. If a constant fraction of the buckets that $i$ lands in contain no other element of $I$, then a constant fraction of the norm of $y_i$ will be represented in the output. We say in this case that the element $i$ is \em good \em (see Definition~\ref{def:good}).

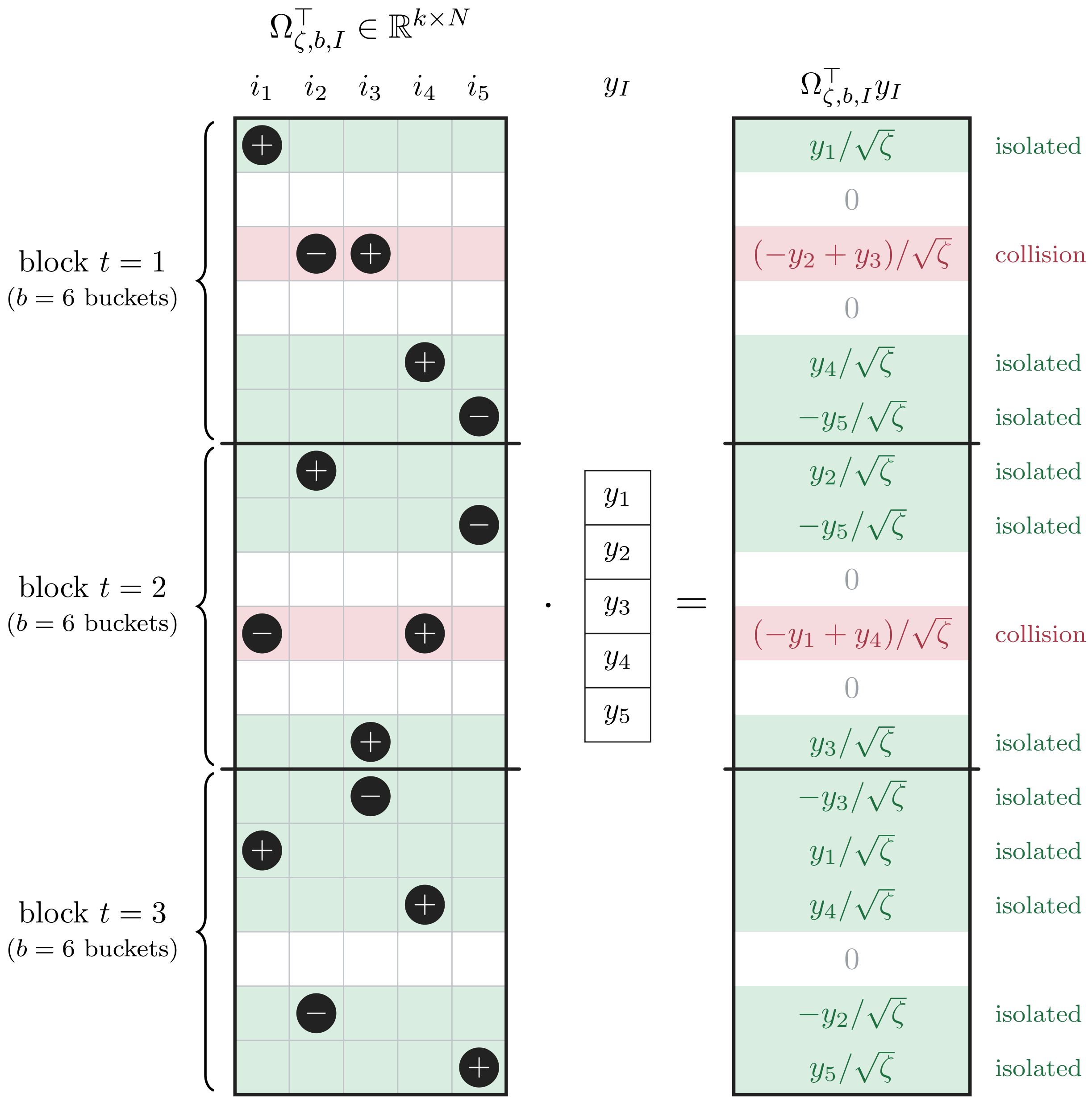
\begin{figure}[htbp]
  \centering
  \vspace{1.5em}
  \resizebox{\textwidth}{!}{%
  \begin{tikzpicture}[font=\small, line cap=round]

  \fill[isoL] (0.0,0.0) rectangle (3.1,-0.62);
  \fill[isoL] (5.7,0.0) rectangle (8.4,-0.62);
  \node[iso!70!black] at (7.050000000000001,-0.31) {$y_{1}/\sqrt{\zeta}$};
  \node[iso!70!black,font=\scriptsize,anchor=west] at (8.55,-0.31) {isolated};
  \fill[white] (0.0,-0.62) rectangle (3.1,-1.24);
  \fill[white] (5.7,-0.62) rectangle (8.4,-1.24);
  \node[cmp] at (7.050000000000001,-0.9299999999999999) {$0$};
  \fill[colL] (0.0,-1.24) rectangle (3.1,-1.8599999999999999);
  \fill[colL] (5.7,-1.24) rectangle (8.4,-1.8599999999999999);
  \node[col!80!black] at (7.050000000000001,-1.55) {$(-y_{2}+y_{3})/\sqrt{\zeta}$};
  \node[col!80!black,font=\scriptsize,anchor=west] at (8.55,-1.55) {collision};
  \fill[white] (0.0,-1.8599999999999999) rectangle (3.1,-2.48);
  \fill[white] (5.7,-1.8599999999999999) rectangle (8.4,-2.48);
  \node[cmp] at (7.050000000000001,-2.17) {$0$};
  \fill[isoL] (0.0,-2.48) rectangle (3.1,-3.1);
  \fill[isoL] (5.7,-2.48) rectangle (8.4,-3.1);
  \node[iso!70!black] at (7.050000000000001,-2.79) {$y_{4}/\sqrt{\zeta}$};
  \node[iso!70!black,font=\scriptsize,anchor=west] at (8.55,-2.79) {isolated};
  \fill[isoL] (0.0,-3.1) rectangle (3.1,-3.72);
  \fill[isoL] (5.7,-3.1) rectangle (8.4,-3.72);
  \node[iso!70!black] at (7.050000000000001,-3.41) {$-y_{5}/\sqrt{\zeta}$};
  \node[iso!70!black,font=\scriptsize,anchor=west] at (8.55,-3.41) {isolated};
  \fill[isoL] (0.0,-3.7199999999999998) rectangle (3.1,-4.34);
  \fill[isoL] (5.7,-3.7199999999999998) rectangle (8.4,-4.34);
  \node[iso!70!black] at (7.050000000000001,-4.029999999999999) {$y_{2}/\sqrt{\zeta}$};
  \node[iso!70!black,font=\scriptsize,anchor=west] at (8.55,-4.029999999999999) {isolated};
  \fill[isoL] (0.0,-4.34) rectangle (3.1,-4.96);
  \fill[isoL] (5.7,-4.34) rectangle (8.4,-4.96);
  \node[iso!70!black] at (7.050000000000001,-4.6499999999999995) {$-y_{5}/\sqrt{\zeta}$};
  \node[iso!70!black,font=\scriptsize,anchor=west] at (8.55,-4.6499999999999995) {isolated};
  \fill[white] (0.0,-4.96) rectangle (3.1,-5.58);
  \fill[white] (5.7,-4.96) rectangle (8.4,-5.58);
  \node[cmp] at (7.050000000000001,-5.27) {$0$};
  \fill[colL] (0.0,-5.58) rectangle (3.1,-6.2);
  \fill[colL] (5.7,-5.58) rectangle (8.4,-6.2);
  \node[col!80!black] at (7.050000000000001,-5.89) {$(-y_{1}+y_{4})/\sqrt{\zeta}$};
  \node[col!80!black,font=\scriptsize,anchor=west] at (8.55,-5.89) {collision};
  \fill[white] (0.0,-6.2) rectangle (3.1,-6.82);
  \fill[white] (5.7,-6.2) rectangle (8.4,-6.82);
  \node[cmp] at (7.050000000000001,-6.51) {$0$};
  \fill[isoL] (0.0,-6.82) rectangle (3.1,-7.44);
  \fill[isoL] (5.7,-6.82) rectangle (8.4,-7.44);
  \node[iso!70!black] at (7.050000000000001,-7.13) {$y_{3}/\sqrt{\zeta}$};
  \node[iso!70!black,font=\scriptsize,anchor=west] at (8.55,-7.13) {isolated};
  \fill[isoL] (0.0,-7.4399999999999995) rectangle (3.1,-8.059999999999999);
  \fill[isoL] (5.7,-7.4399999999999995) rectangle (8.4,-8.059999999999999);
  \node[iso!70!black] at (7.050000000000001,-7.749999999999999) {$-y_{3}/\sqrt{\zeta}$};
  \node[iso!70!black,font=\scriptsize,anchor=west] at (8.55,-7.749999999999999) {isolated};
  \fill[isoL] (0.0,-8.06) rectangle (3.1,-8.68);
  \fill[isoL] (5.7,-8.06) rectangle (8.4,-8.68);
  \node[iso!70!black] at (7.050000000000001,-8.370000000000001) {$y_{1}/\sqrt{\zeta}$};
  \node[iso!70!black,font=\scriptsize,anchor=west] at (8.55,-8.370000000000001) {isolated};
  \fill[isoL] (0.0,-8.68) rectangle (3.1,-9.299999999999999);
  \fill[isoL] (5.7,-8.68) rectangle (8.4,-9.299999999999999);
  \node[iso!70!black] at (7.050000000000001,-8.99) {$y_{4}/\sqrt{\zeta}$};
  \node[iso!70!black,font=\scriptsize,anchor=west] at (8.55,-8.99) {isolated};
  \fill[white] (0.0,-9.3) rectangle (3.1,-9.92);
  \fill[white] (5.7,-9.3) rectangle (8.4,-9.92);
  \node[cmp] at (7.050000000000001,-9.610000000000001) {$0$};
  \fill[isoL] (0.0,-9.92) rectangle (3.1,-10.54);
  \fill[isoL] (5.7,-9.92) rectangle (8.4,-10.54);
  \node[iso!70!black] at (7.050000000000001,-10.23) {$-y_{2}/\sqrt{\zeta}$};
  \node[iso!70!black,font=\scriptsize,anchor=west] at (8.55,-10.23) {isolated};
  \fill[isoL] (0.0,-10.54) rectangle (3.1,-11.159999999999998);
  \fill[isoL] (5.7,-10.54) rectangle (8.4,-11.159999999999998);
  \node[iso!70!black] at (7.050000000000001,-10.85) {$y_{5}/\sqrt{\zeta}$};
  \node[iso!70!black,font=\scriptsize,anchor=west] at (8.55,-10.85) {isolated};
  \draw[cmp!60,step=0.62] (0.0,-11.16) grid (3.1,0.0);
  \draw[cmp!60] (5.7,-11.16) rectangle (8.4,0.0);
  \draw[ink,line width=1.1pt] (-0.15,-3.7199999999999998) -- (3.25,-3.7199999999999998);
  \draw[ink,line width=1.1pt] (5.6000000000000005,-3.7199999999999998) -- (8.5,-3.7199999999999998);
  \draw[ink,line width=1.1pt] (-0.15,-7.4399999999999995) -- (3.25,-7.4399999999999995);
  \draw[ink,line width=1.1pt] (5.6000000000000005,-7.4399999999999995) -- (8.5,-7.4399999999999995);
  \draw[ink,line width=1.1pt] (0.0,-11.16) rectangle (3.1,0.0);
  \draw[ink,line width=1.1pt] (5.7,-11.16) rectangle (8.4,0.0);
  \node[circle,fill=ink,inner sep=0pt,minimum size=13pt,text=white,font=\bfseries\small] at (0.31,-0.31) {$+$};
  \node[circle,fill=ink,inner sep=0pt,minimum size=13pt,text=white,font=\bfseries\small] at (0.9299999999999999,-1.55) {$-$};
  \node[circle,fill=ink,inner sep=0pt,minimum size=13pt,text=white,font=\bfseries\small] at (1.55,-1.55) {$+$};
  \node[circle,fill=ink,inner sep=0pt,minimum size=13pt,text=white,font=\bfseries\small] at (2.17,-2.79) {$+$};
  \node[circle,fill=ink,inner sep=0pt,minimum size=13pt,text=white,font=\bfseries\small] at (2.79,-3.41) {$-$};
  \node[circle,fill=ink,inner sep=0pt,minimum size=13pt,text=white,font=\bfseries\small] at (0.31,-5.89) {$-$};
  \node[circle,fill=ink,inner sep=0pt,minimum size=13pt,text=white,font=\bfseries\small] at (0.9299999999999999,-4.029999999999999) {$+$};
  \node[circle,fill=ink,inner sep=0pt,minimum size=13pt,text=white,font=\bfseries\small] at (1.55,-7.13) {$+$};
  \node[circle,fill=ink,inner sep=0pt,minimum size=13pt,text=white,font=\bfseries\small] at (2.17,-5.89) {$+$};
  \node[circle,fill=ink,inner sep=0pt,minimum size=13pt,text=white,font=\bfseries\small] at (2.79,-4.6499999999999995) {$-$};
  \node[circle,fill=ink,inner sep=0pt,minimum size=13pt,text=white,font=\bfseries\small] at (0.31,-8.370000000000001) {$+$};
  \node[circle,fill=ink,inner sep=0pt,minimum size=13pt,text=white,font=\bfseries\small] at (0.9299999999999999,-10.23) {$-$};
  \node[circle,fill=ink,inner sep=0pt,minimum size=13pt,text=white,font=\bfseries\small] at (1.55,-7.749999999999999) {$-$};
  \node[circle,fill=ink,inner sep=0pt,minimum size=13pt,text=white,font=\bfseries\small] at (2.17,-8.99) {$+$};
  \node[circle,fill=ink,inner sep=0pt,minimum size=13pt,text=white,font=\bfseries\small] at (2.79,-10.85) {$+$};
  \node at (0.31,0.35) {$i_{1}$};
  \node at (0.9299999999999999,0.35) {$i_{2}$};
  \node at (1.55,0.35) {$i_{3}$};
  \node at (2.17,0.35) {$i_{4}$};
  \node at (2.79,0.35) {$i_{5}$};
  \node[font=\normalsize] at (1.55,1.0) {$\Omega_{\zeta,b,I}^\top \in \mathbb{R}^{k\times N}$};
  \draw[decorate,decoration={brace,amplitude=5pt,mirror},thick] (-0.25,-0.05) -- (-0.25,-3.6700000000000004) node[midway,left=7pt,align=center] {block $t=1$\\[-1pt]\scriptsize ($b=6$ buckets)};
  \draw[decorate,decoration={brace,amplitude=5pt,mirror},thick] (-0.25,-3.7699999999999996) -- (-0.25,-7.390000000000001) node[midway,left=7pt,align=center] {block $t=2$\\[-1pt]\scriptsize ($b=6$ buckets)};
  \draw[decorate,decoration={brace,amplitude=5pt,mirror},thick] (-0.25,-7.489999999999999) -- (-0.25,-11.109999999999998) node[midway,left=7pt,align=center] {block $t=3$\\[-1pt]\scriptsize ($b=6$ buckets)};
  \node[font=\Large] at (3.58,-5.58) {$\cdot$};
  \draw[ink] (4.0,-4.03) rectangle (4.75,-4.65);
  \node at (4.375,-4.34) {$y_{1}$};
  \draw[ink] (4.0,-4.65) rectangle (4.75,-5.2700000000000005);
  \node at (4.375,-4.96) {$y_{2}$};
  \draw[ink] (4.0,-5.2700000000000005) rectangle (4.75,-5.890000000000001);
  \node at (4.375,-5.58) {$y_{3}$};
  \draw[ink] (4.0,-5.890000000000001) rectangle (4.75,-6.51);
  \node at (4.375,-6.2) {$y_{4}$};
  \draw[ink] (4.0,-6.51) rectangle (4.75,-7.130000000000001);
  \node at (4.375,-6.819999999999999) {$y_{5}$};
  \node at (4.375,0.35) {$y_I$};
  \node[font=\Large] at (5.220000000000001,-5.58) {$=$};
  \node at (7.050000000000001,0.35) {$\Omega_{\zeta,b,I}^\top y_I$};
  \end{tikzpicture}
  }
  \caption{Isolated buckets: a lone coordinate is copied to the output; colliding ones may cancel.}
  \label{fig:isolated-buckets}
  \vspace{1em}
\end{figure}

For each $i$, it is easy to show using standard concentration tools that with probability $1 - e^{-\Omega(\zeta)}$, element $i$ is good. Unfortunately, this is not large enough to guarantee via a union bound that all elements in $I$ are good. Thus it is still possible that the mass of $y_I$ concentrates on the elements that are not good, and that this mass gets canceled by other elements of $I$.

To bypass this union-bound barrier, we divide the $\zeta$ blocks of the second half into two disjoint groups, Group 1 and Group 2, of $\zeta_0 = \zeta/2 = O(\sqrt{\log r})$ blocks each:
\begin{enumerate}[label=(\arabic*)]
  \item \emph{Group 1:} Let $G \subseteq I$ be the \textit{good} coordinates isolated in at least $\frac{11}{12}\zeta_0$ blocks of Group 1, and let $B = I \setminus G$ be the remaining \textit{bad} coordinates. Each coordinate of $I$ falls into $B$ with probability $e^{-\Omega(\zeta)}$, so by Markov's inequality the bad set shrinks to $|B| \le N e^{-\Omega(\zeta)}$ with high probability, while all coordinates in $G$ are already preserved by Group 1.
  \item \emph{Group 2:} Conditioned on Group 1, the good coordinates $G$ are already preserved by Group 1, so we can absorb $G$ into the complement $I^c$, whose cancellation effect on $B$ is negligible by the exact same argument from Lemma~\ref{lem:cross-gram} as for $I^c$! This is the decisive factor that makes a union bound possible in Group 2 even though it failed before. Thus, in Group 2, the surviving bad coordinates $B$ only need to avoid colliding with each other. Since only $|B| \le N e^{-\Omega(\zeta)}$ bad coordinates remain across the $b$ buckets, the per-block collision probability within $B$ drops from $O(1)$ to $|B|/b \le e^{-\Omega(\zeta)}/16$. Thus, a coordinate $i \in B$ fails to be isolated in $\frac{11}{12}\zeta_0$ blocks of Group 2 with probability at most
  \[
    2^{\zeta_0} \paren{\frac{e^{-\Omega(\zeta)}}{16}}^{\zeta_0/12} = e^{-\Omega(\zeta^2)} = e^{-\Omega(\log r)} \ll \frac{1}{r},
  \]
  which is now small enough to union bound over all coordinates in $B$.
\end{enumerate}

Notice that the argument above bounding the cancellation from the complement only uses $\sum_{j=1}^n \|q_j\|^2 = \|Q\|_F^2 = r$ and never required the target coordinate set to carry a constant fraction of the total norm. This is why the exact same argument applies to Group 2 even though $B$ is not guaranteed to carry a constant fraction of the norm on its own. The condition that $I$ carries a large fraction of the norm is only used at the very end when we sum the squared sketch norms from Group 1, which preserves $(Qx)_G$, and Group 2, which preserves $(Qx)_B$, to recover $\|(Qx)_G\|^2 + \|(Qx)_B\|^2 = \|(Qx)_I\|^2$. This yields the following end-to-end guarantee for the second half of the sketch.

\begin{restatable}{theorem}{thmrepair}\label{thm:repair}
Fix any constants $t_0 \in (0, 1/4)$ and $\delta_{\mathrm{rep}} \in (0, 1)$. There exists a constant $\beta = \Omega(1)$ such that for every fixed isometry $Q \in \R^{n \times r}$ with $r \ge 2$, every integer pair $\zeta = \Omega(\sqrt{\log r})$ and $k = \zeta b = \Omega(r)$, and every coordinate set $I \subseteq [n]$ of size $\abs{I} \le \min\set{r,\, b/16}$, the sketch $Y = (\Omega_{\zeta, b}^{(2)})^\top Q \in \R^{k \times r}$ satisfies, with probability at least $1 - \delta_{\mathrm{rep}}$,
\[
  \norm{Yx}^2 \ge \beta \norm{(Qx)_I}^2 - t_0 \norm{(Qx)_{I^c}}^2 \qquad \text{for all } x \in \R^r,
\]
or equivalently, $Y^\top Y + (\beta + t_0) Q_{I^c}^\top Q_{I^c} \succeq \beta \Id_r$.
\end{restatable}

With Theorem~\ref{prop:energy-cert} from the first half $X$ and Theorem~\ref{thm:repair} from the second half $Y$ in hand, the proof of Theorem~\ref{thm:main} concludes in three lines: Theorem~\ref{thm:repair} gives $Y^\top Y + (\beta + t_0) Q_{I^c}^\top Q_{I^c} \succeq \beta \Id_r$, while Theorem~\ref{prop:energy-cert} bounds the norm on $I^c$ by $Q_{I^c}^\top Q_{I^c} \preceq u \Id_r + v X^\top X$ with $u = \frac{8}{3}\eps^2 < \frac{\beta}{\beta + t_0}$. Substituting the latter into the former immediately gives $Y^\top Y + (\beta + t_0) v X^\top X \succeq \paren{\beta - (\beta + t_0) u}\Id_r = \Omega(\Id_r)$. Since the full sketch $A = \Omega_{2\zeta, b}^\top Q \in \R^{2k \times r}$ is formed by stacking the two halves $X$ and $Y$ via $A = \frac{1}{\sqrt{2}}\begin{bmatrix} X \\ Y \end{bmatrix}$ so that $A^\top A = \frac{1}{2}(X^\top X + Y^\top Y)$, this positive lower bound on a weighted sum of $X^\top X$ and $Y^\top Y$ immediately implies $A^\top A \succeq \Omega(\Id_r)$, completing the proof.

\begin{remark}
Because our analysis divides $\Omega_{\zeta, b}$ into two halves, we are only able to show a constant lower bound on the smallest singular value of $\Omega_{\zeta, b}^{\top}Q$, and not a $1- \eps$ lower bound. We believe this to be an artifact of the proof, and not a fundamental obstruction. 
\end{remark}

\paragraph{Why $\zeta = O(\sqrt{\log r})$ is the natural scale.}
With the full two-stage architecture in view, we can now see why $\zeta = O(\sqrt{\log r})$ is the natural scale for constant injectivity. In the first half, spiky vectors in $\col(Q)$ are localized onto a single coordinate set $I$ of size $N = O(r/\zeta)$. In the second half, dividing the $\zeta$ blocks into Group~1 and Group~2 ensures that most coordinates of $I$ avoid collisions in Group~1, leaving only $|B| \le N e^{-\Omega(\zeta)}$ un-isolated coordinates. In Group~2, because the number of surviving coordinates has shrunk to $|B| \le N e^{-\Omega(\zeta)}$, their mutual collision probability per block drops from $O(1)$ to $|B|/b \le e^{-\Omega(\zeta)}/16$, suppressing the failure probability of Group~2 to $e^{-\Omega(\zeta^2)}$. Setting the row sparsity to $\zeta = O(\sqrt{\log r})$ (specifically $\zeta \ge C\sqrt{\log(er)}$) is precisely the scale where this failure probability becomes $e^{-\Omega(\log r)} \ll 1/r$, allowing the union bound over all surviving coordinates in $B$ to succeed. 

Another lens to see this is the cross term argument of Section~\ref{sec:coupling}. Here we argued that the cross term, governing cancellations between $I$ and $I^c$, was small because the operator norm of the matrix $\Omega_{\zeta, b, I}^{(2)}(\Omega_{\zeta, b, I^c}^{(2)})^{\top}Q$ was bounded by a sufficiently small constant. This used the fact that the entries of $\Omega_{\zeta, b, I}^{(2)}(\Omega_{\zeta, b, I^c}^{(2)})^{\top}$ were almost i.i.d., with non-zero probabilty of $\frac{\zeta^2}{r}$, which is above the universality threshold $\frac{\log(r)}{r}$ whenever $\zeta = \Omega(\sqrt{\log(r)})$.


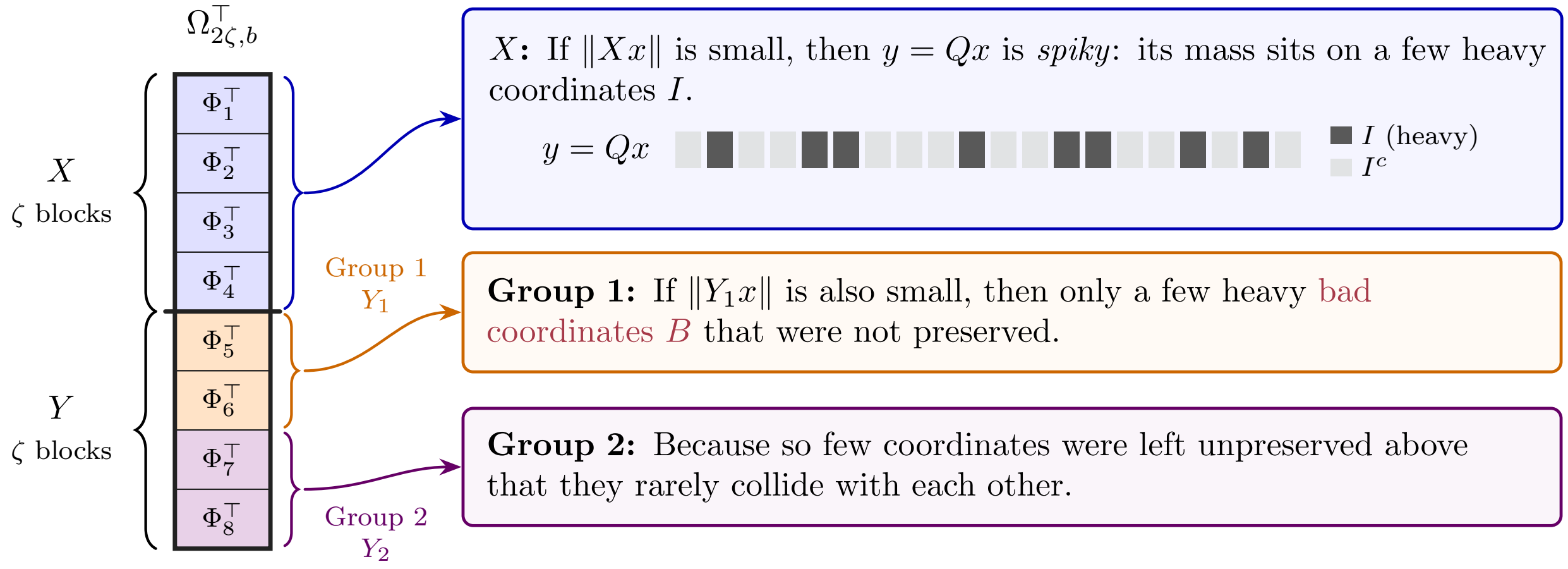
\begin{figure}[htbp]
  \centering
  \vspace{1.5em}
  \resizebox{\textwidth}{!}{%
  \begin{tikzpicture}[font=\small, line cap=round]
  \fill[blue!12] (0.0,1.6) rectangle (1.0,0.9800000000000001);
  \draw[ink] (0.0,1.6) rectangle (1.0,0.9800000000000001);
  \node[font=\scriptsize] at (0.5,1.29) {$\Phi_{1}^\top$};
  \fill[blue!12] (0.0,0.9800000000000001) rectangle (1.0,0.3600000000000001);
  \draw[ink] (0.0,0.9800000000000001) rectangle (1.0,0.3600000000000001);
  \node[font=\scriptsize] at (0.5,0.6700000000000002) {$\Phi_{2}^\top$};
  \fill[blue!12] (0.0,0.3600000000000001) rectangle (1.0,-0.2599999999999999);
  \draw[ink] (0.0,0.3600000000000001) rectangle (1.0,-0.2599999999999999);
  \node[font=\scriptsize] at (0.5,0.0500000000000001) {$\Phi_{3}^\top$};
  \fill[blue!12] (0.0,-0.2599999999999998) rectangle (1.0,-0.8799999999999998);
  \draw[ink] (0.0,-0.2599999999999998) rectangle (1.0,-0.8799999999999998);
  \node[font=\scriptsize] at (0.5,-0.5699999999999998) {$\Phi_{4}^\top$};
  \fill[orange!22] (0.0,-0.8799999999999999) rectangle (1.0,-1.5);
  \draw[ink] (0.0,-0.8799999999999999) rectangle (1.0,-1.5);
  \node[font=\scriptsize] at (0.5,-1.19) {$\Phi_{5}^\top$};
  \fill[orange!22] (0.0,-1.5) rectangle (1.0,-2.12);
  \draw[ink] (0.0,-1.5) rectangle (1.0,-2.12);
  \node[font=\scriptsize] at (0.5,-1.81) {$\Phi_{6}^\top$};
  \fill[violet!18] (0.0,-2.1199999999999997) rectangle (1.0,-2.7399999999999998);
  \draw[ink] (0.0,-2.1199999999999997) rectangle (1.0,-2.7399999999999998);
  \node[font=\scriptsize] at (0.5,-2.4299999999999997) {$\Phi_{7}^\top$};
  \fill[violet!18] (0.0,-2.7399999999999998) rectangle (1.0,-3.36);
  \draw[ink] (0.0,-2.7399999999999998) rectangle (1.0,-3.36);
  \node[font=\scriptsize] at (0.5,-3.05) {$\Phi_{8}^\top$};
  \draw[ink,line width=1.3pt] (0.0,1.6) rectangle (1.0,-3.36);
  \draw[ink,line width=1.3pt] (-0.1,-0.8799999999999999) -- (1.1,-0.8799999999999999);
  \node at (0.5,2.1) {$\Omega_{2\zeta,b}^\top$};
  \draw[decorate,decoration={brace,amplitude=6pt,mirror},thick] (-0.2,1.6) -- (-0.2,-0.8799999999999999) node[midway,left=9pt,align=center] {$X$\\\scriptsize $\zeta$ blocks};
  \draw[decorate,decoration={brace,amplitude=6pt,mirror},thick] (-0.2,-0.8799999999999999) -- (-0.2,-3.36) node[midway,left=9pt,align=center] {$Y$\\\scriptsize $\zeta$ blocks};
  \draw[decorate,decoration={brace,amplitude=5pt},thick,blue!70!black] (1.15,1.57) -- (1.15,-0.8499999999999999) coordinate[midway,xshift=6pt] (xx);
  \draw[decorate,decoration={brace,amplitude=4pt},thick,orange!80!black] (1.15,-0.9099999999999999) -- (1.15,-2.09) coordinate[midway,xshift=6pt] (g1);
  \draw[decorate,decoration={brace,amplitude=4pt},thick,violet!80!black] (1.15,-2.1499999999999995) -- (1.15,-3.33) coordinate[midway,xshift=6pt] (g2);
  \node[draw,rounded corners=3pt,line width=0.9pt,inner sep=6pt,anchor=north west,draw=blue!70!black,fill=blue!4,minimum width=11.5cm,minimum height=2.3cm] (BX) at (3.0,2.3) {};
  \node[anchor=north west,text width=11.1cm,align=left] at (3.15,2.15) {\textbf{$X$:} If $\|Xx\|$ is small, then $y=Qx$ is \emph{spiky}: its mass sits on a few heavy coordinates $I$.};
  \begin{scope}[xshift=0.81cm, local bounding box=strip]
  \fill[cmp!30] (4.430000000000001,0.9999999999999998) rectangle (4.7,0.6199999999999998);
  \fill[ink!75] (4.760000000000001,0.9999999999999998) rectangle (5.03,0.6199999999999998);
  \fill[cmp!30] (5.090000000000001,0.9999999999999998) rectangle (5.36,0.6199999999999998);
  \fill[cmp!30] (5.420000000000001,0.9999999999999998) rectangle (5.69,0.6199999999999998);
  \fill[ink!75] (5.750000000000001,0.9999999999999998) rectangle (6.0200000000000005,0.6199999999999998);
  \fill[ink!75] (6.080000000000001,0.9999999999999998) rectangle (6.3500000000000005,0.6199999999999998);
  \fill[cmp!30] (6.410000000000001,0.9999999999999998) rectangle (6.680000000000001,0.6199999999999998);
  \fill[cmp!30] (6.740000000000001,0.9999999999999998) rectangle (7.010000000000001,0.6199999999999998);
  \fill[cmp!30] (7.070000000000001,0.9999999999999998) rectangle (7.340000000000001,0.6199999999999998);
  \fill[ink!75] (7.400000000000001,0.9999999999999998) rectangle (7.670000000000001,0.6199999999999998);
  \fill[cmp!30] (7.730000000000001,0.9999999999999998) rectangle (8.000000000000002,0.6199999999999998);
  \fill[cmp!30] (8.06,0.9999999999999998) rectangle (8.33,0.6199999999999998);
  \fill[ink!75] (8.389999999999999,0.9999999999999998) rectangle (8.660000000000002,0.6199999999999998);
  \fill[ink!75] (8.72,0.9999999999999998) rectangle (8.99,0.6199999999999998);
  \fill[cmp!30] (9.049999999999999,0.9999999999999998) rectangle (9.320000000000002,0.6199999999999998);
  \fill[cmp!30] (9.38,0.9999999999999998) rectangle (9.65,0.6199999999999998);
  \fill[ink!75] (9.709999999999999,0.9999999999999998) rectangle (9.980000000000002,0.6199999999999998);
  \fill[cmp!30] (10.040000000000001,0.9999999999999998) rectangle (10.31,0.6199999999999998);
  \fill[ink!75] (10.37,0.9999999999999998) rectangle (10.640000000000002,0.6199999999999998);
  \fill[cmp!30] (10.700000000000001,0.9999999999999998) rectangle (10.97,0.6199999999999998);
  \node[anchor=east] at (4.300000000000001,0.8099999999999998) {$y=Qx$};
  \node[anchor=west,align=left,font=\scriptsize] at (11.15,0.8099999999999998) {\tikz\fill[ink!75](0,0)rectangle(.22,.18); $I$ (heavy)\\\tikz\fill[cmp!30](0,0)rectangle(.22,.18); $I^c$};
  \end{scope}
  \draw[-{Stealth},thick,blue!70!black] (xx) to[out=0,in=180] (BX.west);
  \node[draw,rounded corners=3pt,line width=0.9pt,inner sep=7pt,anchor=north west,text width=11.0cm,align=left,execute at begin node={\hyphenpenalty=10000},draw=orange!80!black,fill=orange!4] (B1) at (3.0,-0.25) {\textbf{Group 1:} If $\|Y_1x\|$ is also small, then only a few heavy {\color{col!80!black}bad coordinates $B$} that were not preserved.};
  \draw[-{Stealth},thick,orange!80!black] (g1) to[out=0,in=180] node[pos=0.45,above=6pt,font=\scriptsize,orange!80!black,align=center] {Group 1\\$Y_1$} (B1.west);
  \node[draw,rounded corners=3pt,line width=0.9pt,inner sep=7pt,anchor=north west,text width=11.0cm,align=left,execute at begin node={\hyphenpenalty=10000},draw=violet!80!black,fill=violet!4] (B2) at ($(B1.south west)+(0,-0.35)$) {\textbf{Group 2:} Because so few coordinates were left unpreserved above that they rarely collide with each other.};
  \draw[-{Stealth},thick,violet!80!black] (g2) to[out=0,in=180] node[pos=0.45,below=4pt,font=\scriptsize,violet!80!black,align=center] {Group 2\\$Y_2$} (B2.west);
  \end{tikzpicture}
  }
  \caption{Proof structure: $X$ handles flat vectors, and the two groups of $Y$ handle spiky ones.}
  \label{fig:two-stage}
  \vspace{1em}
\end{figure}

\subsection{Lower Bound Overview}\label{subsec:lower-overview}

Our starting point is the breakthrough framework of Huang, Rudelson, and Tikhomirov~\cite{HRT26}, denoted by HRT, which demonstrated that constant row sparsity $\zeta = O(1)$ cannot achieve an OSI. However, their analysis was capped at $\zeta \lesssim (\log r)^{1/7}$ and relied on an artificial entry-magnitude regularity assumption. In this work, we build on HRT's high-level combinatorial architecture, but fundamentally overhaul the underlying linear-algebraic cancellation technique to achieve the tight $\zeta \le O(\sqrt{\log r / \log \log r})$ lower bound without entry-magnitude assumptions.

\subsubsection{Preliminaries}
Recall that an isotropic random sketching matrix $\Omega \in \R^{n \times k}$ with $k \le Kr$ for some constant $K$ is an $(r, \alpha)$-OSI if $\Prb_\Omega[\smin(\Omega^\top Q)^2 \ge \alpha] \ge 19/20$ for every matrix $Q \in \R^{n \times r}$.
To show that $\Omega$ fails this property, it suffices to find a single $r$-dimensional subspace $Q$ where injectivity is broken under the random sketch with high probability.
Even if $Q$ is restricted to \emph{coordinate subspaces} spanned by $r$ standard basis vectors $\{e_i\}_{i \in R}$, where $Q_R = I_R$ for an index subset $R \subset [n]$ with $|R| = r$, showing that injectivity fails on one such candidate is already sufficient.
Furthermore, if we select $R \subset [n]$ of size $r$ uniformly at random, showing that injectivity fails with high probability on average is also sufficient:
\begin{equation}\label{eq:minimax-average}
\begin{aligned}
  \min_{R^* \in \binom{[n]}{r}} \Prb_\Omega\bracket{\smin(\Omega_{R^*}^\top)^2 \ge \alpha} 
  &\le \E_{R \sim \text{Unif}\binom{[n]}{r}}\bracket{\Prb_\Omega\bracket{\smin(\Omega_R^\top)^2 \ge \alpha}} \\
  &= \E_\Omega\bracket{\Prb_{R \sim \text{Unif}\binom{[n]}{r}}\bracket{\smin(\Omega_R^\top)^2 \ge \alpha}}.
\end{aligned}
\end{equation}

By isotropy, $\E_\Omega[\|\Omega\|_F^2] = \sum_{i=1}^n \E[\|s_i\|_2^2] = n$. Hence, by Markov's inequality, the bad event of a large Frobenius norm satisfies $\Prb_\Omega[\|\Omega\|_F^2 > 10n] \le 1/10$, and by assumption $\Omega$ also satisfies the \hyperref[def:condition-h]{overlap condition} with high probability (say with failure probability at most $1/10$).
Therefore, we can reduce our objective to proving the following:

\begin{quote}
\textit{For any fixed matrix $S \in \R^{n \times k}$ with row sparsity $\zeta$ and $\|S\|_F^2 \le 10n$ satisfying the overlap condition, prove that a random coordinate subspace $R \sim \text{Unif}\binom{[n]}{r}$ satisfies $\Prb[\smin(S_R^\top)^2 \ge \alpha] \le 1/4$.}
\end{quote}
Once this statement is established, adding the bad event probability and the sketch success probability on a random subspace, the overall success probability on the subspace $R^*$ is at most $9/20 < 19/20$, which refutes the injectivity and completes the reduction.

In the bipartite graph representation $G = ([n], [k], E)$ of an $n \times k$ matrix $S$, an edge connects row $i$ and column $j$ whenever $S_{ij} \ne 0$, and two distinct rows $u, v \in [n]$ share two or more columns in their support if and only if they participate in a 4-cycle $C_4$. The overlap condition bounds the number of such row pairs sharing two or more columns by $\eta n^2 / k$ with overlap parameter $\eta \le (C_H \zeta^6)^{-1}$, ensuring that 4-cycles are sparse across $S$. Under our contradiction hypothesis $\zeta \le c \sqrt{\log r / \log \log r}$, this threshold satisfies
\[
  \frac{1}{C_H \zeta^6} \ge \frac{1}{C_H c^6} \left( \frac{\log \log r}{\log r} \right)^3 = \Omega\left( \left( \frac{\log \log r}{\log r} \right)^3 \right),
\]
so the overlap condition allows an overlap parameter $\eta = \Omega((\log \log r / \log r)^3)$ that decays with $r$. This regularity ensures that the density of 4-cycles across $S$ is sufficiently small that subsampling an intermediate subset of $n_0 = \Theta(r \zeta^6 / \alpha)$ rows.

\subsubsection{The Tree Cancellation Mechanism.}
To understand the core intuition of the HRT framework, consider the simplest illustrative setting.
Suppose that every row $s_i$ satisfies $\|s_i\|_2 = 1$, and all non-zero entries are identically $\pm 1/\sqrt{\zeta}$.
Furthermore, suppose that the locations of the non-zero entries are chosen uniformly at random for each row.

Breaking injectivity on a coordinate subspace simply means that among the selected rows of $S$, there exists a non-trivial linear combination whose resulting norm is tiny. 
If we single out one row from this combination, the remaining rows must collectively \emph{cancel} all of its non-zero entries, while simultaneously leaving negligible norm across all other coordinates. 
To formalize this mechanism, suppose we pick an arbitrary row as a \emph{center row} $s_{i^*}$ and wish to cancel all $\zeta$ of its non-zero entries.
Then for each column $j_a \in \supp(s_{i^*})$ with $a \in [\zeta]$, one would naturally hope to bring another row $s_\ell$ sharing the same column $j_a$ in its support and take a difference such as $s_{i^*} - s_\ell$.
However, doing so introduces an obstacle because row $s_\ell$ possesses $\zeta - 1$ other non-zero entries on columns outside $j_a$.
We refer to any non-zero coordinate outside the specific target column intended for cancellation as an \emph{outer entry} or \emph{outer column}.
These additional outer non-zero entries generally carry substantial norm, so simply subtracting one row per column increases the overall norm.

Instead, suppose we select $q$ auxiliary rows for each column.
Averaging these $q$ rows preserves a large entry on the target column, while on all outer columns, their mutually disjoint supports dilute the entries by a factor of $1/q$.
Using these averaged vectors to cancel each column of the center row leaves only tiny residuals across all other coordinates.
Consequently, a collection of just $1 + \zeta q$ rows yields a linear combination that breaks injectivity.

Formally, we select $q$ rows $L_a = \{\ell_{a, 1}, \ldots, \ell_{a, q}\}$ for each column $j_a \in \supp(s_{i^*})$ with $a \in [\zeta]$, and define the \emph{cancellation vector}:
\begin{equation}\label{eq:cancellation-vector-def}
  x := e_{i^*} - \sum_{a=1}^\zeta \sum_{\ell \in L_a} \frac{S_{i^*, j_a}}{q S_{\ell, j_a}} e_\ell \in \R^r.
\end{equation}
While this cancellation vector is an element of $\R^n$, we treat $x$ as an element of $\R^r$: its restriction to the sampled rows indexed by $R$ yields a vector in $\R^r$.

For convenience in this example, let us assume that all $q\zeta$ leaf rows have mutually non-overlapping supports outside the center row support $\supp(s_{i^*})$.
Under this disjointness assumption, the center coordinates cancel identically via $(S_R^\top x)_{j_a} = S_{i^*, j_a} - q \cdot \frac{S_{i^*, j_a}}{q} = 0$.
Moreover, on outer columns $c \notin \supp(s_{i^*})$, we obtain:
\begin{equation}\label{eq:outer-energy-pythagoras}
  \sum_{c \notin \supp(s_{i^*})} |(S_R^\top x)_c|^2 
  = \sum_{a=1}^\zeta \sum_{\ell \in L_a} \left(\frac{1}{q}\right)^2 \sum_{c \ne j_a} |S_{\ell, c}|^2 
  \le \sum_{a=1}^\zeta q \cdot \frac{1}{q^2} \cdot \|s_\ell\|_2^2 
  = \frac{\zeta}{q}.
\end{equation}
Because $\|x\|_2^2 = 1 + \zeta/q \ge 1$, setting $q = 2\zeta/\alpha$ yields $\smin(S_R^\top)^2 \le \|S_R^\top x\|_2^2 / \|x\|_2^2 < \alpha$, breaking injectivity using a single collection of $1 + \zeta q = \Theta(\zeta^2/\alpha)$ rows.

\paragraph{Bipartite Graph Modeling.}
We can view this cancellation structure naturally in bipartite graphs.
Let $G = ([n], [k], E)$ be the bipartite graph between the $n$ rows and $k$ columns, with an edge between row $i$ and column $j$ whenever $S_{ij} \ne 0$.
The cancellation pattern above is simply a rooted tree in $G$.
This tree consists of a center row, $\zeta$ \emph{branch columns} connected to it, and $q$ \emph{leaf rows} attached to each branch column, with $1 + \zeta q$ rows in total.
To make the outer cancellation work, the leaf rows must have disjoint outer supports.

Now suppose we can find $T$ such vertex-disjoint trees in $G$.
If a random coordinate subspace $Q$ contains every row of even one tree, injectivity fails on $Q$.
If the random sample $R$ captures each tree with probability at least $p$, the chance of missing all $T$ trees is at most $(1-p)^T$, which we want to show is tiny.
This is precisely where the lower bound on $\zeta$ originates: smaller $\zeta$ shrinks each tree to fewer rows, which makes the capture probability $p$ much larger.

While this tree cancellation structure resembles that of HRT, it features a crucial difference.
In HRT, leaves were only required to have disjoint outer entries within the same branch.
In contrast, we also require leaf rows in different branches to have disjoint outer entries.
This gives a much tighter bound on the outer residual, which permits choosing a smaller value of $q$.
In turn, reducing $q$ increases the capture probability $p$, directly improving the lower bound.

\paragraph{Relative Weights and Candidate Sets.}
In actual matrices, entry values need not be uniform.
HRT assumed that all non-zero entries have magnitude $O(1/\sqrt{\zeta})$ to handle this issue, but our approach completely drops this assumption.
The idea is to pick leaf rows whose entries on the $\supp(s_{i^*})$ are relatively large.
Specifically, for row $i$ and column $j$, we define the relative weight $w_{ij} = |S_{ij}|^2 / \|s_i\|_2^2$.
For each column $j$, we sort the rows containing $j$ by their relative weights $w_{ij}$ and place the top $D$ rows into a leaf candidate set $F_j$.
When $D$ is small enough, then some rows never appear in any candidate set, and we choose center rows exclusively from these unassigned rows.

Assuming we can find leaves whose supports do not overlap outside the center support $\supp(s_{i^*})$, defining the cancellation vector $x$ as in Equation~\eqref{eq:cancellation-vector-def} bounds the outer entries:
\begin{equation}\label{eq:outer-entry-bound}
  \left(\frac{S_{i^*, j}}{q S_{\ell, j}}\right)^2 \sum_{c \ne j} S_{\ell, c}^2 
  \le \frac{w_{i^*, j} \|s_{i^*}\|_2^2}{q^2 w_{\ell, j} \|s_\ell\|_2^2} \|s_\ell\|_2^2 
  = \frac{w_{i^*, j}}{w_{\ell, j}} \frac{\|s_{i^*}\|_2^2}{q^2} 
  \le \frac{\|s_{i^*}\|_2^2}{q^2}.
\end{equation}
Whenever the center row satisfies $\|s_{i^*}\|_2^2 \le M$ for a constant $M$, summing this over all $q$ leaves on the $\zeta$ branches gives a total squared norm on the outer columns of at most $\zeta M / q$.
Choosing $q = \Theta(\zeta/\alpha)$ then keeps this residual squared norm below $\alpha / 4$.

\subsubsection{Graph Preprocessing and Tree Searching.}
A few technical details remain.
We need two main conditions:
first, the center row must have norm bounded by a constant;
second, we must be able to pick $\zeta q$ leaf rows from our candidate sets $F_j$ such that their outer entries do not overlap.
To handle these issues, we first prune the graph $G$.
Because our fixed matrix $S$ satisfies $\sum_{i=1}^n \|s_i\|_2^2 = \|S\|_F^2 \le 10n$, at most $10n / M$ rows can have $\|s_i\|_2^2 > M$.
Setting $M = 160$ and simply removing these heavy rows deterministically leaves at least $15n/16$ rows in $S$.


Also, we use the overlap condition here to get a good property for a graph $G$.
In $G$, two rows sharing two or more columns correspond to a 4-cycle $C_4$.
Suppose we initially sample $n_0 = \Theta(r \cdot \text{poly}(\zeta))$ rows from $S$ and build $G$ on these $n_0$ rows.
Under the overlap condition, the full matrix has at most $\eta n^2 / k$ row pairs sharing two or more columns, so when sampling $n_0$ rows, each pair is chosen with probability roughly $(n_0/n)^2$, which bounds the expected number of 4-cycles in $G$ by $\eta \frac{n^2}{k} \cdot \frac{n_0^2}{n^2} = \eta \frac{n_0}{k} \cdot n_0 = o(n_0)$.
Deleting the rows in these 4-cycles removes very few vertices, leaving a $C_4$-free subgraph $G_0$ on at least $5n_0 / 8$ rows.
On this clean graph $G_0$, we can now form the leaf candidate sets $F_j$ and find a tree of size $1 + \zeta q$ as before.
Because $G_0$ is $C_4$-free, leaves selected in the same branch automatically cannot share any outer columns.

What remains is ensuring that leaves in different branches do not collide outside the center support $\supp(s_{i^*})$.
Consider a single tree first.
We want to pick $q$ rows from each candidate set $F_{j_a}$, but a row could potentially appear in multiple candidate sets, so we must examine whether we can always choose fresh leaf rows.
When we pick leaves branch by branch, the first $a - 1$ branches take up at most $q(\zeta - 1)(a - 1) \le q(\zeta - 1)^2$ outer entries.
Now we choose $q$ rows of $a$-th branch.
Because $G_0$ is $C_4$-free, none of the already chosen leaf rows can have a non-zero entry at column $j_a$, as that would mean sharing two columns with the center row, which is a contradiction.
Moreover, for each outer entries occupied by leaves from the preceding $a - 1$ branches, the $C_4$-free condition guarantees that at most one row in $F_{j_a}$ can contain that column, because If there were two or more, they would share that column in addition to $j_a$.
Therefore, as long as $D := \abs{F_{j_a}} \ge q + q(\zeta - 1)^2 =: L$, at most $q(\zeta - 1)^2$ rows can collide with previously occupied columns, allowing us to select $q$ collision-free rows.
This gives a single tree whose leaves are completely disjoint on all outer columns.

Now, how do we find $T$ such disjoint trees?
We first select the center rows.
To prevent different trees from interfering with each other, we require their supports to be completely pairwise disjoint.
To do so, we conceptually consider a conflict graph whose vertices are rows, with an edge between any two rows that share at least one column.
Because $G_0$ is $C_4$-free and each row has bounded degree, the average degree in this conflict graph is small.
Tur\'an's theorem then yields an independent set of size $T = \Theta(r / \text{poly}(\zeta))$, providing $T$ center rows with completely disjoint column supports.

What remains is to find pairwise disjoint subsets of size $L$ from the candidate sets $F_j$ corresponding to all $T \zeta$ branch columns.
To establish the existence of such disjoint subsets, we apply Hall's Marriage Theorem.
We define an auxiliary bipartite assignment graph as follows:
the left vertices are the branch columns across all $T$ centers, each cloned $L$ times to represent the $L$ required reservoir slots;
the right vertices are the candidate rows in $\bigcup_j F_j$, where each row is connected to the slots of column $j$ whenever it belongs to $F_j$.
Then every left slot has degree $D$, while every right row has degree at most $\zeta L$.
Therefore, we require at least $D \ge \zeta L$ to obtain the desired assignment, and under this condition Hall's condition holds:
for any subset of left slots $A$, counting the edges between $A$ and its neighborhood $N(A)$ gives $|A| D = e(A, N(A)) \le |N(A)| \zeta L$, which yields $|N(A)| \ge \frac{D}{\zeta L} |A| \ge |A|$.
This matching assigns mutually disjoint leaf candidate sets of size $L$ to all branches of all $T$ centers simultaneously.
Then as in the single tree case, it gives $T$ pairwise vertex-disjoint witness trees $T_1, \ldots, T_T$, each having $B_{\mathrm{tree}} = 1 + \zeta q = \Theta(\zeta^2 / \alpha)$ rows.

\paragraph{Tree Capture Probability.}
When sampling $r$ rows from these $n_0$ rows, each row of $G_0$ is selected with probability roughly $r / n_0 = 1 / \text{poly}(\zeta)$.
Because each tree has size $B_{\mathrm{tree}} = 1 + q \zeta = \Theta(\zeta^2 / \alpha)$, the sample captures any specific tree with probability
\begin{equation}\label{eq:single-tree-capture}
  p \ge \left( \frac{1}{\text{poly}(\zeta)} \right)^{B_{\mathrm{tree}}} = \exp\left( -\Theta\left( \frac{\zeta^2}{\alpha} \log \zeta \right) \right).
\end{equation}
Because the $T$ trees are mutually vertex-disjoint, the events of capturing different trees are essentially independent.
Thus the probability of missing all $T$ trees simultaneously is bounded by $(1 - p)^T \le \exp(-Tp)$.
To ensure that at least one tree is captured with high probability, we only require $T p \ge \Omega(1)$, which is equivalent to $\log T \gtrsim \frac{\zeta^2}{\alpha} \log \zeta$ by our parameter setting.
Since we extracted $T = \Theta(r / \text{poly}(\zeta))$ center rows, we have $\log T = \log r - O(\log \zeta)$.
Substituting this into the balance condition yields
\begin{equation}\label{eq:asymptotic-derivation}
  \frac{\zeta^2}{\alpha} \log \zeta \lesssim \log r \implies \zeta \le O\left( \sqrt{\frac{\log r}{\log \log r}} \right).
\end{equation}
Whenever $\zeta$ falls below this threshold, a random sample captures at least one witness tree with high probability.
The corresponding cancellation vector from Equation~\eqref{eq:cancellation-vector-def} then breaks injectivity with $\smin(S_R^\top)^2 < \alpha$, concluding the lower bound proof.

\paragraph{Organization.}
The remainder of the paper is organized as follows:
\begin{itemize}[leftmargin=*,itemsep=1pt]
  \item \textbf{Section~\ref{sec:localization}:} Preserves all flat unit vectors under the first-half sketch (Theorem~\ref{thm:localization}).
  \item \textbf{Section~\ref{sec:linearization}:} Extracts a common support $I$ capturing all spiky vectors (Theorem~\ref{prop:energy-cert}).
  \item \textbf{Section~\ref{sec:coupling}:} Bounds cross-term cancellation between $I$ and $I^c$ via coupling.
  \item \textbf{Section~\ref{sec:repair}:} Preserves the coordinates on $I$ using the second-half sketch (Theorem~\ref{thm:repair}).
  \item \textbf{Section~\ref{sec:main-proof}:} Combines both sketch halves to complete the proof of Theorem~\ref{thm:main}.
  \item \textbf{Section~\ref{sec:graph-pruning}:} Preprocesses the sketching matrix to construct a $C_4$-free bipartite graph $G_0$.
  \item \textbf{Section~\ref{sec:tree-cancellation}:} Extracts a forest of disjoint cancellation trees in $G_0$ (Theorem~\ref{thm:witness-forest-existence}).
  \item \textbf{Section~\ref{sec:lower-bound-proof}:} Concludes the lower bound proof for Theorem~\ref{thm:intro-lower-bound}.
\end{itemize}

\section{Preserving Flat Vectors}\label{sec:localization}

Recall from Section~\ref{sec:overview} that our overall strategy splits the $2\zeta$-block sketch $\Omega_{2\zeta, b} \in \R^{n \times 2k}$ into two independent halves $X = (\Omega_{\zeta, b}^{(1)})^\top Q \in \R^{k \times r}$ and $Y = (\Omega_{\zeta, b}^{(2)})^\top Q \in \R^{k \times r}$ of $\zeta = O(\sqrt{\log r})$ blocks and embedding dimension $k = \zeta b$ each, and begins by showing that the first half $X$ simultaneously preserves the norm of all \emph{flat} unit vectors $y = Qx$ for $x \in S^{r-1}$. Our goal in this section is to prove Theorem~\ref{thm:localization}, which states that $\|Xx\| \ge \lambda = \Omega(\eps)$ simultaneously for all flat unit vectors $y = Qx$ with $x \in S^{r-1}$ with high probability.

We first show that for any \emph{single} fixed flat vector $y = Qx$ satisfying $d_m(y) \ge \eps$, the sketch preserves $\|\Omega_{\zeta, b}^\top y\| = \Omega(\eps)$ with exponentially small failure probability (Lemma~\ref{lem:small-ball}). To do so, we decompose $y = h + az$ into the vector $h$ supported on its $m$ largest coordinates in magnitude and its tail $az$ supported on the remaining coordinates, where $a = d_m(y) \ge \eps$ and $\|z\| = 1$.
Because every coordinate of the tail $az$ is smaller in magnitude than each of the $m$ coordinates in $h$, the unit tail $z$ has bounded entries $\|z\|_\infty^2 \le \frac{1}{m\eps^2} =: D^{-1}$. 

We establish $\|\Omega_{\zeta, b}^\top y\| \ge \Omega(\eps)$ by revealing the random column indices $\{s_{it}\}$ and the random signs $\{\varrho_{it}\}$ in turn. First, in each block $t \in [\zeta]$, the column indices partition the unit squared norm $\sum_{i=1}^n z_i^2 = 1$ among the $b$ buckets, so the total unnormalized tail weight summed across all $\zeta$ blocks is $\zeta$. We call a bucket $u$ is \emph{light} if it receives at most $v_{\max} := 20/D$ of $z$'s squared norm (i.e., for a block $t$, $\sum_{i: s_{it}=u} z_i^2 \le 20/D$), and \emph{heavy} otherwise. 

Here, we show that with probability at least $1 - e^{-\Omega(\zeta D)}$, the vast majority of the tail weight ($\ge \frac{7}{8}\zeta$) is contained in light buckets. We can prove that only a small fraction of the total weight $\zeta$ lands in heavy buckets in expectation. Then, we prove that for each light bucket $u$ in block $t$, its unnormalized squared entry $\bigl(\sum_{i:\,s_{it}=u}\varrho_{it}y_i\bigr)^2$ retains at least $\Omega\bigl(a^2 \sum_{i:\,s_{it}=u} z_i^2\bigr)$ with constant probability $p_0 \ge 4/27$. Because each light bucket has small weight $\le 20/D$, summing these independent contributions across all light buckets gives $\Omega(a^2 \zeta)$ with failure probability $e^{-\Omega(\zeta D)}$, yielding $\|\Omega_{\zeta, b}^\top y\| \ge \Omega(a) \ge \Omega(\eps)$.

\begin{fact}[McDiarmid's bounded-differences inequality~\cite{McD89}]\label{fact:standard-prob}
If $f$ is a function of independent random variables $Z_1, \dots, Z_N$ such that changing $Z_i$ changes the value of $f$ by at most $c_i$, then for every $t \ge 0$,
\begin{equation}\label{eq:mcdiarmid}
  \Prb\bracket{f - \E[f] \ge t} \le \exp\!\paren{-\frac{2t^2}{\sum_{i=1}^N c_i^2}}.
\end{equation}
\end{fact}

\begin{lemma}\label{lem:small-ball}
There exist universal constants $c_0, c_{\mathrm{sb}} > 0$, such that for a given vector $y = h + az$, where $h, z \in \R^n$ have disjoint supports, $a > 0$, $\|z\| = 1$, and
\[
  1 \le D \le b, \qquad \norm{z}_\infty^2 \le D^{-1},
\]
a SparseStack sketch $\Omega_{\zeta, b} \in \R^{n \times k}$ with $k = \zeta b$ satisfies
\[
  \Prb\bracket{\norm{\Omega_{\zeta, b}^\top y} < c_0 a} \le 2 e^{-c_{\mathrm{sb}} \zeta D}.
\]
\end{lemma}

\begin{proof}
Write $p_i = z_i^2$, so that $\sum_{i=1}^n p_i = 1$ and $\sum_{i=1}^n p_i^2 \le \max_i p_i \le 1/D$. For each block $t \in [\zeta]$ and bucket $u \in [b]$, define the bucket tail weight $W_{tu}$ and the excess weight $\mathcal{L}$ by
\[
  W_{tu} = \sum_{i:\, s_{it} = u} p_i, \qquad \mathcal{L} = \sum_{t=1}^\zeta \sum_{u=1}^b \max\set{W_{tu} - \frac{10}{D},\, 0}.
\]
For a fixed block $t$, expanding $W_{tu}^2$ and using independence of the column indices $\{s_{it}\}_{i=1}^n$ gives
\[
  \E\bracket{\sum_{u=1}^b W_{tu}^2} = \sum_{i=1}^n p_i^2 + \frac{1}{b}\paren{1 - \sum_{i=1}^n p_i^2} \le \frac{1}{D} + \frac{1}{b} \le \frac{2}{D}.
\]
Using the elementary pointwise inequality $\max\{x - A, 0\} \le x^2 / (4A)$ for $x \ge 0$ and $A = 10/D > 0$, we obtain
\[
  \E[\mathcal{L}] \le \frac{D}{40} \sum_{t=1}^\zeta \E\bracket{\sum_{u=1}^b W_{tu}^2} \le \frac{\zeta}{20}.
\]
Observe that changing a single column index $s_{it}$ moves weight $p_i$ from one bucket in block $t$ to another, which changes $\mathcal{L}$ by at most $p_i$. With the fact that $\sum_{t=1}^\zeta \sum_{i=1}^n p_i^2 \le \zeta/D$, applying Fact~\ref{fact:standard-prob}~\eqref{eq:mcdiarmid} gives
\[
  \Prb\bracket{\mathcal{L} > \frac{\zeta}{16}} \le \exp\!\paren{-2\paren{\frac{1}{16} - \frac{1}{20}}^2 \zeta D} \le e^{-c_{\mathrm{sb}} \zeta D}
\]
for a universal constant $c_{\mathrm{sb}} > 0$.
On the event $\{\mathcal{L} \le \zeta/16\}$, since $W_{tu} \le 2\max\{W_{tu} - 10/D, 0\}$ whenever $W_{tu} > 20/D$, the heavy buckets with $W_{tu} > 20/D$ carry total tail weight at most $2\mathcal{L} \le \zeta/8$. For light bucket $j = (t, u)$, let its tail weight be $v_j = W_{tu}$. Thus, if we retain only the light buckets $(t, u)$ satisfying $W_{tu} \le 20/D$, their total tail weight and maximum bucket weight $v_{\max}$ satisfy
\[
  \sum_j v_j \ge \zeta - \frac{\zeta}{8} = \frac{7\zeta}{8}, \qquad v_{\max} \le \frac{20}{D}.
\]

Now condition on all column indices $\{s_{it}\}$ on the event $\{\mathcal{L} \le \zeta/16\}$ and on all signs $\{\varrho_{it}\}_{i \in \mathrm{supp}(h)}$ in the support of $h$. Because $\mathrm{supp}(h) \cap \mathrm{supp}(z) = \varnothing$, the signs $\{\varrho_{it}\}_{i \in \mathrm{supp}(z)}$ remain independent and uniform on $\{-1, 1\}$. For a light bucket $j = (t, u)$, its unnormalized output entry is $h_j + X_{0,j}$, where $h_j$ is the deterministic contribution from $h$ and $X_{0,j} = a \sum_{i \in \mathrm{supp}(z): s_{it}=u} \varrho_{it} z_i$ is a Rademacher sum of variance $a^2 v_j$. By fourth-moment expansion for Rademacher sums, $\E[X_{0,j}^2] = a^2 v_j$ and $\E[X_{0,j}^4] \le 3 a^4 v_j^2$, so the non-negative random variable $Z_j = (h_j + X_{0,j})^2$ satisfies
\[
  \E\bracket{Z_j} = h_j^2 + a^2 v_j \ge a^2 v_j, \qquad \E\bracket{Z_j^2} = h_j^4 + 6h_j^2 a^2 v_j + \E\bracket{X_{0,j}^4} \le 3\paren{h_j^2 + a^2 v_j}^2 = 3\paren{\E[Z_j]}^2.
\]
Decomposing $Z_j = Z_j \1_{\{Z_j < \frac{1}{3}\E[Z_j]\}} + Z_j \1_{\{Z_j \ge \frac{1}{3}\E[Z_j]\}}$ and applying Cauchy--Schwarz to the second term gives
\[
  \E[Z_j] \le \frac{1}{3}\E[Z_j] + \sqrt{\E\bracket{Z_j^2}}\,\sqrt{\Prb\bracket{Z_j \ge \frac{1}{3}\E[Z_j]}},
\]
so rearranging and using $\E[Z_j^2] \le 3(\E[Z_j])^2$ and $\frac{1}{3}\E[Z_j] \ge a^2 v_j / 3$ yields
\[
  \Prb\bracket{(h_j + X_{0,j})^2 \ge \frac{a^2 v_j}{3}} \ge \Prb\bracket{Z_j \ge \frac{1}{3}\E[Z_j]} \ge \frac{4}{9}\,\frac{(\E[Z_j])^2}{\E\bracket{Z_j^2}} \ge \frac{4}{27}.
\]
Furthermore, since distinct buckets $(t, u)$ within each block receive disjoint sets of $z$-coordinates and the signs across blocks are independent, the indicators $B_j = \1_{\{(h_j + X_{0,j})^2 \ge a^2 v_j / 3\}}$ are mutually independent across all light buckets $j$, with $\Prb[B_j = 1] \ge p_0 := 4/27$.

To establish that the sum of these independent contributions concentrates around its expectation, we bound the lower tail of the weighted sum $\sum_j v_j B_j$. For each light bucket $j$, normalize its weight by setting $u_j = v_j / v_{\max} \in [0, 1]$. Since $B_j \in \{0, 1\}$ with $\Prb[B_j = 1 \mid \{s_{it}\},\, \varrho_{\mathrm{supp}(h)}] \ge p_0$, we have:
\[
  \E\bracket{\exp(-u_j B_j) \;\Big|\; \{s_{it}\},\ \varrho_{\mathrm{supp}(h)}} = 1 - \Prb\bracket{B_j = 1}(1 - e^{-u_j}) \le 1 - p_0(1 - e^{-u_j}).
\]
Using the concavity of $f(u) = 1 - e^{-u}$, $1 - e^{-u} \ge (1 - e^{-1})u$ holds for all $u \in [0, 1]$. Then we have:
\[
  \E\bracket{\exp(-u_j B_j) \;\Big|\; \{s_{it}\},\ \varrho_{\mathrm{supp}(h)}} \le 1 - p_0(1 - e^{-1})u_j \le \exp\!\paren{-p_0(1 - e^{-1})u_j}.
\]
By conditional mutual independence of the indicators $\{B_j\}$ across all light buckets, the joint moment-generating function factorizes as
\begin{align*}
  \E\bracket{\exp\!\paren{-\sum_j \frac{v_j B_j}{v_{\max}}} \;\Bigg|\; \{s_{it}\},\ \varrho_{\mathrm{supp}(h)}}
  &= \prod_j \E\bracket{\exp(-u_j B_j) \;\Big|\; \{s_{it}\},\ \varrho_{\mathrm{supp}(h)}} \\
  &\le \prod_j \exp\!\paren{-p_0(1 - e^{-1})u_j} = \exp\!\paren{-2c_1 \frac{\sum_j v_j}{v_{\max}}},
\end{align*}
where we set $c_1 = \frac{1}{2}p_0(1 - e^{-1}) > 0$.
We now apply Markov's inequality to the non-negative random variable $\exp(-\sum_j v_j B_j / v_{\max})$. We obtain
\begin{align*}
  \Prb\bracket{\sum_j v_j B_j < c_1 \sum_j v_j \;\Bigg|\; \{s_{it}\},\ \varrho_{\mathrm{supp}(h)}}
  &\le \frac{\E\bracket{\exp\!\paren{-\sum_j \frac{v_j B_j}{v_{\max}}} \;\Bigg|\; \{s_{it}\},\ \varrho_{\mathrm{supp}(h)}}}{\exp\!\paren{-c_1 \frac{\sum_j v_j}{v_{\max}}}} \\
  &\le \exp\!\paren{-c_1 \frac{\sum_j v_j}{v_{\max}}}.
\end{align*}
Recalling that $\sum_j v_j \ge \frac{7}{8}\zeta$ and $v_{\max} \le \frac{20}{D}$ on the event $\{\mathcal{L} \le \zeta/16\}$, the exponent is bounded by
\[
  c_1 \frac{\sum_j v_j}{v_{\max}} \ge c_1 \frac{7\zeta/8}{20/D} = \frac{7c_1}{160} \zeta D \ge c_{\mathrm{sb}} \zeta D,
\]
with $c_{\mathrm{sb}} := \min(\frac{1}{3200}, \frac{7c_1}{160})$.
Thus,
\[
  \Prb\bracket{\sum_j v_j B_j < c_1 \sum_j v_j \;\Bigg|\; \{s_{it}\},\ \varrho_{\mathrm{supp}(h)}} \le e^{-c_{\mathrm{sb}} \zeta D}.
\]

Outside this conditional failure event, we lower bound the squared Euclidean norm of $\Omega_{\zeta, b}^\top y$. By the normalization factor $\zeta^{-1/2}$ on each block, the squared norm is the sum of squared bucket entries across all $\zeta$ blocks. Retaining only the non-negative contributions from the light buckets and applying the lower bound $(h_j + X_{0,j})^2 \ge \frac{a^2 v_j}{3} B_j$ gives
\begin{align*}
  \norm{\Omega_{\zeta, b}^\top y}^2
  &= \frac{1}{\zeta} \sum_{t=1}^\zeta \sum_{u=1}^b (h_{tu} + X_{0,tu})^2
  \ge \frac{1}{\zeta} \sum_{j \in \mathrm{light}} (h_j + X_{0,j})^2
  \ge \frac{1}{\zeta} \sum_j \frac{a^2 v_j}{3} B_j \\
  &\ge \frac{a^2}{3\zeta} c_1 \sum_j v_j
  \ge \frac{a^2}{3\zeta} c_1 \paren{\frac{7}{8}\zeta} \ge c_0^2 a^2,
\end{align*}
where $c_0 = \sqrt{c_1/4} > 0$. Taking square roots yields $\|\Omega_{\zeta, b}^\top y\| \ge c_0 a$.

Finally, by a union bound, the event $\|\Omega_{\zeta, b}^\top y\| < c_0 a$ can occur only if either the column indices fail to allocate enough weight to light buckets ($\mathcal{L} > \zeta/16$, which occurs with probability at most $e^{-c_{\mathrm{sb}} \zeta D}$) or the conditional sign realization fails ($\sum_j v_j B_j < c_1 \sum_j v_j$, which occurs with conditional probability at most $e^{-c_{\mathrm{sb}} \zeta D}$). The total failure probability is therefore at most
\[
  \Prb\bracket{\norm{\Omega_{\zeta, b}^\top y} < c_0 a} \le e^{-c_{\mathrm{sb}} \zeta D} + e^{-c_{\mathrm{sb}} \zeta D} = 2 e^{-c_{\mathrm{sb}} \zeta D},
\]
which completes the proof.
\end{proof}

Having shown in Lemma~\ref{lem:small-ball} that $\Omega_{\zeta, b}^\top$ preserves the norm of any \emph{single} flat vector with failure probability $e^{-\Omega(\zeta D)} = e^{-\Omega(r)}$ when $m = \Theta(r/\zeta)$ and $k = \zeta b = \Theta(r)$, we now turn to the second step: extending this pointwise lower bound simultaneously to all flat unit vectors $y = Qx$ ($x \in S^{r-1}$).

Lemma~\ref{lem:unbiased-lattice} constructs a deterministic finite grid $\mathcal{N}_\eta$ of size at most $(16/\eta)^r$ such that for any matrix $T$ with $r$ columns and any unit vector $x \in S^{r-1}$, randomly rounding each coordinate of $x$ in an unbiased manner to an adjacent multiple of $h = \eta/\sqrt{r}$ guarantees the existence of a grid point $y \in \mathcal{N}_\eta$ satisfying both $\|x - y\| \le \eta$ and $\|T(x - y)\|^2 \le \frac{\eta^2}{4r}\|T\|_F^2$.

\begin{lemma}\label{lem:unbiased-lattice}
For every $0 < \eta < 1$, there exists a deterministic finite set $\mathcal{N}_\eta \subset (1+\eta) B_2^r$ of cardinality
\[
  \abs{\mathcal{N}_\eta} \le \paren{\frac{16}{\eta}}^r
\]
such that, for every real matrix $T$ with $r$ columns and every unit vector $x \in S^{r-1}$, there is some $y \in \mathcal{N}_\eta$ satisfying
\[
  \norm{x - y} \le \eta, \qquad \norm{T(x - y)}^2 \le \frac{\eta^2}{4r} \norm{T}_F^2.
\]
\end{lemma}

\begin{proof}
Set the grid spacing to be $h = \eta / \sqrt{r}$. Given any unit vector $x \in S^{r-1}$, we construct a random lattice point $y \in h\Z^r$ by independently rounding each coordinate $x_k$ for $1 \le k \le r$ to one of its two neighboring multiples of $h$, choosing the rounding probabilities so that $\E[y_k] = x_k$. Since $|x_k - y_k| \le h$ for every coordinate, every realization of $y$ satisfies
\[
  \norm{x - y} \le \sqrt{r h^2} = \eta \qquad \text{and hence} \qquad \norm{y} \le 1 + \eta.
\]
Moreover, the coordinate rounding errors $x_k - y_k$ are independent, mean-zero, and supported in intervals of length $h$, so each coordinate error has variance $\E[(x_k - y_k)^2] \le h^2/4$. Consequently, for any deterministic matrix $T$ with columns $t_1, \dots, t_r$,
\[
  \E\bracket{\norm{T(x - y)}^2} = \sum_{k=1}^r \E\bracket{(x_k - y_k)^2} \norm{t_k}^2 \le \frac{h^2}{4} \sum_{k=1}^r \norm{t_k}^2 = \frac{\eta^2}{4r} \norm{T}_F^2.
\]
By the probabilistic method, at least one realization $y$ in the deterministic finite lattice set $\mathcal{N}_\eta = h\Z^r \cap (1+\eta)B_2^r$ achieves $\|T(x-y)\|^2 \le \frac{\eta^2}{4r}\|T\|_F^2$.

To bound the cardinality $|\mathcal{N}_\eta|$, note that the open hypercubes of side length $h$ centered at the points of $\mathcal{N}_\eta$ are pairwise disjoint and have Euclidean diameter $\sqrt{r}h = \eta$, so they are all contained in the Euclidean ball $(1 + 3\eta/2)B_2^r$. Using the standard formula for the volume of an $r$-dimensional Euclidean ball,
\[
  \vol(B_2^r) = \frac{\pi^{r/2}}{\Gamma(r/2 + 1)} \le \paren{\frac{2\pi e}{r}}^{r/2} < \paren{\frac{5}{\sqrt{r}}}^r.
\]
By volume comparison,
\[
  \abs{\mathcal{N}_\eta} \le \frac{\vol\paren{(1 + 3\eta/2)B_2^r}}{h^r} < \paren{\frac{5(1 + 3\eta/2)}{\eta}}^r \le \paren{\frac{16}{\eta}}^r,
\]
as claimed.
\end{proof}

We now combine Lemma~\ref{lem:small-ball} and Lemma~\ref{lem:unbiased-lattice} to establish Theorem~\ref{thm:localization}.

\thmlocalization*

\begin{proof}
Suppose $k = \zeta b \ge C_{\mathrm{loc}} r$ and $m \ge Mr/\zeta$ for sufficiently large constants $C_{\mathrm{loc}}, M > 0$ depending on $\eps$ and $\delta_{\mathrm{loc}}$. Since each row of $\Omega_{\zeta, b}^{(1)}$ has unit norm and $\E[\Omega_{\zeta, b}^{(1)} (\Omega_{\zeta, b}^{(1)})^\top] = \Id_n$, we have $\E[\|X\|_F^2] = \tr(Q^\top Q) = r$. By Markov's inequality, with probability at least $1 - \delta_{\mathrm{loc}}/2$,
\begin{equation}\label{eq:frob-event}
  \norm{X}_F^2 \le \frac{2}{\delta_{\mathrm{loc}}}\,r.
\end{equation}
Choose the net resolution $\eta = c'\eps$ with a sufficiently small constant $c' > 0$ depending on $\delta_{\mathrm{loc}}$ so that $\frac{\eta}{\sqrt{2\delta_{\mathrm{loc}}}} \le \frac{1}{4} c_0 \eps$, and set $\lambda = \frac{1}{2} c_0 \eps$, where $c_0$ is the constant from Lemma~\ref{lem:small-ball}.

On the event~\eqref{eq:frob-event}, suppose there exists a unit vector $x \in S^{r-1}$ satisfying $d_m(Qx) \ge \eps$ and $\|Xx\| < \lambda$. Applying Lemma~\ref{lem:unbiased-lattice} to $T = X$, there exists a lattice point $y \in \mathcal{N}_\eta$ such that $\|x - y\| \le \eta$ and $\|X(x-y)\| \le \frac{\eta}{\sqrt{2\delta_{\mathrm{loc}}}} \le \frac{1}{4} c_0 \eps$. Since $Q$ is an isometry and $d_m$ is $1$-Lipschitz, this lattice point $y$ satisfies
\[
  d_m(Qy) \ge d_m(Qx) - \norm{Q(x-y)} \ge \eps - \eta \ge \frac{7}{8}\eps,
\]
and
\[
  \norm{Xy} \le \norm{Xx} + \norm{X(x-y)} < \lambda + \frac{1}{4}c_0\eps = \frac{3}{4}c_0\eps \le c_0(\eps - \eta).
\]

Fix any deterministic lattice point $y \in \mathcal{N}_\eta$ satisfying $d_m(Qy) \ge \eps - \eta$. Let $I_y \subseteq [n]$ be the indices of the $m$ largest coordinates of $Qy$ in magnitude, and decompose $Qy = h + az$ where $h = (Qy)_{I_y}$, $a = \|(Qy)_{I_y^c}\| = d_m(Qy) \ge \eps - \eta$, and $z = a^{-1}(Qy)_{I_y^c}$ has unit norm $\|z\| = 1$ and support disjoint from $h$. Since every coordinate of $(Qy)_{I_y^c}$ is bounded in magnitude by each of the $m$ coordinates in $I_y$, we have
\[
  \norm{z}_\infty^2 \le \frac{\norm{Qy}^2}{m a^2} \le \frac{(1+\eta)^2}{m(\eps-\eta)^2}.
\]
Setting
\[
  D = \min\set{b,\, m\frac{(\eps-\eta)^2}{(1+\eta)^2}},
\]
we have $1 \le D \le b$ and $\|z\|_\infty^2 \le 1/D$. Moreover, since $\zeta \le r$, $m \ge Mr/\zeta$, and $b\zeta \ge C_{\mathrm{loc}} r$, choosing the constants $M, C_{\mathrm{loc}} > 0$ sufficiently large ensures that
\[
  \zeta D = \min\set{b\zeta,\, m\zeta\frac{(\eps-\eta)^2}{(1+\eta)^2}} \ge \min\set{C_{\mathrm{loc}},\, M\frac{(\eps-\eta)^2}{(1+\eta)^2}} r \ge \frac{2\log(16/\eta)}{c_{\mathrm{sb}}}\,r.
\]
Applying Lemma~\ref{lem:small-ball} to $Qy = h + az$ shows that
\[
  \Prb\bracket{\norm{Xy} < \frac{3}{4}c_0\eps} \le \Prb\bracket{\norm{(\Omega_{\zeta, b}^{(1)})^\top Qy} < c_0 a} \le 2 e^{-c_{\mathrm{sb}} \zeta D} \le 2\paren{\frac{\eta}{16}}^{2r}.
\]
Taking a union bound over all $|\mathcal{N}_\eta| \le (16/\eta)^r$ deterministic lattice points $y \in \mathcal{N}_\eta$ and adding the Frobenius failure probability $\delta_{\mathrm{loc}}/2$ from~\eqref{eq:frob-event}, the total probability that such a unit vector $x$ exists is at most
\[
  \frac{\delta_{\mathrm{loc}}}{2} + \paren{\frac{16}{\eta}}^r \cdot 2\paren{\frac{\eta}{16}}^{2r} = \frac{\delta_{\mathrm{loc}}}{2} + 2\paren{\frac{\eta}{16}}^r \le \delta_{\mathrm{loc}}
\]
for $r$ sufficiently large.
\end{proof}

\section{Extracting a Common Support for Spiky Vectors}\label{sec:linearization}

In Section~\ref{sec:localization}, we showed that with high probability the first half of the sketch $(\Omega_{\zeta, b}^{(1)})^{\top} \in \R^{k \times n}$ preserves the norm of all flat unit vectors. Consequently, any unit vector $y = Qx$ on which injectivity fails under the first half $\|(\Omega_{\zeta, b}^{(1)})^{\top}y\| < \lambda$) must be \emph{spiky}, satisfying $d_m(y) < \eps$. Our goal in this section is to prove Theorem~\ref{prop:energy-cert}, showing that all such spiky vectors must concentrate their norm on a \emph{single, common} set $I \subseteq [n]$ of at most $8m$ coordinates.

\paragraph{Step 1: Expected tail norm of random linear combinations.}
Recall from Section~\ref{sec:overview} that a random linear combination of vectors supported on mostly disjoint $m$-coordinate sets spreads its entries across their union and becomes flat. To make this precise, for a matrix $V \in \R^{n \times r}$ and Gaussian coefficients $x \sim N(0, R)$ with covariance $R \succeq 0$, the random linear combination $g = Vx \in \R^n$ is a centered Gaussian vector whose $i$-th coordinate has marginal variance $a_i = \E[g_i^2]$.

Sorting these variances as $a_1 \ge a_2 \ge \cdots \ge a_n$, Lemma~\ref{lem:gaussian-trunc} shows that the expected squared tail norm $\E[d_m(g)^2]$ after deleting the $m$ largest coordinates of $g$ is at least $\frac{3}{4}$ of the total variance $\sum_{i > 4m} a_i$ outside the top $4m$ coordinates. In other words, whenever a random linear combination places noticeable variance outside its top $4m$ coordinates, the resulting vector is guaranteed to have large tail norm $d_m(g)$ on average and thus behaves like a flat vector rather than a spiky one.

\begin{lemma}\label{lem:gaussian-trunc}
Let $g \in \R^n$ be a centered Gaussian vector with arbitrary coordinate correlations, and let its coordinate variances be sorted in non-increasing order as $a_1 \ge \cdots \ge a_n$. If $n > 4m$, then
\[
  \E\bracket{d_m(g)^2} \ge \frac{3}{4} \sum_{i > 4m} a_i.
\]
\end{lemma}

\begin{proof}
For every vector $z \in \R^n$ and every truncation threshold $T \ge 0$, removing at most $m$ coordinates from the sum of truncated squares $\sum_{i=1}^n \min\{z_i^2, T\}$ reduces the sum by at most $mT$. Hence
\begin{equation}\label{eq:trunc-lower}
  d_m(z)^2 = \min_{|I| \le m} \sum_{i \notin I} z_i^2 \ge \min_{|I| \le m} \sum_{i \notin I} \min\set{z_i^2, T} \ge \sum_{i=1}^n \min\set{z_i^2, T} - mT.
\end{equation}
For a standard normal random variable $G \sim N(0, 1)$, we have $\E[G^2] = 1$ and $\E[G^4] = 3$. Applying the elementary inequality $\max\{x - A, 0\} \le x^2/(4A)$ with $A = 3$ and $x = G^2$ gives
\[
  \E\bracket{\min\set{G^2, 3}} = \E\bracket{G^2} - \E\bracket{\max\set{G^2 - 3, 0}} \ge 1 - \frac{\E\bracket{G^4}}{12} = 1 - \frac{3}{12} = \frac{3}{4}.
\]
Now set $j = 4m$ and $a = a_j$. If $a = 0$, then $a_i = 0$ for all $i > 4m$ and the lemma holds trivially. Otherwise, apply~\eqref{eq:trunc-lower} to $z = g$ with threshold $T = 3a > 0$. Since each marginal is distributed as $g_i \sim \sqrt{a_i}\,G$, monotonicity of $u \mapsto \min\{u, 3a\}$ implies that for $i \le j$, where $a_i \ge a$,
\[
  \E\bracket{\min\set{g_i^2, 3a}} \ge \E\bracket{\min\set{a G^2, 3a}} = a\,\E\bracket{\min\set{G^2, 3}} \ge \frac{3}{4}a,
\]
whereas for $i > j$, where $a_i \le a$,
\[
  \E\bracket{\min\set{g_i^2, 3a}} \ge \E\bracket{\min\set{a_i G^2, 3a_i}} = a_i\,\E\bracket{\min\set{G^2, 3}} \ge \frac{3}{4}a_i.
\]
Taking expectations in~\eqref{eq:trunc-lower} and noting that $\frac{3}{4}j a - m(3a) = 3ma - 3ma = 0$ yields
\[
  \E\bracket{d_m(g)^2} \ge \sum_{i=1}^j \frac{3}{4}a + \sum_{i > 4m} \frac{3}{4}a_i - 3ma = \frac{3}{4} \sum_{i > 4m} a_i. \qedhere
\]
\end{proof}

\paragraph{Step 2: Extracting a single common support via minimax duality.}
In our setting, every unit vector $x \in S^{r-1}$ satisfies either $d_m(Qx)^2 < \eps^2$ when $\|Xx\| < \lambda$, or $d_m(Qx)^2 \le \|Qx\|^2 = 1 \le \lambda^{-2}\|Xx\|^2$ when $\|Xx\| \ge \lambda$. Combining the two cases yields the quadratic upper bound
\begin{equation}\label{eq:dm-quadratic}
  d_m(Qx)^2 \le \eps^2 \norm{x}^2 + \lambda^{-2}\norm{Xx}^2 = x^\top H x, \qquad \text{where } H = \eps^2 \Id_r + \lambda^{-2} X^\top X.
\end{equation}
More generally, whenever $V \in \R^{n \times r}$ and $H \succeq 0$ satisfy $d_m(Vx)^2 \le x^\top H x$ for all $x \in \R^r$, taking expectations over $x \sim N(0, R)$ and applying Lemma~\ref{lem:gaussian-trunc} implies that for \emph{every} covariance $R \succeq 0$, there exists a set of $4m$ coordinates—potentially depending on $R$—outside of which the expected squared norm of $Vx$ is at most $\frac{4}{3}\E[x^\top H x]$.

The goal of Step 2 is to lift this distributional guarantee to a \emph{single, common} coordinate set $I$ that works for all $x \in \R^r$ simultaneously. 
As outlined in Section~\ref{sec:overview}, we relax the choice of a $4m$-element subset to a fractional weight vector $w \in [0, 1]^n$ with $\sum_{i=1}^n w_i \le 4m$, which forms a compact convex set $\mathcal{W}$. We will show that by minimax duality, there exists a single weight vector $w \in \mathcal{W}$ that bounds the weighted tail norm for all $R \succeq 0$ at once. Rounding these weights at $w_i \ge 1/2$ then yields a single coordinate set $I \subseteq [n]$ of size $|I| \le 8m$ satisfying $\|(Vx)_{I^c}\|^2 \le \frac{8}{3}x^\top H x$ for all $x \in \R^r$, or equivalently $V_{I^c}^\top V_{I^c} \preceq \frac{8}{3}H$. We formally state the result and the proof below.

\begin{lemma}\label{thm:linearization}
Let $V \in \R^{n \times r}$, let $H \succeq 0$ be an $r \times r$ positive-semidefinite matrix, and let $m \ge 1$ be an integer. Suppose that $d_m(Vx)^2 \le x^\top H x$ for all $x \in \R^r$. Then there exists a coordinate subset $I \subseteq [n]$ satisfying
\[
  \abs{I} \le 8m, \qquad V_{I^c}^\top V_{I^c} \preceq \frac{8}{3} H.
\]
\end{lemma}

\begin{proof}
The case $r = 0$ is trivial, and if $n \le 4m$ we may simply take $I = [n]$. Assume $n > 4m$, put $j = 4m$, and write $v_i = V^\top e_i \in \R^r$ for the $i$-th row of $V$.

For any positive-semidefinite matrix $R \succeq 0$ with $\tr R = 1$, let $x \sim N(0, R)$. Then $g = Vx \in \R^n$ is a centered Gaussian vector whose $i$-th coordinate has variance $a_i(R) = \E[(v_i^\top x)^2] = v_i^\top R v_i$. By the hypothesis of the lemma, $\E[d_m(Vx)^2] \le \E[x^\top H x] = \tr(HR)$. On the other hand, Lemma~\ref{lem:gaussian-trunc} shows that $\E[d_m(Vx)^2]$ is at least $\frac{3}{4}$ times the sum of all $v_i^\top R v_i$ excluding the $4m$ largest values:
\begin{equation}\label{eq:min-T-trace}
  \min_{|T| \le 4m} \sum_{i \notin T} v_i^\top R v_i \le \frac{4}{3} \tr(HR).
\end{equation}
Define the compact convex weight polytope $\mathcal{W}$ and the linear matrix map $D(w)$ by
\[
  \mathcal{W} = \set{w \in [0, 1]^n : \sum_{i=1}^n w_i \le 4m}, \qquad D(w) = \sum_{i=1}^n (1 - w_i) v_i v_i^\top.
\]
For any fixed $R \succeq 0$, the linear functional $w \mapsto \tr(R D(w)) = \sum_{i=1}^n (1 - w_i) v_i^\top R v_i$ is minimized over $\mathcal{W}$ by placing unit weight $w_i = 1$ on the $4m$ indices with the largest values of $v_i^\top R v_i$. Therefore,~\eqref{eq:min-T-trace} is equivalent to the inequality
\begin{equation}\label{eq:minimax-trace}
  \min_{w \in \mathcal{W}} \tr\paren{R \left(D(w) - \frac{4}{3}H\right)} \le 0 \qquad \text{for all } R \succeq 0 \text{ with } \tr R = 1.
\end{equation}

\begin{claim}\label{claim:minimax}
There exists a single weight vector $w \in \mathcal{W}$ such that $D(w) \preceq \frac{4}{3}H$. 
\end{claim}
\begin{proof}
Let $\mathcal{R} = \{R : R \succeq 0, \tr(R) = 1\}$, such that both $\mathcal{R}$ and $\mathcal{W}$ are convex and compact.  By a standard minimax theorem for compact convex sets and bilinear payoffs~\cite{du1995minimax}, since the function $\tr\paren{R \left(D(w) - \frac{4}{3}H\right)}$ is bilinear in $w$ and $R$, we have that
\begin{equation}
    \min_{w \in \mathcal{W}} \max_{R \in \mathcal{R}} \tr\paren{R \left(D(w) - \frac{4}{3}H\right)} = \max_{R \in \mathcal{R}} \min_{w \in \mathcal{W}} \tr\paren{R \left(D(w) - \frac{4}{3}H\right)} \leq 0.
\end{equation}

Now for any symmetric matrix $A$, we have that 
\begin{equation}
    \max_{R \in \mathcal{R}} \tr(RA) = \lambda_{max}(A).
\end{equation}
Thus 
\begin{align}
    \min_{w \in \mathcal{W}} \lambda_{max}\left(D(w) - \frac{4}{3}H\right) \leq 0,
\end{align}
meaning that there exists some $w \in \mathcal{W}$ such that $D(w) \preceq \frac{4}{3}H$. This proves the claim.
\end{proof}



Now we round the $w$ attained by  Claim~\ref{claim:minimax}. Define the thresholded coordinate set $I = \{i \in [n] : w_i \ge 1/2\}$. By Markov's inequality on the weights, $|I| \le 2\sum_{i=1}^n w_i \le 2j = 8m$, and for every $i \in I^c$ we have $1 \le 2(1 - w_i)$. Hence
\[
  V_{I^c}^\top V_{I^c} = \sum_{i \in I^c} v_i v_i^\top \preceq 2 \sum_{i=1}^n (1 - w_i) v_i v_i^\top = 2 D(w) \preceq \frac{8}{3}H,
\]
completing the proof.
\end{proof}

Finally, we apply Lemma~\ref{thm:linearization} with $V = Q$ and $H = \eps^2 \Id_r + \lambda^{-2} X^\top X$ to obtain Theorem~\ref{prop:energy-cert}. As discussed above, the matrix inequality $Q_{I^c}^\top Q_{I^c} \preceq \frac{8}{3}\eps^2 \Id_r + \frac{8}{3\lambda^2} X^\top X$ states that every spiky unit vector $y = Qx$ with small $\|Xx\|$ concentrates at least $1 - \frac{8}{3}\eps^2$ of its squared norm on the single coordinate set $I$ of size $|I| \le 8m$.

\propenergycert*

\begin{proof}
By~\eqref{eq:dm-quadratic}, the hypothesis implies $d_m(Qx)^2 \le x^\top (\eps^2 \Id_r + \lambda^{-2} X^\top X) x$ for all $x \in \R^r$. Applying Lemma~\ref{thm:linearization} with $V = Q$ and $H = \eps^2 \Id_r + \lambda^{-2} X^\top X$ guarantees the existence of $I \subseteq [n]$ with $|I| \le 8m$ satisfying $Q_{I^c}^\top Q_{I^c} \preceq \frac{8}{3}\eps^2 \Id_r + \frac{8}{3\lambda^2} X^\top X$.
\end{proof}

\section{Bounding Cancellation from the Complementary Coordinates}\label{sec:coupling}

In Section~\ref{sec:linearization}, we proved that every spiky vector $y \in \operatorname{span}(Q)$ whose norm under the first half $(\Omega_{\zeta, b}^{(1)})^{\top}$ is small must concentrate its norm on a single common coordinate set $I \subseteq [n]$ of size $N = |I| \le 8m = O(r/\zeta)$. It remains to show that with high probability, any unit vector $y \in \operatorname{span}(Q)$ with sufficiently large mass on the coordinates $y_I$, satisfies that $\|(\Omega_{\zeta, b}^{(2)})^{\top}y\| \geq \Omega(1)$. 

Throughout this section, we will use the shorthand $\Omega^{\top} := (\Omega_{\zeta, b}^{(2)})^{\top} \in \R^{k \times n}$. We will decompose $y = Qx$ into its heavy coordinates $h = (Qx)_I$ and complementary coordinates $(Qx)_{I^c} = Wf$ (with $W \in \R^{|I^c| \times d}$ an orthonormal basis for $\col(Q_{I^c})$ and $\|f\| = \|(Qx)_{I^c}\|$). 

With this notation, we can expand the squared sketch norm $\|\Omega_{\zeta, b}^{(2)}y\|^2$ as follows: 

\[
 \|\Omega_{\zeta, b}^{(2)}y\|^2 =  \norm{\Omega_I^{\top} h + \Omega_{I^c}^{\top} Wf}^2 = \norm{\Omega_I^{\top}h}^2 + 2h^\top (\Omega_I \Omega_{I^c}^{\top}Wf) + \norm{\Omega_{I^c}^{\top}Wf}^2.
\]

For reasons that will become clear in Section~\ref{subsec:cross-term-isolated}, it will be easer to lower bound this norm over a subset of coordinates $T \subseteq [k]$, which correspond to rows of $\Omega_{I}^{\top}$ in which there is exactly one non-zero entry. Let $\widetilde{\Omega}$ denote the restriction of $\Omega$ to rows in $T$. Then we have 

\[
  \norm{\Omega^{\top}y}^2 \geq \norm{\widetilde{\Omega}^{\top}y}^2 = \norm{\widetilde{\Omega}_I^{\top}h}^2 + 2h^\top (\widetilde{\Omega}_I \widetilde{\Omega}_{I^c}^{\top}W) f + \norm{\widetilde{\Omega}_{I^c}^{\top}Wf}^2.
\]

In Section~\ref{sec:repair}, we will claim that there is not too much cancellation within $I$, and thus $\norm{\widetilde{\Omega}_I^{\top}h}^2$ is a sufficiently large constant. 

In this section, we will show that in expectation, the cross term is sufficiently small, by bounding the operator norm $\|\widetilde{\Omega}_I \widetilde{\Omega}_{I^c}^{\top}W\|$. The following is the main result of Section~\ref{sec:coupling}.

\crossgram*

Before proving this result, let us unpack its meaning. When applied with $N = \Theta(r/\zeta)$ and $d = \Theta(r)$, $\zeta = O(\sqrt{\log(r)})$, and $k = \Theta(r)$, this result says that $\E\bracket{\norm{\widetilde{\Omega}_I\widetilde{\Omega}_{I^c}^{\top}W} \;\Big|\; \Omega_{\zeta, b, I}} = O(1)$. As noted before in Section~\ref{subsec:embvsinj} (see the discussion of ``coupon-collector argument''), in this regime, the operator norm of $\Omega_I$ or $\Omega_{I^c}$ is likely to be as large as $\Theta(\sqrt{\log r}/\log\log r)$, and so it is crucial to our proof that this cross-term cancels out the super-constant norm of $\Omega_I$ or $\Omega_{I^c}$ alone. 

The key idea in proving Lemma~\ref{lem:cross-gram} is to compare $\widetilde{\Omega}_I\widetilde{\Omega}_{I^c}^{\top}W$ to a matrix $XW$, where $X$ has i.i.d. entries, with the same entry-wise marginal distribution as $\widetilde{\Omega}_I\widetilde{\Omega}_{I^c}^{\top}$. It is easy to check that each entry $(\widetilde{\Omega}_I\widetilde{\Omega}_{I^c}^{\top})_{i, j}$, is non-zero with probabilty roughly $\zeta/b = \Theta(\zeta^2/r) = \Omega(\log r /r)$ in the regime of our main theorem. To control the operator norm of this matrix $XW$, we are able to leverage the recent work of \cite{Heidary26}, which shows that SparseStack is an OSE in this regime of $\Theta(\log r/r)$ sparsity. We prove this i.i.d. result in Section~\ref{subsec:independent}. 

In Section~\ref{subsec:cross-term-isolated}, we show how to use Jensen's inequality to transfer this result from the i.i.d. matrix to $\widetilde{\Omega}_I\widetilde{\Omega}_{I^c}^{\top}W$.

Finally in Section~\ref{subsec:full-sketch-moment}, we give an additional estimate for high order moments of the norm of $\Omega_{\zeta, b}Q$, also based off of \cite{Heidary26}. Note that this is only used in the high-sparsity regime $\zeta = \Omega(\log r)$, but enables our result to hold for the full spectrum of $\zeta$ above the $\Theta(\sqrt{\log r})$ threshold.

\subsection{Operator-norm estimate for matrices with independent entries}\label{subsec:independent}
To bound the cross-term $\widetilde{\Omega}_I\widetilde{\Omega}_{I^c}^{\top}W$, we first analyze the simpler model where the $XW$, where $X$ has independent entries. For a random matrix $X \in \R^{m \times n}$ with independent centered entries and a fixed isometry $W \in \R^{n \times d}$, we control $\E[\|XW\|^2] = \E[\|(XW)^\top (XW) \|]$ via decomposing it into its diagonal and off-diagonal parts:
\[
  (XW)^\top (XW) = \sum_{\ell=1}^m \sum_{j=1}^n X_{\ell j}^2 w_j w_j^\top + \sum_{\ell=1}^m \sum_{j \neq j'} X_{\ell j} X_{\ell j'} w_j w_{j'}^\top.
\]
This separation is essential because the two terms exhibit fundamentally different probabilistic behaviors:
\begin{itemize}
  \item \emph{Diagonal term ($j = j'$):} The entries $X_{\ell j}^2$ are non-negative with non-zero mean. Centering yields a sum of independent positive-semidefinite rank-one random matrices with zero mean, which is naturally controlled by the Matrix Bernstein inequality (Fact~\ref{lem:matrix-bernstein}).
  \item \emph{Off-diagonal term ($j \neq j'$):} This is a hollow symmetric matrix whose entries are mean-zero products of independent coordinates with no diagonal bias. Its operator norm concentrates sharply via high-order trace moments (Fact~\ref{lem:hollow-moment}).
\end{itemize}

\begin{fact}[\cite{Heidary26}]\label{lem:hollow-moment}
Let $B \ge 2$ and $a \ge 1$ be integers, put $m = aB$, and let $X \in \R^{m \times n}$ have independent entries distributed as
\[
  \Prb\bracket{X_{\ell j} = 1} = \Prb\bracket{X_{\ell j} = -1} = \frac{1}{2B}, \qquad \Prb\bracket{X_{\ell j} = 0} = 1 - \frac{1}{B}.
\]
Let $W \in \R^{n \times d}$ ($d \ge 1$) be a fixed isometry satisfying $W^\top W = \Id_d$, with rows $w_j^\top$. For the off-diagonal matrix
\begin{equation}\label{eq:hollow-F}
  F = \frac{1}{a} \sum_{\ell=1}^m \sum_{j \neq j'} X_{\ell j} X_{\ell j'} w_j w_{j'}^\top
\end{equation}
and every integer $p \ge 1$, set $D_p = d + 2p + 1$. Then
\begin{equation}\label{eq:hollow-moment-bound}
  \E\bracket{\tr(F^{2p})} \le d \bracket{3(3 + \sqrt{2})\paren{\sqrt{\frac{D_p}{m}} + \frac{D_p}{m} + \frac{2p+1}{a}}}^{2p}.
\end{equation}
\end{fact}

\begin{fact}[\cite{Tropp15}]\label{lem:matrix-bernstein}
If $R_j$ are independent, centered, symmetric $d \times d$ random matrices satisfying
\[
  \norm{R_j} \le \ell \quad \text{almost surely} \qquad \text{and} \qquad \sum_j \E\bracket{R_j^2} \preceq v \Id_d,
\]
then
\begin{equation}\label{eq:matrix-bernstein}
  \E\bracket{\norm{\sum_j R_j}} \le \sqrt{2v \log(2d)} + \frac{\ell}{3} \log(2d).
\end{equation}
\end{fact}

Combining Fact~\ref{lem:hollow-moment} for the off-diagonal sum and Fact~\ref{lem:matrix-bernstein} for the centered diagonal sum yields Lemma~\ref{lem:indep-norm}. This lemma states that when an $m \times n$ independent-entry matrix $X$ with non-zero probability $1/B$ is applied to a $d$-dimensional isometry $W$, the expected squared operator norm $\E[\|XW\|^2]$ is bounded by the expected column sparsity $a = m/B$, the dimension-to-sparsity ratio $D_p/B$, and the moment order $p \ge \log(2d)$.

\begin{lemma}\label{lem:indep-norm}
In the independent-entry model of Fact~\ref{lem:hollow-moment}, if $p \ge \max\{1, \log(2d)\}$ is an integer and $D_p = d + 2p + 1$, then there is an absolute constant $C_{\mathrm{ind}} > 0$ such that
\[
  \E\bracket{\norm{XW}^2} \le C_{\mathrm{ind}} \paren{a + \frac{D_p}{B} + p}.
\]
\end{lemma}

\begin{proof}
We separate $a^{-1}(XW)^\top (XW)$ into its diagonal and off-diagonal parts:
\[
  a^{-1}(XW)^\top (XW) = D + F, \qquad D = \frac{1}{a} \sum_{\ell=1}^m \sum_{j=1}^n X_{\ell j}^2 w_j w_j^\top,
\]
where $F$ is defined in~\eqref{eq:hollow-F}. Since $\E[X_{\ell j}^2] = B^{-1}$, we have $\E[D] = \frac{m}{aB} \sum_{j=1}^n w_j w_j^\top = \Id_d$. Writing $D - \Id_d = \sum_{\ell, j} R_{\ell j}$, the summands
\[
  R_{\ell j} = \frac{1}{a}\paren{X_{\ell j}^2 - B^{-1}} w_j w_j^\top
\]
are independent, centered, symmetric $d \times d$ matrices. Since $\|w_j\|^2 \le 1$, each summand satisfies $\|R_{\ell j}\| \le 1/a$, and the variance proxy satisfies
\[
  \sum_{\ell, j} \E\bracket{R_{\ell j}^2} \preceq \frac{m}{B a^2} \sum_{j=1}^n \norm{w_j}^2 w_j w_j^\top \preceq \frac{1}{a} \Id_d.
\]
Applying Fact~\ref{lem:matrix-bernstein} with $v = 1/a$, $\ell = 1/a$, and $\log(2d) \le p$ yields
\[
  \E[\norm{D}] \le 1 + \E[\norm{D - \Id_d}] \le 1 + \sqrt{\frac{2p}{a}} + \frac{p}{3a}.
\]
For the off-diagonal part $F$, since $\|F\|^{2p} \le \tr(F^{2p})$ and $d^{1/(2p)} \le \sqrt{e} < 2$ for $p \ge \log(2d)$ and $m = aB$, Jensen's inequality and Fact~\ref{lem:hollow-moment} give
\[
  \E[\norm{F}] \le \paren{\E\bracket{\tr(F^{2p})}}^{1/(2p)} \le O(1)\paren{\sqrt{\frac{D_p}{aB}} + \frac{D_p}{aB} + \frac{2p+1}{a}}.
\]
Multiplying the bound on $\E[\|D\|] + \E[\|F\|]$ by $a$ and applying the Cauchy--Schwarz inequalities $\sqrt{2ap} \le a + p/2$ and $\sqrt{a D_p / B} \le \frac{1}{2}(a + D_p/B)$ yields
\[
  \E\bracket{\norm{XW}^2} = a\,\E[\norm{D + F}] \le a\paren{\E[\norm{D}] + \E[\norm{F}]} \le O\!\paren{a + \frac{D_p}{B} + p},
\]
completing the proof.
\end{proof}

\subsection{Coupling argument to bound the cross-term}\label{subsec:cross-term-isolated}
Our goal is to reduce the analysis of the expected operator norm of $\widetilde{\Omega}_I\widetilde{\Omega}_{I^c}^{\top}W$ to the independent-entry setting of Lemma~\ref{lem:indep-norm}. To do this, observe that we can write

\begin{equation}
  (\widetilde{\Omega}_I\widetilde{\Omega}_{I^c}^{\top})_{ij} = \frac{1}{\zeta} \sum_{t=1}^\zeta \alpha_{ti} \varrho_{it} \xi_{jt}[s_{it}],
\end{equation}
where $\alpha_{ti}$ is the indicator that coordinate $i$ occupies a bucket in block $t$ with no other element of $I$, and $\xi_{j t} = \varrho_{j t}e_{u_{j t}} \in \R^b$. (Recall that $u_{\ell t} \sim \textsc{Uniform}[b]$, $\varrho_{\ell t} \in \{-1, 1\}$ uniform.) Note that conditional on $\Omega_{I}$ (which fixes the $\alpha_{ti}$ and the $\xi_{it}$), even though the $\xi_{jt}$ are independent for each $j$, the entries $(\widetilde{\Omega}_I\widetilde{\Omega}_{I^c}^{\top})_{ij}$ are not independent, since the entries of $\xi_{jt}$ are anti-correlated.

However, if we replaced the $\xi_{jt}$ with some random vectors $\eta_{jt}$, where each entry of $\eta_{jt}$ was i.i.d. with distribution equal to the marginal distribution of $\xi_{jt}$, we would have that, conditional on $\Omega_{\zeta, b, I}$, the entries 
\begin{equation}
\frac{1}{\zeta} \sum_{t=1}^\zeta \alpha_{it}\xi_{it}^{\top} \eta_{jt} = \frac{1}{\zeta} \sum_{t=1}^\zeta \alpha_{it}\varrho_{it}\eta_{jt}[s_{it}]
\end{equation}
are independent. Indeed, each $s_{it}$, for which $\alpha_{it} = 1$, is distinct, and thus the values $(x_{tij} := \eta_{jt}[s_{it}])_{i \in I, j \in J}$ are independent! Summing over $t$ retains independence.

To bound the expected operator norm of $\widetilde{\Omega}_I\widetilde{\Omega}_{I^c}^{\top} W$, we will use Jensen's argument to reduce to the independent entry case. To set this up, consider the following coupling of a $\xi \in \R^b$ with an independent-entry vector $\eta \in \{-1, 0, 1\}^b$:
\begin{itemize}
  \item Draw $\eta \in \{-1, 0, 1\}^b$ with independent entries: $\Prb[\eta_u = \pm 1] = \frac{1}{2b}$ and $\Prb[\eta_u = 0] = 1 - \frac{1}{b}$.
  \item If $\eta \neq 0$, pick $u^*$ uniformly from $\mathrm{supp}(\eta) = \{u \in [b] : \eta_u \neq 0\}$ and set $\xi = \eta_{u^*} e_{u^*}$; if $\eta = 0$, pick $\xi \in \{\pm e_1, \dots, \pm e_b\}$ uniformly.
\end{itemize}
By symmetry, $\xi$ has the exact marginal distribution of a CountSketch column. We will use the following claim.
\begin{claim}\label{claim:xi_to_eta}
In the coupling above, we have
\begin{equation}\label{eq:selector-identity}
  \E\bracket{\eta \;\big|\; \xi} = c_b^{-1} \xi, \qquad \text{where } c_b = \bracket{1 - \paren{1 - \frac{1}{b}}^b}^{-1} \in \bracket{1, \frac{1}{1 - e^{-1}}} \subset [1, 2).
\end{equation}
\end{claim}
\begin{proof}
Conditioned on $\xi = \varrho e_u$, the coordinate $\eta_u$ equals $\varrho$ whenever $\eta \neq 0$ (which occurs with probability $c_b^{-1} := 1 - (1 - 1/b)^b$), while all other coordinates $u' \neq u$ have conditional mean zero. 
\end{proof}

Using the claim above and conditional Jensen's inequality yields the following lemma.

\begin{lemma}\label{lemma:jensen}
With all variables defined as in Lemma~\ref{lem:cross-gram}, and $c_b$ as in Claim~\ref{claim:xi_to_eta}, we have
\begin{equation}
\E[\|\widetilde{\Omega}_I\widetilde{\Omega}_{I^c}^{\top}W\| \mid \Omega_{\zeta, b, I}] \leq c_b \E[\|UW\|],
\end{equation}
 where the $N \times |J|$ matrix $U$ has independent entries $U_{ij} = \frac{1}{\zeta}\sum_{t=1}^\zeta x_{tij}$, where all $x_{tij}$ are independently drawn as
 \begin{equation}
 x_{tij} = \begin{cases}
  1 & \text{with probability } \frac{1}{2b} \\
   -1 & \text{with probability } \frac{1}{2b} \\
   0 &  \text{with probability } 1 - \frac{1}{b}
 \end{cases}
 \end{equation}
\end{lemma}
\begin{proof}
Condition on $\Omega_{\zeta, b, I}$ throughout the proof. 


First define the intermediate matrix $\widetilde{U}$ to be the matrix with entries $\widetilde{U}_{ij} = \frac{1}{\zeta} \sum_{t=1}^\zeta \alpha_{it}\varrho_{it}\eta_{jt}[s_{it}]$, where each vector $\eta_{jt}$ is coupled with $\xi_{jt}$ as in Claim~\ref{claim:xi_to_eta}. Notice that each entry $\widetilde{U}_{ij}$ is independently drawn, but their distributions depend on the $\alpha_{it}$.


The operator norm is convex. So by Jensen's inequality, additionally conditioning on the $\xi_{jt}$, and applying Claim~\ref{claim:xi_to_eta}, we have 
\begin{equation}
c_b^{-1}\|\widetilde{\Omega}_I\widetilde{\Omega}_{I^c}^{\top}W\| \leq \mathbb{E}_{\eta_{jt}} [\|\widetilde{U}W\| \mid \{\xi_{tj}\}] .
\end{equation}
Taking expectation over the $\xi_{tj}$ on both sides, yields
\begin{equation}\label{eq:jenson1}
 \mathbb{E}\left[\|\widetilde{\Omega}_I\widetilde{\Omega}_{I^c}^{\top}W\|\right] \leq c_b\mathbb{E}_{\eta_{jt}, \xi_{jt}} [\|\widetilde{U}W\|] = c_b\mathbb{E}_{\tilde{U}} [\|\widetilde{U}W\|]
\end{equation}
Now we will apply Jensen's inequality again to reduce this to the matrix $U$ from the claim which has i.i.d.\ entries. Observe that $\widetilde{U}$ has entries $\widetilde{U}_{ij} = \frac{1}{\zeta}\sum_{t = 1}^{\zeta} \tilde{x}_{tij}$, where
 \begin{equation}
 \tilde{x}_{tij} = \begin{cases}
  \alpha_{ti} & \text{with probability } \frac{1}{2b} \\
   -\alpha_{ti} & \text{with probability } \frac{1}{2b} \\
   0 &  \text{with probability } 1 - \frac{1}{b}
 \end{cases}
 \end{equation}
 Couple the variables $\tilde{x}_{tij}$ and $x_{tij}$ such that 
 \begin{align}
     \tilde{x}_{tij} = \alpha_{it}x_{tij} = \begin{cases}
         \mathbb{E}[x_{tij}] & \alpha_{ti} = 0 \\
         x_{tij} & \alpha_{ti} = 1.
     \end{cases}
 \end{align}
 Thus we have from Jenson's inequality again that conditional on all the $x_{tij}$ where $\alpha_{ti} = 1$,
\begin{align}
    \|\widetilde{U}W\| \leq \mathbb{E}_{x_{tij}: \alpha_{ti} = 0}\|UW\|.
\end{align}
Taking expectation over all the $x_{tij}$ where $\alpha_{ti} = 1$ yields
\begin{align}
     \mathbb{E}_{\tilde{U}}\|\widetilde{U}W\| \leq  \mathbb{E}_{U}\|UW\|.
\end{align}
Combining with \eqref{eq:jenson1} yields the lemma.
\end{proof}

We are now ready to prove Lemma~\ref{lem:cross-gram}, which we restate here for the reader's convenience. 

\crossgram*

\begin{proof}
Fix $\Omega_{I}$.
By Lemma~\ref{lemma:jensen},
\begin{equation}\label{eq:cross-gram-U}
  \E\bracket{\norm{\widetilde{\Omega}_I\widetilde{\Omega}_{I^c}^{\top}W} \;\Big|\; \Omega_{I}} \le c_b\,\E\bracket{\norm{\frac{1}{\zeta} U W}}, \qquad \text{where } U_{ij} = \sum_{t=1}^\zeta x_{tij} \quad (i \in I,\ j \in J),
\end{equation}
where all variables $x_{tij}$ for $t \in [\zeta], i \in I, j \in J$ are mutually independent and distributed as $\{-1, 0, 1\}$ with probabilities $\frac{1}{2b}, 1 - \frac{1}{b}, \frac{1}{2b}$. The entries $U_{ij}$ of the $N \times |J|$ matrix $U$ are mutually independent and symmetric. For each pair $(i, j) \in I \times J$, let $c_{ij} = \sum_{t=1}^\zeta \1_{\{x_{tij} \neq 0\}} \sim \mathrm{Bin}(\zeta, 1/b)$ be the number of blocks in which $(i, j)$ is active, and decompose $U$ into low- and high-multiplicity parts:
\[
  U^{\mathrm{lo}}_{ij} = U_{ij} \1_{\{c_{ij} \le 2\}}, \qquad U^{\mathrm{hi}}_{ij} = U_{ij} \1_{\{c_{ij} \ge 3\}}, \qquad U = U^{\mathrm{lo}} + U^{\mathrm{hi}}.
\]

\paragraph{High-multiplicity part $U^{\mathrm{hi}}$.}
The entries of $U^{\mathrm{hi}}$ are mutually independent and centered, and conditional on $c_{ij} = c$, $U_{ij}$ is a sum of $c$ independent Rademacher signs so $\E[U_{ij}^2 \mid c_{ij} = c] = c$. Using the elementary inequality $c \1_{\{c \ge 3\}} \le 3\binom{c}{3}$ for integers $c \ge 0$ and $\E\bracket{\binom{c_{11}}{3}} = \binom{\zeta}{3} b^{-3}$, together with $\sum_{j \in J} \|w_j\|^2 = \tr(W^\top W) = d$, we bound the expected squared Frobenius norm of $U^{\mathrm{hi}} W / \zeta$:
\begin{equation}\label{eq:U-hi-bound}
  \E\bracket{\norm{\frac{1}{\zeta}U^{\mathrm{hi}} W}_F^2} = \frac{1}{\zeta^2} \sum_{i \in I} \sum_{j \in J} \E\bracket{(U^{\mathrm{hi}}_{ij})^2} \norm{w_j}^2 = \frac{Nd}{\zeta^2}\,\E\bracket{c_{11}\1_{\{c_{11} \ge 3\}}} \le \frac{3Nd}{\zeta^2 b^3} \binom{\zeta}{3}.
\end{equation}

\paragraph{Low-multiplicity part $U^{\mathrm{lo}}$.}
Our goal will be to use Lemma~\ref{lem:indep-norm} to control the operator norm of $U^{\mathrm{lo}}W$, but since the entries of $U^{\mathrm{lo}}$ are distributed differently than those of \ref{lem:hollow-moment}, we will need to leverage convexity again to reduce to this case.  Put $B = b/\zeta \ge 2$, using $b \ge 2\zeta$. Each entry $V = U^{\mathrm{lo}}_{ij}$ is symmetric, bounded in magnitude by $|V| \le 2$, and satisfies $\E[|V|] \le \E\bracket{\sum_{t=1}^\zeta |x_{tij}|} = \zeta/b = 1/B$. Let $\eta_B \in \{-1, 0, 1\}$ take values $\pm 1$ with probability $1/(2B)$ each and $0$ with probability $1 - 1/B$. We seek to control the $\|U^{\mathrm{lo}}W\|$ by replacing each copy of $V$ with a copy of $2\eta_{B}$. 

\begin{claim}\label{claim:convex_dom}
For any convex function $\phi$, we have 
\begin{equation}
    \mathbb{E}[\phi(V)] \leq \mathbb{E}[\phi(2\eta_B)].
\end{equation}
\end{claim}
\begin{proof}
Let $\psi$ be the even part of $\phi$, namely $\frac{1}{2}(\phi(u) + \phi(-u))$. Observe that $\psi$ is convex because the average of two convex function is convex. Further, its convexity implies that it is non-decreasing on $[0, 2]$, and satisfies for any symmetric $X$,
\begin{align}\label{eq:symexp}
    \mathbb{E}_X[\psi(X)] = \mathbb{E}_X[\phi(X)],
\end{align}
Interpolating linearly on $[0, 2]$ gives
\begin{align}
      \E[\psi(V)] &\le \paren{1 - \frac{\E[|V|]}{2}}\psi(0) + \frac{\E[|V|]}{2}\psi(2) \\
      &\le \paren{1 - \frac{1}{B}}\psi(0) + \frac{1}{B}\psi(2) = \E[\psi(2\eta_B)],
\end{align}
where the second line follows from the fact that $\psi$ is non-decreasing on $[0, 2]$ and the bound $\mathbb{E}[|V|] \leq 1/B$. Finally, applying \eqref{eq:symexp} twice with the symmetric distributions of $V$ and $2\eta_B$ respectively, yields the desired claim.
\end{proof}

Applying Claim~\ref{claim:convex_dom} entry by entry to the convex function $U^{\mathrm{lo}} \mapsto \|U^{\mathrm{lo}} W\|$ yields
\[
  \E\bracket{\norm{U^{\mathrm{lo}} W}} \le 2\,\E\bracket{\norm{X_N W}},
\]
where $X_N \in \{-1, 0, 1\}^{N \times |J|}$ has independent entries with non-zero probability $1/B$. Letting $a = N/B + 1 \ge 1$ and $m = aB \ge N$, we may append $m - N$ independent rows to $X_N$ to form an $m \times |J|$ matrix $X_m$, which can only increase the operator norm $\|X_N W\| \le \|X_m W\|$. Applying Lemma~\ref{lem:indep-norm} to $X_m W$ with $B = b/\zeta$ gives
\begin{equation}\label{eq:U-lo-bound}
  \E\bracket{\norm{\frac{1}{\zeta} U^{\mathrm{lo}} W}} \le \frac{2}{\zeta}\paren{\E\bracket{\norm{X_m W}^2}}^{1/2} \le \frac{O(1)}{\zeta}\sqrt{a + \frac{D_p}{B} + p} \le O(1)\sqrt{\frac{N + D_p}{k} + \frac{p+1}{\zeta^2}}.
\end{equation}
Combining~\eqref{eq:cross-gram-U},~\eqref{eq:U-hi-bound}, and~\eqref{eq:U-lo-bound} via the triangle inequality and Cauchy--Schwarz ($\E[\|U^{\mathrm{hi}} W / \zeta\|] \le (\E[\|U^{\mathrm{hi}} W / \zeta\|_F^2])^{1/2}$) completes the proof of Lemma~\ref{lem:cross-gram}.
\end{proof}

\subsection{High-moment bound for the full sketch}\label{subsec:full-sketch-moment}

In this subsection, we establish a high-order ($2p$-th) moment bound for the full sketch $G = U^\top \Omega_{\zeta, b} \Omega_{\zeta, b}^\top U$ of an arbitrary isometry $U \in \R^{n \times D}$. This bound is the main tool in proving Theorem~\ref{thm:main} for the high-sparsity regime $\zeta = \Omega(\log(r))$.
In this regime, the sketch possesses enough independent blocks that it acts directly as an OSE on the entire subspace. The entrywise coupling~\eqref{eq:selector-identity} expresses the off-diagonal error $G - \Id_D$ as the conditional expectation of the hollow matrix $F$, directly transferring the moment bound of Fact~\ref{lem:hollow-moment} to the SparseStack sketch.

\begin{lemma}\label{lem:stacked-moment}
There exists an absolute constant $C_{\mathrm{mom}} > 0$ such that for any fixed isometry $U \in \R^{n \times D}$ ($D \ge 1$), any $b \ge 2$, $\zeta \ge 1$, and any integer $p \ge 1$, the matrix $G = U^\top \Omega_{\zeta, b} \Omega_{\zeta, b}^\top U$ for a $\zeta$-block SparseStack $\Omega_{\zeta, b} \in \R^{n \times k}$ ($k = \zeta b$) satisfies
\begin{equation}\label{eq:stacked-moment}
  \E\bracket{\norm{G - \Id_D}^{2p}} \le D \bracket{C_{\mathrm{mom}}\paren{\sqrt{\frac{D + 2p + 1}{k}} + \frac{D + 2p + 1}{k} + \frac{2p+1}{\zeta}}}^{2p}.
\end{equation}
\end{lemma}

\begin{proof}
Since every row of $\Omega_{\zeta, b}$ has deterministic unit Euclidean norm and $U^\top U = \Id_D$, the diagonal part of $U^\top \Omega_{\zeta, b} \Omega_{\zeta, b}^\top U = \frac{1}{\zeta} \sum_{t=1}^\zeta U^\top \Phi_t \Phi_t^\top U$ is deterministically equal to $\sum_{i=1}^n u_i u_i^\top = \Id_D$. Thus $G - \Id_D$ is a purely off-diagonal sum over distinct coordinates $i \neq i'$:
\[
  G - \Id_D = \frac{1}{\zeta} \sum_{t=1}^\zeta \sum_{i \neq i'} \ip{\xi_{ti}, \xi_{ti'}} u_i u_{i'}^\top,
\]
where $\xi_{ti} = \varrho_{ti} e_{s_{ti}} \in \R^b$ is the $i$-th column vector of the sketching block $\Phi_t^\top \in \R^{b \times n}$. Applying the coupling~\eqref{eq:selector-identity} independently to each $(\eta_{ti}, \xi_{ti})$, independence across distinct coordinates $i \neq i'$ gives
\[
  G - \Id_D = \E\bracket{c_b^2 F \;\Big|\; \{\xi_{ti}\}},
\]
where $F$ is the off-diagonal matrix of Fact~\ref{lem:hollow-moment} with $B = b$, $a = \zeta$, and $m = k = \zeta b$. By conditional Jensen's inequality for the convex function $M \mapsto \|M\|^{2p} \le \tr(M^{2p})$, together with $c_b < 2$ and Fact~\ref{lem:hollow-moment}, we obtain~\eqref{eq:stacked-moment} with $C_{\mathrm{mom}} = 12(3 + \sqrt{2})$.
\end{proof}

\section{Preserving the Heavy Coordinates}\label{sec:repair}

Our goal in this section is to prove Theorem~\ref{thm:repair}, showing that the second-half sketch $(\Omega_{\zeta, b}^{(2)})^\top \in \R^{k \times n}$ ($k = \zeta b$) preserves the norm on the heavy coordinates $y_I$, for $I \subseteq [n]$ of size $|I| \le \min\{r, b/16\}$. For any $y = Qx$, decomposing $y$ into its heavy coordinates $h = (Qx)_I$ and complementary coordinates $(Qx)_{I^c} = Wf$ (with $W$ an orthonormal basis for $\col(Q_{I^c})$ and $\|f\| = \|(Qx)_{I^c}\|$) splits the sketch as
\begin{equation}\label{eq:Y-split-HZ}
  (\Omega_{\zeta, b}^{(2)})^\top Q x = (\Omega_{\zeta, b, I}^{(2)})^\top h + (\Omega_{\zeta, b, I^c}^{(2)})^\top Wf,
\end{equation}

Our main approach in showing that the norm of $y_I$ is preserved by $(\Omega_{\zeta, b, I}^{(2)})^\top$ will be finding rows in $(\Omega_{\zeta, b, I}^{(2)})^\top$ which only have a single non-zero entry. To formalize this idea, we introduce the following definitions.

\begin{definition}[Isolated bucket]\label{def:isolated}
For a matrix $\Omega^{\top} \in \R^{k \times N}$, we say that column $i \in [N]$ lands in bucket $r$ if $\Omega^{\top}_{ri} \neq 0$. We say bucket $r$ is \em isolated \em if the $r$th row of $\Omega^{\top}$ has a single non-zero entry. 
\end{definition}
In $\Omega_{\zeta, b, I}^{(2)}$, if $i \in I$ lands in an isolated bucket $r$, then $((\Omega_{\zeta, b, I}^{(2)})^\top y_I)_r = \pm \frac{y_i}{\sqrt{\zeta}}$. Thus up to constants, the mass of coordinate $y_i$ will be preserved by $(\Omega_{\zeta, b, I}^{(2)})^\top$ whenever a constant fraction of the rows in buckets $i$ lands in are isolated. We formalize this in the following definition.

\begin{definition}[Good coordinate]\label{def:good}
For a matrix $\Omega^{\top} \in \R^{k \times N}$, we say that a coordinate $i \in [N]$ is \em good \em if in at least a $\frac{11}{12}$ fraction of the buckets $i$ lands in are isolated.
\end{definition}

Fix some $i \in I$. When $b \ge 16|I|$, in each block $\Phi$ of $(\Omega_{\zeta, b, I}^{(2)})^\top$, with probability at least $15/16$, $i$ lands in an isolated bucket. Thus with probability  $1 - e^{-\Omega(\zeta)}$, coordinate $i$ is good.

At sparsity $\zeta = O(\sqrt{\log r})$, however, the failure probability $e^{-\Omega(\zeta)}$ is far too large for a union bound over all $|I| \approx r/\zeta$ coordinates. We overcome this by splitting $((\Omega_{\zeta, b, I}^{(2)})^\top$ into two independent groups of $\zeta_1 = \zeta/2 $ blocks. Our strategy is as follows:
\begin{enumerate}
    \item Using the first group, we will argue that a $1 - e^{-\Omega(\zeta)}$ fraction of the coordinates in $I$ are good, and thus their norm is mostly preserved (Lemma~\ref{lem:single-group-lower}).
    \item For the remaining coordinates $B \subset I$ which are not good for the first group of block, we will consider the second group of blocks, restricted to the columns in $B$. Since $|B| \lesssim \frac{r}{\zeta}e^{-\Omega(\zeta)}$, on this new block, the coordinates in $B$ will have fewer collisions, and thus, we will be able to show that with high probability, all elements of $B$ are good in this submatrix. We show this in Lemma~\ref{lem:two-group-isolation}. We can then again use Lemma~\ref{lem:single-group-lower} to show that their norm is preserved.
    \item Finally, we put together these estimates to prove Theorem~\ref{thm:repair}.
\end{enumerate}

The following lemma analyzes this single-group behavior for an arbitrary coordinate set $S \subseteq [n]$ of size $|S| \le r$ and complementary isometry $W \in \R^{|S^c| \times d}$.

\begin{lemma}\label{lem:single-group-lower}
Fix any constants $t, \delta_{\mathrm{rep}} > 0$, a set $S \subseteq [n]$, and an isometry  $W \in \R^{|S^c| \times d}$ for $d \le 2r$. Suppose $\Omega^{\top} \in \R^{k \times n}$ is a SparseStack matrix for $\zeta = \Omega(\sqrt{\log r})$ and $k = \zeta b = \Omega(r)$. With probability at least $1 - \delta_{\mathrm{rep}}/4$ over $\Omega$, we have simultaneously for all $h \in \R^{|S|}$ and $f \in \R^d$:
\begin{equation}\label{eq:single-group-quad}
  \norm{\Omega_{S}^{\top} h + \Omega_{S^c}^{\top} W f}^2 + t \norm{f}^2 \ge \paren{\frac{11}{12} - t}\norm{h_G}^2 - t \norm{h_B}^2,
\end{equation}
where $G$ contains all good coordinates in $S$ (for the matrix $\Omega_{\zeta, b, S}^{\top}$), and $B = S \setminus G$.
\end{lemma}

\begin{proof}
Set $L = \log(er) \ge 1$. Let $T \subseteq [k]$ be the set of all isolated buckets of $\Omega_{S}^{\top}$, and define
$\widetilde{\Omega}_S^{\top}$ and $\widetilde{\Omega}_{S^c}^{\top}$ respectively be the restrictions of $\Omega_S^{\top}$ and $\Omega_{S^c}^{\top}$ to rows in $T$. We apply Lemma~\ref{lem:cross-gram} with block count $\zeta$, $|S| \le r$, $d \le 2r$, and $p = \lceil \log(4er) \rceil = \Theta(L)$. Lemma~\ref{lem:cross-gram} yields
\[
  \E\bracket{\norm{\widetilde{\Omega}_S \widetilde{\Omega}_{S^c}^{\top}W} \;\Big|\; \Omega_{S}} \le O\!\paren{\sqrt{\frac{r}{k} + \frac{L}{\zeta^2}} + \sqrt{\frac{r^2 \zeta}{b^3}}}.
\]
Choosing the implied constants in $\zeta = \Omega(\sqrt{L})$ and $k = \Omega(r)$ sufficiently large relative to $t, \delta_{\mathrm{rep}}$, Markov's inequality ensures that conditionally on $\Omega_S$, the event $\left\{\norm{\widetilde{\Omega}_S \widetilde{\Omega}_{S^c}^{\top}W} \le t\right\}$ occurs with probability at least $1 - \delta_{\mathrm{rep}}/4$.

Furthermore, because $\widetilde{\Omega}_S^{\top}$ retains only the isolated buckets, every row of $\widetilde{\Omega}_S^{\top}$ contains at most one non-zero entry, so distinct columns of $\widetilde{\Omega}_S^{\top}$ have pairwise disjoint supports. Hence, the columns are mutually orthogonal, which implies $\|\widetilde{\Omega}_S^{\top} h\|^2 = \sum_{i \in S} h_i^2 \frac{a_i}{\zeta}$, where $a_i$ is the number of isolated buckets of $\Omega_S^{\top}$ that $i$ lands in. Thus by definition of good, $a_i \geq \frac{11}{12}\zeta$ for $i \in G$.

On the event  $\left\{\norm{\widetilde{\Omega}_S \widetilde{\Omega}_{S^c}^{\top}W} \le t\right\}$, dropping coordinates outside $T$ and completing the square with $2\|h\|\|f\| \le \|h\|^2 + \|f\|^2 = \|h_G\|^2 + \|h_B\|^2 + \|f\|^2$ yields:
\begin{align*}
  \norm{\Omega_{S}^{\top} h + \Omega_{S^c}^{\top} W f}^2 + t \norm{f}^2
  &\ge \norm{\widetilde{\Omega}_S^{\top} h + \widetilde{\Omega}_{S^c}^{\top} W f}^2 + t \norm{f}^2 \\
  &= \sum_{i \in S} \frac{a_i}{\zeta} h_i^2 + 2 h^\top \widetilde{\Omega}_S\widetilde{\Omega}_{S^c}^{\top}W f + \norm{\widetilde{\Omega}_{S^c}^{\top}W f}^2 + t \norm{f}^2 \\
  &\ge \frac{11}{12} \norm{h_G}^2 - 2 t \norm{h} \norm{f} + t \norm{f}^2 \\
  &\ge \paren{\frac{11}{12} - t}\norm{h_G}^2 - t \norm{h_B}^2,
\end{align*}
completing the proof.
\end{proof}

It remains to preserve the norm of the coordinates that fail to land in sufficiently many isolated buckets in the first group.
Among the coordinates of $I$, the set of coordinates that fail to achieve isolation in at least $\frac{11}{12}\zeta_1$ blocks of Group~1, denoted by $B$, has expected size $\E[|B|] \le |I| e^{-\Omega(\zeta)}$.
Our goal is to ensure that their norms are fully preserved in the second group. While this was impossible for the full set $I$, the size of $B$ is now small enough that a union bound over all coordinates in $B$ becomes applicable. The following lemma formalizes this via a balls-into-bins analysis across the two groups.

\begin{lemma}\label{lem:two-group-isolation}
For each $i \in I$, let $a_i$ be the number of blocks in Group~1 where $i$ lands in a bucket containing no other element of $I$. For $\zeta = \Omega(\sqrt{\log r})$ and $|I| \le \min\{r, b/16\}$, two independent $\zeta_1$-block groups satisfy:
\begin{enumerate}[label=(\roman*)]
  \item In Group~1, the bad set $B = \{i \in I : a_i < \frac{11}{12}\zeta_1\}$ satisfies $|B| \le |I| e^{-\Omega(\zeta)}$ with probability at least $1 - \delta_{\mathrm{rep}}/4$.
  \item Conditionally on any such Group~1, with probability at least $1 - \delta_{\mathrm{rep}}/4$ over Group~2, \emph{every} coordinate in $B$ lands in a bucket containing no other element of $B$ in at least $\frac{11}{12}\zeta_1$ blocks of Group~2.
\end{enumerate}
\end{lemma}

\begin{proof}
Set $c_1 = 1/2304$. For part~(i), since $|I| \le b/16$, in each block of Group~1 a coordinate $i \in I$ avoids colliding with the other $|I| - 1$ elements of $I$ with probability at least $1 - (|I|-1)/b \ge 15/16$. Thus $\E[a_i] \ge \frac{15}{16}\zeta_1$.
Changing a single column index in block $t$ alters $a_i$ by at most $1$. Applying McDiarmid's inequality (Fact~\ref{fact:standard-prob}) gives
\[
  \Prb\bracket{i \in B} = \Prb\bracket{a_i < \frac{11}{12} \zeta_1} \le \exp\!\paren{-2\paren{\frac{1}{48}}^2 \zeta_1} = e^{-2c_1 \zeta_1}.
\]
Thus $\E[|B|] \le |I| e^{-2c_1 \zeta_1}$, and Markov's inequality yields $\Pr[\abs{B} > |I| e^{-c_1 \zeta_1}] \le e^{-c_1 \zeta_1} \le \delta_{\mathrm{rep}}/4$ for $\zeta = \Omega(\sqrt{\log r})$ with sufficiently large constant.

For part~(ii), condition on any realization of Group~1 satisfying $|B| \le |I| e^{-c_1 \zeta_1}$. For any fixed $i \in B$ and any block $t$ of Group~2, the probability that $i$ collides with another element of $B$ in block $t$ is at most
\[
  \frac{\abs{B}}{b} \le \frac{|I|}{b} e^{-c_1 \zeta_1} \le \frac{1}{16} e^{-c_1 \zeta_1}.
\]
Across the $\zeta_1$ independent blocks of Group~2, coordinate $i \in B$ fails to achieve isolation from $B \setminus \{i\}$ in at least $\frac{11}{12}\zeta_1$ blocks only if at least $u = \zeta_1/12$ collisions occur. Union-bounding over all $\binom{\zeta_1}{u} \le 2^{\zeta_1}$ subsets of blocks of size $u$, there is an absolute constant $c_2 > 0$ such that the probability of failing isolation in $\ge \zeta_1/12$ blocks is at most
\[
  2^{\zeta_1} \left(\frac{\abs{B}}{b}\right)^{\zeta_1/12} \le 2^{\zeta_1} \paren{\frac{e^{-c_1 \zeta_1}}{16}}^{\zeta_1/12} \le e^{-c_2 \zeta_1^2}.
\]
Taking a union bound over all $|B| \le r$ coordinates in $B$ gives
\[
  \Prb\bracket{\exists i \in B \text{ isolated in } < \frac{11}{12}\zeta_1 \text{ blocks of Group 2} \;\Big|\; \text{Group 1}} \le r e^{-c_2 \zeta_1^2} \le \frac{\delta_{\mathrm{rep}}}{4}
\]
for $\zeta = \Omega(\sqrt{\log r})$ with sufficiently large implied constant, completing the proof.
\end{proof}

We are now ready to combine Lemmas~\ref{lem:single-group-lower} and~\ref{lem:two-group-isolation} to prove Theorem~\ref{thm:repair}.

\thmrepair*

\begin{proof}
Let $I^c = [n] \setminus I$, and let $W \in \R^{|I^c| \times d}$ with $d = \mathrm{rank}(Q_{I^c}) \le r$ be an orthonormal basis for $\col(Q_{I^c})$, so that $(Qx)_{I^c} = Wf$ where $f \in \R^d$ satisfies $\|f\| = \|(Qx)_{I^c}\|$. Write $h = (Qx)_I \in \R^{|I|}$, so that $\|h\|^2 + \|f\|^2 = \|Qx\|^2 = \|x\|^2$.

Partition the $\zeta$ blocks of $\Omega_{\zeta, b}^{(2)}$ into two disjoint groups, Group~1 and Group~2, of $\zeta_1 = \zeta/2$ blocks, with corresponding sketching matrices $\Omega_1, \Omega_2 \in \R^{n \times k_1}$ ($k_1 = \zeta_1 b = \frac{\zeta}{2} b$). Then
\[
  \norm{Yx}^2 = \frac{\zeta_1}{\zeta}\paren{\norm{\Omega_{1, I}^\top h + \Omega_{1, I^c}^\top W f}^2 + \norm{\Omega_{2, I}^\top h + \Omega_{2, I^c}^\top W f}^2}.
\]

In Group~1, we apply Lemma~\ref{lem:single-group-lower} with target set $S = I$ and isometry $W$. By Lemma~\ref{lem:two-group-isolation}(i), Group~1 isolates a good set $G \subseteq I$, leaving a bad set $B = I \setminus G$ of size $|B| \le |I| e^{-c_1 \zeta_1}$ outside a failure event of probability at most $\delta_{\mathrm{rep}}/4$. By Lemma~\ref{lem:single-group-lower} with $t = t_0$, outside another failure event of probability at most $\delta_{\mathrm{rep}}/4$, for all $x \in \R^r$:
\begin{equation}\label{eq:group1-lower}
  \norm{\Omega_{1, I}^\top h + \Omega_{1, I^c}^\top W f}^2 + t_0 \norm{f}^2 \ge \paren{\frac{11}{12} - t_0}\norm{h_G}^2 - t_0 \norm{h_B}^2.
\end{equation}

Next, condition on Group~1 on the event where $|B| \le |I| e^{-c_1 \zeta_1}$ and~\eqref{eq:group1-lower} hold. In Group~2, we set $B$ as the target set ($S = B$) and absorb $G$ into the complementary coordinates via the augmented isometry
\[
  W_+ = \diag(\Id_G, W) \in \R^{\abs{G \cup I^c} \times (\abs{G} + d)},
\]
which has rank $|G| + d \le |I| + d \le 2r$ and is fixed conditionally on Group~1. The complementary vector becomes $f_+ = (h_G, f)$, which satisfies $\|f_+\|^2 = \|h_G\|^2 + \|f\|^2$. By Lemma~\ref{lem:two-group-isolation}(ii), every coordinate in $B$ is isolated in Group~2 with conditional probability at least $1 - \delta_{\mathrm{rep}}/4$, so the bad set in Group~2 is empty. Applying Lemma~\ref{lem:single-group-lower} to Group~2 with target set $S = B$, isometry $W_+$, and $t = t_0$ shows that outside a conditional failure event of probability at most $\delta_{\mathrm{rep}}/4$, for all $x \in \R^r$:
\begin{equation}\label{eq:group2-lower}
  \norm{\Omega_{2, I}^\top h + \Omega_{2, I^c}^\top W f}^2 + t_0\paren{\norm{h_G}^2 + \norm{f}^2} \ge \paren{\frac{11}{12} - t_0}\norm{h_B}^2.
\end{equation}

Adding~\eqref{eq:group1-lower} and~\eqref{eq:group2-lower} cancels $-t_0\|h_G\|^2$ and $-t_0\|h_B\|^2$ against the $(\frac{11}{12} - t_0)$ terms, giving
\begin{align*}
  \norm{\Omega_{1, I}^\top h + \Omega_{1, I^c}^\top W f}^2 &+ \norm{\Omega_{2, I}^\top h + \Omega_{2, I^c}^\top W f}^2 + 2t_0 \norm{f}^2 \\
  &\ge \paren{\frac{11}{12} - 2t_0}\paren{\norm{h_G}^2 + \norm{h_B}^2} = \paren{\frac{11}{12} - 2t_0}\norm{h}^2.
\end{align*}
Therefore,
\[
  \norm{Yx}^2 + t_0 \norm{f}^2 \ge \beta \norm{h}^2
\]
with $\beta = (\frac{11}{24} - t_0) > 0$. Substituting $\|h\|^2 = \|x\|^2 - \|f\|^2$ and $\|f\|^2 = \|(Qx)_{I^c}\|^2$ gives
\[
  \norm{Yx}^2 + (\beta + t_0)\norm{(Qx)_{I^c}}^2 \ge \beta \norm{x}^2 \qquad \text{for all } x \in \R^r.
\]
Union bound over the four failure events ensures that this holds with probability at least $1 - \delta_{\mathrm{rep}}$.
\end{proof}

\section{Proof of the main theorem}\label{sec:main-proof}
We are now ready to prove the main theorem.

\thmmain*

\begin{proof}
We write the total number of blocks as $2\zeta$ and the total embedding dimension as $2k = 2\zeta b \ge Kr$, so that the sketched matrix is $A = \Omega_{2\zeta, b}^\top Q \in \R^{2k \times r}$. 
Partition the $2\zeta$ blocks of $\Omega_{2\zeta, b} \in \R^{n \times 2k}$ into two independent halves: the first half $\Omega_{\zeta, b}^{(1)} \in \R^{n \times k}$ and the second half $\Omega_{\zeta, b}^{(2)} \in \R^{n \times k}$ (where $k = \zeta b$). Let $X = (\Omega_{\zeta, b}^{(1)})^\top Q \in \R^{k \times r}$ and $Y = (\Omega_{\zeta, b}^{(2)})^\top Q \in \R^{k \times r}$. Then $A = \frac{1}{\sqrt{2}}\begin{bmatrix} X \\ Y \end{bmatrix} \in \R^{2k \times r}$.

We set the parameters as follows:
fix $t_0 = \frac{11}{48}$ in Theorem~\ref{thm:repair}, so that $\beta = \frac{11}{24} - t_0 = \frac{11}{48}$ and $\gamma = \beta + t_0 = \frac{11}{24}$;
fix $\eps = \frac{1}{3}$ in Theorem~\ref{thm:localization}, so that $u := \frac{8}{3}\eps^2 = \frac{8}{27} < \frac{1}{2} = \frac{\beta}{\gamma}$;
and let $\lambda = \Omega(\eps)$, $C_{\mathrm{loc}}, M > 0$ be the constants from Theorem~\ref{thm:localization} for $\eps = 1/3$ and failure probability $\delta_{\mathrm{loc}} = \frac{1}{40}$, setting $v = \frac{8}{3\lambda^2}$ (so that $\gamma v \ge 1$).

First, apply Theorem~\ref{thm:localization} to the first half $X = (\Omega_{\zeta, b}^{(1)})^\top Q \in \R^{k \times r}$, which has $k = \zeta b \ge \frac{1}{2} K r \ge C_{\mathrm{loc}} r$ for $K$ sufficiently large. Outside an event of probability at most $\delta_{\mathrm{loc}} = \frac{1}{40}$, Theorem~\ref{prop:energy-cert} supplies a coordinate set $I \subseteq [n]$ determined by $X$, with $m = Mr/\zeta$, such that
\begin{equation}\label{eq:HJ-domination}
  \abs{I} \le 8m, \qquad Q_{I^c}^\top Q_{I^c} \preceq u \Id_r + v X^\top X, \qquad \text{where } I^c = [n] \setminus I.
\end{equation}
Since $b = k/\zeta \ge Kr/(2\zeta)$, choosing $K \ge 256M$ ensures that
\[
  \abs{I} \le 8m = \frac{8Mr}{\zeta} \le \min\set{r,\, \frac{b}{16}},
\]
where $\frac{8Mr}{\zeta} \le r$ holds because $\zeta = \Theta(\sqrt{\log r}) \ge 8M$ for $r$ sufficiently large. Furthermore, $k \ge \frac{1}{2}Kr$ satisfies the embedding dimension requirement of Theorem~\ref{thm:repair} for $K$ sufficiently large with failure probability $\delta_{\mathrm{rep}} = \frac{1}{40}$.

Next, condition on the first half $X$ and hence on the realized set $I$. Because the second half $Y = (\Omega_{\zeta, b}^{(2)})^\top Q$ is independent of $X$, Theorem~\ref{thm:repair} applies conditionally on $X$ and shows that with conditional failure probability at most $\delta_{\mathrm{rep}} = \frac{1}{40}$,
\begin{equation}\label{eq:Y-repair-matrix}
  Y^\top Y + \gamma Q_{I^c}^\top Q_{I^c} \succeq \beta \Id_r,
\end{equation}
where $\gamma = \beta + t_0$. Substituting~\eqref{eq:HJ-domination} into~\eqref{eq:Y-repair-matrix} gives
\[
  Y^\top Y + \gamma v X^\top X \succeq (\beta - \gamma u)\Id_r.
\]
Finally, since $\gamma v \ge 1$ and $\beta - \gamma u > 0$,
\[
  A^\top A = \frac{1}{2}\paren{X^\top X + Y^\top Y} \succeq \frac{1}{2\gamma v}\paren{Y^\top Y + \gamma v X^\top X} \succeq \frac{1}{2v}\paren{\frac{\beta}{\gamma} - u}\Id_r =: \alpha \Id_r,
\]
where $\alpha = \frac{1}{2v}(\frac{\beta}{\gamma} - u) = \frac{11}{108 v} > 0$ is an absolute constant. The total failure probability is at most $\delta_{\mathrm{loc}} + \delta_{\mathrm{rep}} = \frac{1}{40} + \frac{1}{40} = \frac{1}{20}$, completing the proof of Theorem~\ref{thm:main}.
\end{proof}

\section{Bipartite Graph Preprocessing and the $C_4$-Free Graph $G_0$}
\label{sec:graph-pruning}

As we discussed above, we fix a deterministic matrix $S \in \R^{n \times k}$ with row sparsity $\zeta$ (meaning $|\supp(s_i)| \le \zeta$ for all $i \in [n]$) and $\|S\|_F^2 \le 10n$ satisfying the overlap condition, and evaluate its injectivity against a random coordinate subspace $R \sim \mathrm{Unif}\binom{[n]}{r}$.
We represent $S$ as a bipartite graph $G = ([n], [k], E)$ between the $n$ rows and $k$ columns, with an edge between row $i$ and column $j$ whenever $S_{ij} \ne 0$.
In this section, we establish the graph preprocessing pipeline that filters out pathological rows and constructs a large $C_4$-free subgraph $G_0$. All subsequent tree cancellation arguments will be executed directly on $G_0$.

\subsection{Row Preprocessing}
\label{subsec:row-filtering}

We first clarify the independence structure of the preprocessing steps. Because the matrix $S$ is fixed deterministically, properties such as row Euclidean norms and support sizes are fixed deterministic attributes of each row in $S$. 
Specifically, we classify the row set $[n]$ into regular rows and three exceptional sets:
\begin{enumerate}
  \item Heavy rows: $I_{\mathrm{heavy}} = \{i \in [n] : \|s_i\|_2^2 > M\}$, where $M = 160$.
  \item Small-norm rows: $I_{\mathrm{small}} = \{i \in [n] : \|s_i\|_2^2 < \alpha\}$.
  \item One-sparse rows: $I_{\mathrm{one}} = \{i \in [n] : |\supp(s_i)| = 1\}$.
\end{enumerate}

The following lemma shows that heavy rows constitute at most a $1/16$ fraction of all rows.

\begin{lemma}
\label{lem:heavy-row-filter}
Let $M = 160$. In any matrix $S \in \R^{n \times k}$ with $\|S\|_F^2 \le 10n$, the set of heavy rows satisfies $|I_{\mathrm{heavy}}| \le n / 16$.
\end{lemma}

\begin{proof}
By definition of the Frobenius norm, we have
\[
  10n \ge \|S\|_F^2 = \sum_{i=1}^n \|s_i\|_2^2 \ge \sum_{i \in I_{\mathrm{heavy}}} \|s_i\|_2^2 > M |I_{\mathrm{heavy}}| = 160|I_{\mathrm{heavy}}|.
\]
\end{proof}

We next show that if either small-norm rows or one-sparse rows are abundant in $S$, then injectivity breaks directly on a random coordinate subspace $R$ without requiring tree structures.

\begin{lemma}
\label{lem:trivial-refutations}
For sufficiently large $r$ and $n \ge \Omega(r \log^3 r)$, in any matrix $S \in \R^{n \times k}$ with $k \le Kr$, if either $|I_{\mathrm{small}}| \ge n / 16$ or $|I_{\mathrm{one}}| \ge n / 16$, then a uniformly random coordinate subspace $R \sim \mathrm{Unif}\binom{[n]}{r}$ satisfies
\begin{equation}\label{eq:trivial-refutation-prob}
  \Prb_{R \sim \mathrm{Unif}\binom{[n]}{r}}\bracket{\smin(S_R^\top)^2 < \alpha} \ge \frac{9}{10}.
\end{equation}
\end{lemma}

\begin{proof}
We consider each condition separately.

Case 1: $|I_{\mathrm{small}}| \ge n / 16$. When drawing $r$ indices uniformly without replacement from $[n]$, the probability of missing all elements of $I_{\mathrm{small}}$ is at most
\[
  \prod_{t=0}^{r-1} \left( 1 - \frac{|I_{\mathrm{small}}|}{n - t} \right) \le \left( 1 - \frac{1}{16} \right)^r \le \exp\left( -\frac{r}{16} \right) \le \frac{1}{10}
\]
for all $r \ge 37$. Whenever $R$ contains an index $i \in I_{\mathrm{small}}$, the standard basis vector $e_i \in \R^r$ satisfies $\|S_R^\top e_i\|_2^2 = \|s_i\|_2^2 < \alpha$, which breaks injectivity with $\smin(S_R^\top)^2 < \alpha$.

Case 2: $|I_{\mathrm{one}}| \ge n / 16$. Partition $I_{\mathrm{one}} = \bigsqcup_{j=1}^k J_j$ according to the unique non-zero column of each row, where $J_j = \{i \in I_{\mathrm{one}} : \supp(s_i) = \{j\}\}$ and $m_j = |J_j|$. Call a column $j \in [k]$ \emph{heavy} if $m_j \ge \frac{n}{32Kr}$, and let $U = \bigsqcup_{j:\, m_j \ge n/(32Kr)} J_j$ be the union of the heavy column buckets. Since $k \le Kr$, all light columns together contain fewer than $k \cdot \frac{n}{32Kr} \le \frac{n}{32}$ rows, so $|U| \ge |I_{\mathrm{one}}| - \frac{n}{32} \ge \frac{n}{32}$.

We expose the $r$ draws of $R \sim \mathrm{Unif}\binom{[n]}{r}$ sequentially in two phases of $r_1 = r/2$ and $r_2 = r - r_1 = r/2$ draws. Since $|U| \ge n/32$ and $r_1 \le n/64$, each of the first $r_1$ sequential draws lands in $U$ with conditional probability at least $(|U| - r_1)/n \ge 1/64$ given the previous draws. Consequently, by Chernoff's bound, at least $M_1 \ge r_1 / 128$ draws land in $U$ with probability at least $19/20$ for $r$ sufficiently large. If two of these draws already fall into the same bucket $J_j$, a collision has occurred; otherwise, the first phase hits at least $M_1 \ge r_1 / 128$ distinct heavy buckets $J_j$, each of which retains at least $m_j - 1 \ge \frac{n}{64Kr}$ unselected rows. Consequently, in each of the $r_2 \ge r/2$ draws of the second phase, conditioned on no collision having occurred yet, the next draw hits one of the already-visited heavy buckets with probability at least
\[
  \frac{M_1 \cdot \frac{n}{64Kr}}{n} \ge \frac{r_1}{128} \cdot \frac{1}{64Kr} \ge \frac{1}{16384 K},
\]
so the conditional probability of avoiding a collision across all $r_2$ draws of the second phase is at most
\[
  \left( 1 - \frac{1}{16384 K} \right)^{r/2} \le \exp\left( -\frac{r}{32768 K} \right) \le \frac{1}{20}
\]
for $r \ge C_0 K$. Thus, with probability at least $1 - \frac{1}{20} - \frac{1}{20} = \frac{9}{10}$, $R$ contains two distinct rows $u, v \in J_{j^*}$ from the same column $j^* \in [k]$. Setting $x_u = S_{v, j^*}$, $x_v = -S_{u, j^*}$, and zero elsewhere produces a non-zero vector $x \in \R^r$ with $S_R^\top x = 0$. Thus injectivity breaks with $\smin(S_R^\top)^2 = 0 < \alpha$, completing the proof.
\end{proof}

By Lemma~\ref{lem:heavy-row-filter} and Lemma~\ref{lem:trivial-refutations}, we may assume that $|I_{\mathrm{heavy}} \cup I_{\mathrm{small}} \cup I_{\mathrm{one}}| \le 3n / 16$. We define the set of clean rows by
\begin{equation}\label{eq:clean-rows-def}
  V_{\mathrm{clean}} = [n] \setminus (I_{\mathrm{heavy}} \cup I_{\mathrm{small}} \cup I_{\mathrm{one}}).
\end{equation}
The set $V_{\mathrm{clean}}$ is a fixed deterministic subset of size at least $13n / 16$. Every row $i \in V_{\mathrm{clean}}$ deterministically satisfies
\[
  \alpha \le \|s_i\|_2^2 \le M \quad \text{and} \quad 2 \le |\supp(s_i)| \le \zeta.
\]

\subsection{Construction of the $C_4$-Free Subgraph $G_0$}
\label{subsec:condition-h-c4}

We now employ the overlap condition to eliminate 4-cycles. By the definition of the bipartite graph $G$, two rows sharing two or more columns in their support form a 4-cycle in $G$.

To construct the $C_4$-free subgraph $G_0$, we subsample a row set $V_0 \subset [n]$ of size $n_0 = C_{\mathrm{samp}} r \zeta^6 / \alpha$ uniformly at random from $[n]$. This intermediate subsampling step couples cleanly with the ultimate coordinate sample $R$: choosing $V_0 \subset [n]$ of size $n_0$ uniformly and then drawing $r$ rows $R \subset V_0$ uniformly from $V_0$ produces a marginal distribution identical to drawing $R \sim \mathrm{Unif}\binom{[n]}{r}$ directly from $[n]$. To restrict to clean rows and eliminate all 4-cycles, we form the pruned row set $V(G_0) \subseteq V_0 \cap V_{\mathrm{clean}}$ by discarding the non-clean rows $V_0 \setminus V_{\mathrm{clean}}$ and removing one row from each pair $\{u, v\} \subset V_0 \cap V_{\mathrm{clean}}$ sharing two or more columns ($|\supp(s_u) \cap \supp(s_v)| \ge 2$), and define $G_0 = (V(G_0), [k], E_0)$ to be the bipartite subgraph of $G$ induced by $V(G_0)$ and $[k]$.

\begin{lemma}
\label{lem:c4-free-construction}
For sufficiently large $r$ and $n \ge \Omega(r \log^3 r)$, let $S \in \R^{n \times k}$ be a matrix with row sparsity $\zeta$ satisfying the overlap condition and $|V_{\mathrm{clean}}| \ge 13n / 16$. Let $n_0 = C_{\mathrm{samp}} r \zeta^6 / \alpha$ for a sufficiently large constant $C_{\mathrm{samp}} > 0$, and let $V_0 \subset [n]$ be subsampled uniformly at random among all subsets of $[n]$ of size $n_0$. Then with probability at least $7/8$, the set $V_0$ contains at least $3n_0 / 4$ clean rows and at most $n_0 / 16$ pairs sharing two or more columns, and the resulting induced subgraph $G_0 = (V(G_0), [k], E_0)$ satisfies:
\begin{enumerate}
  \item The row set has size $|V(G_0)| \ge 5n_0 / 8$.
  \item The graph $G_0$ is completely $C_4$-free: no two rows share more than one column.
  \item Every row $i \in V(G_0)$ satisfies $\alpha \le \|s_i\|_2^2 \le M$ and $2 \le \deg_{G_0}(i) \le \zeta$.
\end{enumerate}
\end{lemma}

\begin{proof}
First, let $N_{\mathrm{bad}} = |V_0 \setminus V_{\mathrm{clean}}|$ denote the number of non-clean rows in $V_0$. Since $|[n] \setminus V_{\mathrm{clean}}| \le 3n / 16$, when drawing $n_0$ rows uniformly without replacement from $[n]$, the hypergeometric random variable $N_{\mathrm{bad}}$ has mean $\E[N_{\mathrm{bad}}] \le 3n_0 / 16$ and variance $\mathrm{Var}(N_{\mathrm{bad}}) \le 3n_0 / 16$. By Chebyshev's inequality, for $C_{\mathrm{samp}}$ sufficiently large we have
\[
  \Prb\bracket{N_{\mathrm{bad}} > \frac{n_0}{4}} \le \Prb\bracket{N_{\mathrm{bad}} - \E[N_{\mathrm{bad}}] > \frac{n_0}{16}} \le \frac{3n_0 / 16}{(n_0 / 16)^2} = \frac{48}{n_0} \le \frac{1}{16}.
\]

Next, let $N_{\mathrm{coll}}$ denote the number of pairs $\{u, v\} \subset V_0$ such that $|\supp(s_u) \cap \supp(s_v)| \ge 2$. In the bipartite graph $G$, every such pair corresponds to a 4-cycle $C_4$. When drawing $n_0$ rows uniformly without replacement from $[n]$, each pair of rows from $[n]$ is chosen with probability
\[
  \frac{\binom{n-2}{n_0-2}}{\binom{n}{n_0}} = \frac{n_0(n_0 - 1)}{n(n - 1)} \le \frac{n_0^2}{n^2}.
\]
Using the overlap condition, the expected number of such pairs in $V_0$ satisfies
\[
  \E[N_{\mathrm{coll}}] \le \frac{n_0^2}{n^2} \sum_{1 \le u < v \le n} \mathbf{1}_{\{|\supp(s_u) \cap \supp(s_v)| \ge 2\}} \le \frac{n_0^2}{n^2} \left( \eta \frac{n^2}{k} \right) = \eta \frac{n_0^2}{k}.
\]
Since $k < r$ trivially implies $\smin(S_R^\top) = 0$, we may assume $k \ge r$. Substituting $\eta \le 1 / (C_H \zeta^6)$ and $n_0 = C_{\mathrm{samp}} r \zeta^6 / \alpha$ gives
\[
  \E[N_{\mathrm{coll}}] \le \frac{1}{C_H \zeta^6} \frac{n_0 \cdot (C_{\mathrm{samp}} r \zeta^6 / \alpha)}{r} = \frac{C_{\mathrm{samp}}}{C_H \alpha} n_0.
\]
Choosing $C_H \ge 256 C_{\mathrm{samp}} / \alpha$ ensures that $\E[N_{\mathrm{coll}}] \le n_0 / 256$. By Markov's inequality, we have
\[
  \Prb\bracket{N_{\mathrm{coll}} > \frac{n_0}{16}} \le \frac{\E[N_{\mathrm{coll}}]}{n_0 / 16} \le \frac{16}{256} = \frac{1}{16}.
\]
By a union bound, with probability at least $1 - 1/16 - 1/16 = 7/8$, we have $N_{\mathrm{bad}} \le n_0 / 4$ and $N_{\mathrm{coll}} \le n_0 / 16$. Discarding the $N_{\mathrm{bad}}$ non-clean rows and deleting one row from each such pair removes at most $n_0 / 4 + n_0 / 16 = 5n_0 / 16$ vertices and eliminates every 4-cycle, leaving a $C_4$-free subgraph $G_0$ with at least $n_0 - 5n_0 / 16 = 11n_0 / 16 \ge 5n_0 / 8$ rows. All remaining rows belong to $V_{\mathrm{clean}}$ and thus inherit the norm bounds $\alpha \le \|s_i\|_2^2 \le M$ and support bounds $2 \le \deg_{G_0}(i) \le \zeta$.
\end{proof}

\section{Constructing a Forest of Disjoint Trees}
\label{sec:tree-cancellation}

Having constructed the $C_4$-free subgraph $G_0$, where every row satisfies $\alpha \le \|s_i\|_2^2 \le M$ and $2 \le \deg_{G_0}(i) \le \zeta$, our main goal in this section is to prove Theorem~\ref{thm:witness-forest-existence}, which constructs a forest of $T = \Theta(r / \zeta^5)$ mutually vertex-disjoint witness trees on $G_0$ that breaks injectivity. Each tree in $G_0$ consists of a center row $i^* \in V(G_0)$, its branch columns $\supp(s_{i^*}) \subset [k]$, and $q = 4\zeta M / \alpha$ leaf rows in $V(G_0)$ attached to each branch column. 

\subsection{Disjoint Leaves for a Single Tree}
\label{subsec:single-tree-disjoint}

Recall that for any row $i \in V(G_0)$ and column $j \in [k]$ with $S_{ij} \ne 0$, the relative weight is defined by $w_{ij} = |S_{ij}|^2 / \|s_i\|_2^2 \in (0, 1]$. For each column $j \in [k]$, we sort all rows in $V(G_0)$ containing $j$ by non-increasing relative weights and place the top $D$ rows into a candidate set $F_j \subseteq V(G_0)$. Because center rows $i^*$ will be chosen exclusively from rows outside $\bigcup_j F_j$, every branch column $j \in \supp(s_{i^*})$ is guaranteed to have full size $|F_j| = D$ and satisfy $w_{\ell, j} \ge w_{i^*, j}$ for all $\ell \in F_j$.

The following lemma bounds the squared norm on the outer columns for a center row and its leaf rows.
\begin{lemma}
\label{lem:outer-energy-bound}
For a matrix $S \in \R^{n \times k}$ with row sparsity $\zeta$, let $s_{i^*}^\top$ be a row in $G_0$ with branches $\supp(s_{i^*}) = \{j_1, \ldots, j_d\}$ of size $d \le \zeta$. For each branch index $a \in [d]$, let $L_a \subseteq F_{j_a} \setminus \{i^*\}$ be a subset of $q$ leaf rows. Suppose that for every leaf $\ell \in L_a$, the relative weight satisfies $w_{\ell, j_a} \ge w_{i^*, j_a}$, and suppose that the outer supports $\supp(s_\ell) \setminus \{j_a\}$ across all $\ell \in \bigcup_{a=1}^d L_a$ are pairwise disjoint. Define the cancellation vector $x \in \R^n$ by
\begin{equation}\label{eq:cancellation-vector-formula}
  x_{i^*} = 1, \quad x_\ell = -\frac{S_{i^*, j_a}}{q S_{\ell, j_a}} \text{ for each } a \in [d] \text{ and } \ell \in L_a, \quad x_m = 0 \text{ otherwise}.
\end{equation}
Then for every branch column $j_a \in \supp(s_{i^*})$, we have $(S^\top x)_{j_a} = 0$. Moreover, the total squared norm on all outer columns satisfies
\begin{equation}\label{eq:outer-energy-sum}
  \sum_{c \notin \supp(s_{i^*})} |(S^\top x)_c|^2 \le \frac{\zeta \|s_{i^*}\|_2^2}{q}.
\end{equation}
\end{lemma}

\begin{proof}
We first verify the exact cancellation on the active columns of the center row. For any fixed branch $a \in [d]$, the $j_a$-th coordinate of $S^\top x$ receives contributions only from row $i^*$ and the leaf rows in $L_a$, because no other leaf contains $j_a$ in its support. Using the definition of $x$ in Equation~\eqref{eq:cancellation-vector-formula}, we compute
\[
  (S^\top x)_{j_a} = S_{i^*, j_a} x_{i^*} + \sum_{\ell \in L_a} S_{\ell, j_a} x_\ell = S_{i^*, j_a} - \sum_{\ell \in L_a} S_{\ell, j_a} \left( \frac{S_{i^*, j_a}}{q S_{\ell, j_a}} \right) = S_{i^*, j_a} - q \cdot \frac{S_{i^*, j_a}}{q} = 0.
\]
Thus exact cancellation holds on every column in $\supp(s_{i^*})$.

We next evaluate the outer coordinates. By assumption, the outer supports $\supp(s_\ell) \setminus \{j_a\}$ are pairwise disjoint across all leaves $\ell \in \bigcup_{a=1}^d L_a$. Therefore, for every outer column $c \notin \supp(s_{i^*})$, at most one leaf row contributes to $(S^\top x)_c$. Thus we get:
\[
  \sum_{c \notin \supp(s_{i^*})} |(S^\top x)_c|^2 = \sum_{a=1}^d \sum_{\ell \in L_a} \sum_{c \ne j_a} |x_\ell S_{\ell, c}|^2 = \sum_{a=1}^d \sum_{\ell \in L_a} |x_\ell|^2 \sum_{c \ne j_a} |S_{\ell, c}|^2.
\]
By the definition, $\sum_{c \ne j_a} |S_{\ell, c}|^2 \le \|s_\ell\|_2^2$. Substituting the value of $x_\ell$ gives
\[
  |x_\ell|^2 \sum_{c \ne j_a} |S_{\ell, c}|^2 \le \left( \frac{S_{i^*, j_a}}{q S_{\ell, j_a}} \right)^2 \|s_\ell\|_2^2 = \frac{|S_{i^*, j_a}|^2}{q^2 |S_{\ell, j_a}|^2} \|s_\ell\|_2^2.
\]
By the definition of the relative weights $w_{ij}$, we have $|S_{i^*, j_a}|^2 = w_{i^*, j_a} \|s_{i^*}\|_2^2$ and $|S_{\ell, j_a}|^2 = w_{\ell, j_a} \|s_\ell\|_2^2$. Substituting these identities into the bound yields
\[
  |x_\ell|^2 \sum_{c \ne j_a} |S_{\ell, c}|^2 \le \frac{w_{i^*, j_a} \|s_{i^*}\|_2^2}{q^2 w_{\ell, j_a} \|s_\ell\|_2^2} \|s_\ell\|_2^2 = \frac{w_{i^*, j_a}}{w_{\ell, j_a}} \frac{\|s_{i^*}\|_2^2}{q^2}.
\]
Since row $\ell \in F_{j_a}$ was selected from the candidate set with $w_{\ell, j_a} \ge w_{i^*, j_a}$, the ratio satisfies $w_{i^*, j_a} / w_{\ell, j_a} \le 1$. Summing this over all $q$ leaves in branch $a$ gives
\[
  \sum_{\ell \in L_a} |x_\ell|^2 \sum_{c \ne j_a} |S_{\ell, c}|^2 \le q \cdot \frac{\|s_{i^*}\|_2^2}{q^2} = \frac{\|s_{i^*}\|_2^2}{q}.
\]
Finally, summing over all $d$ active branches and using $d \le \zeta$ yields Equation~\eqref{eq:outer-energy-sum}.
\end{proof}

Therefore, when selecting leaf rows, we require their outer supports to be pairwise disjoint. We now show that whenever each leaf candidate set $F_j$ has size at least $L = q + q(\zeta - 1)^2$, such a selection is always guaranteed on any $C_4$-free graph.

\begin{lemma}
\label{lem:single-tree-disjointness}
For a matrix $S \in \R^{n \times k}$ with row sparsity $\zeta$, let $s_{i^*}^\top$ be a center row in $G_0$ with branches $\supp(s_{i^*}) = \{j_1, \ldots, j_d\}$ of size $d \le \zeta$. For each branch $a \in [d]$, let $F_{j_a} \subseteq V(G_0) \setminus \{i^*\}$ be a candidate set of size at least
\begin{equation}\label{eq:L-size-def}
  L = q + q(\zeta - 1)^2.
\end{equation}
Then there exist subsets $L_a \subseteq F_{j_a}$ of size $q$ for all $a \in [d]$ such that the outer supports $\supp(s_\ell) \setminus \{j_a\}$ across all $\ell \in \bigcup_{a=1}^d L_a$ are mutually disjoint.
\end{lemma}

\begin{proof}
We select the subsets $L_1, \ldots, L_d$ sequentially for branches $a = 1, \dots, d$. Suppose we have already selected subsets $L_1, \ldots, L_{a-1}$ satisfying the disjointness condition. The total number of outer coordinates occupied by these previously selected leaves is at most
\[
  \sum_{t=1}^{a-1} \sum_{\ell \in L_t} |\supp(s_\ell) \setminus \{j_t\}| \le (a - 1) \cdot q(\zeta - 1) \le q(\zeta - 1)^2.
\]

We now analyze how many rows in $F_{j_a}$ can conflict with previously selected leaves.
First, no row in $F_{j_a}$ can have been selected in any previous branch $L_t$ for $t < a$. Indeed, if row $\ell$ belonged to both $F_{j_a}$ and $L_t$, it would contain both column $j_a$ and column $j_t$. But the center row $i^*$ also contains both $j_a$ and $j_t$, which would create a 4-cycle between rows $\{i^*, \ell\}$ and columns $\{j_a, j_t\}$, contradicting the fact that $G_0$ is $C_4$-free.

Second, for each occupied column $c \in \bigcup_{t=1}^{a-1} \bigcup_{\ell \in L_t} (\supp(s_\ell) \setminus \{j_t\})$, there is at most one row in $F_{j_a}$ containing column $c$. If there were two distinct rows $\ell_1, \ell_2 \in F_{j_a}$ containing column $c$, they would both contain column $j_a$ and column $c$, forming a 4-cycle between rows $\{\ell_1, \ell_2\}$ and columns $\{j_a, c\}$, which again contradicts $C_4$-freeness.

Consequently, each of the at most $q(\zeta - 1)^2$ occupied columns in previously selected leaves eliminates at most one candidate row in $F_{j_a}$. Since $|F_{j_a}| \ge L = q + q(\zeta - 1)^2$, there remain at least
\[
  |F_{j_a}| - q(\zeta - 1)^2 \ge q
\]
candidate rows in $F_{j_a}$ whose outer supports are completely disjoint from all previously occupied columns. Furthermore, among these remaining rows, no two rows can share any outer column because $G_0$ is $C_4$-free. Picking any $q$ of these rows as $L_a$ maintains the induction hypothesis and completes the proof.
\end{proof}

\subsection{Disjoint Multiple Trees}
\label{subsec:turan-centers}

Having established how to construct a single tree whose leaves have pairwise disjoint outer supports, which by Lemma~\ref{lem:outer-energy-bound} guarantees the existence of a cancellation vector achieving small residual norm, we now discuss how to extract multiple such trees in $G_0$ that are mutually vertex-disjoint.

Before we start, we first cap the maximum column degree by splitting columns into \emph{virtual columns}. Specifically, set the degree threshold $\Delta := \zeta n_0 / k$. For each column $j \in [k]$ with $\deg_{G_0}(j) > \Delta$, we partition the rows of $V(G_0)$ containing $j$ into disjoint groups of size at most $\Delta$, viewing each resulting group across all columns as a virtual column indexed by $j' \in [k']$. Since $\sum_{j=1}^k \deg_{G_0}(j) \le \zeta |V(G_0)| \le \zeta n_0$, splitting adds at most $\sum_{j=1}^k \deg_{G_0}(j) / \Delta \le k$ groups, so the total number of virtual columns is bounded by
\[
  k' \le k + \frac{\zeta n_0}{\Delta} = 2k \le 2Kr.
\]
Since each row $i \in V(G_0)$ has at most $\zeta$ non-zero entries, each row $i$ still belongs to at most $\zeta$ virtual columns. Within each virtual column $j' \in [k']$, we sort its rows by non-increasing relative weight $w_{ij}$ on the underlying column $j \in [k]$ and place the top $D$ rows into its candidate set $F_{j'} \subseteq V(G_0)$.

To construct multiple vertex-disjoint trees, we first seek center rows whose supports (over the virtual columns) are all mutually disjoint. In principle, branch columns could overlap between different trees as long as the vertex sets of the center rows and leaf rows remain disjoint; however, having pairwise disjoint supports for the center rows allows us to draw the leaf rows of each tree from distinct candidate sets $F_{j'}$. Lemma~\ref{lem:turan-centers} establishes how many such center rows with pairwise disjoint supports can be extracted. To prove it, we use the following independent-set version of Turán's theorem, also established by Caro and Wei \cite{Tur41,Car79,Wei81}.

\begin{lemma}
\label{lem:turan-classical}
Let $H = (V_H, E_H)$ be a finite undirected graph with average degree $d_{\mathrm{avg}} = 2|E_H|/|V_H|$. Then the independence number of $H$ satisfies
\begin{equation}\label{eq:turan-bound-statement}
  \alpha(H) \ge \frac{|V_H|}{1 + d_{\mathrm{avg}}}.
\end{equation}
\end{lemma}

\begin{lemma}
\label{lem:turan-centers}
For a matrix $S \in \R^{n \times k}$ with row sparsity $\zeta$, let $U$ be the set of rows in $G_0$ that do not belong to any candidate set $F_{j'}$ for $j' \in [k']$. Let $H = (U, E_H)$ be the conflict graph with an edge between rows $u, v \in U$ whenever $u$ and $v$ share at least one virtual column in $[k']$. Then there exists an independent set in $H$ of size
\begin{equation}\label{eq:T-center-count}
  T = \Theta\left( \frac{r}{\zeta^5} \right),
\end{equation}
providing $T$ center rows whose virtual-column supports are pairwise disjoint.
\end{lemma}

\begin{proof}
We first bound the size of $U$. Each virtual column $j' \in [k']$ places at most $D$ rows into its candidate set $F_{j'}$. Since there are $k' \le 2k \le 2Kr$ virtual columns, the total number of rows belonging to at least one candidate set is at most $D k' \le 2 D K r$. Since $|V(G_0)| \ge 5n_0 / 8$ and $n_0 = C_{\mathrm{samp}} r \zeta^6 / \alpha$, choosing $C_{\mathrm{samp}}$ sufficiently large ensures that
\[
  |U| \ge |V(G_0)| - 2 D K r \ge \frac{n_0}{2}.
\]
We next bound the number of edges in the conflict graph $H$. Two rows $u, v \in U$ share an edge if and only if there is some virtual column $j' \in [k']$ containing both $u$ and $v$. For each virtual column $j' \in [k']$, let $d_{j'}$ denote its degree in $U$. The number of edges in $H$ is bounded by
\[
  |E_H| \le \sum_{j'=1}^{k'} \binom{d_{j'}}{2} = \frac{1}{2} \sum_{j'=1}^{k'} d_{j'}(d_{j'} - 1).
\]
By construction of the virtual columns, every virtual column $j' \in [k']$ has degree $d_{j'} \le \Delta = \zeta n_0 / k$, and since each row in $U$ belongs to at most $\zeta$ virtual columns, the total sum of degrees satisfies $\sum_{j'=1}^{k'} d_{j'} \le \zeta |U|$. Therefore, the overlap sum satisfies
\[
  \sum_{j'=1}^{k'} d_{j'}(d_{j'} - 1) \le \Delta \sum_{j'=1}^{k'} d_{j'} \le \frac{\zeta^2 n_0 |U|}{k}.
\]
Consequently, using $k \ge r$ (as $k < r$ trivially implies $\smin(S_R^\top) = 0$), the average degree $d_{\mathrm{avg}} = 2|E_H| / |U|$ in the conflict graph satisfies
\[
  d_{\mathrm{avg}} \le O\left( \frac{\zeta^2 n_0}{k} \right) \le O\left( \frac{\zeta^2 n_0}{r} \right) = O(\zeta^8).
\]
By Lemma~\ref{lem:turan-classical}, the independence number of $H$ satisfies
\[
  \alpha(H) \ge \frac{|U|}{1 + d_{\mathrm{avg}}} \ge \frac{n_0 / 2}{1 + O(\zeta^8)} = \Theta\left( \frac{r}{\zeta^2} \right).
\]
Selecting any $T = \Theta(r / \zeta^5)$ rows from this independent set yields $T$ center rows whose virtual-column supports are completely pairwise disjoint.
\end{proof}

Having found center rows whose supports over the branch (virtual) columns are pairwise disjoint, we next select leaf rows from the corresponding candidate sets such that their outer supports are pairwise disjoint. In Lemma~\ref{lem:single-tree-disjointness}, we showed that for a single tree, having candidate sets of size at least $L$ guarantees such a leaf selection. Therefore, if we can extract mutually disjoint subsets of size $L$ from the candidate sets $F_{j'}$ for all active branches across the $T$ trees, we can apply the single-tree selection procedure to each tree independently. In the following lemmas, we establish the existence of these disjoint reservoirs via Hall's Marriage Theorem~\cite{Hal35}.

\begin{lemma}
\label{lem:hall-classical}
Let $B = (V_L, V_R, E_B)$ be a finite bipartite graph. There exists a matching in $B$ saturating every vertex in $V_L$ if and only if for every subset $A \subseteq V_L$, the neighborhood $N_B(A) \subseteq V_R$ satisfies the marriage condition
\begin{equation}\label{eq:hall-condition-statement}
  |N_B(A)| \ge |A|.
\end{equation}
\end{lemma}

\begin{lemma}
\label{lem:hall-reservoirs}
For a matrix $S \in \R^{n \times k}$ with row sparsity $\zeta$, let $J_{\mathrm{branch}} \subseteq [k']$ be the set of branch virtual columns across the $T$ center rows in $G_0$, with $|J_{\mathrm{branch}}| \le \zeta T$. Suppose that for every $j' \in J_{\mathrm{branch}}$, the candidate set size satisfies
\begin{equation}\label{eq:D-hall-condition}
  D \ge \zeta L.
\end{equation}
Then there exist mutually disjoint subsets $R_{j'} \subseteq F_{j'}$ of size $|R_{j'}| = L$ for all $j' \in J_{\mathrm{branch}}$.
\end{lemma}

\begin{proof}
We construct an auxiliary bipartite assignment graph $B = (V_L, V_R, E_B)$ as follows:
\begin{enumerate}
  \item The left vertex set $V_L$ consists of all branch virtual columns $j' \in J_{\mathrm{branch}}$, each cloned $L$ times. That is, $V_L = J_{\mathrm{branch}} \times [L]$, so $|V_L| = L |J_{\mathrm{branch}}|$.
  \item The right vertex set $V_R = \bigcup_{j' \in J_{\mathrm{branch}}} F_{j'} \subset V(G_0)$ consists of all candidate rows.
  \item An edge is placed between a left slot $(j', t) \in V_L$ and a right row $\ell \in V_R$ if and only if $\ell \in F_{j'}$.
\end{enumerate}

Every left slot $(j', t) \in V_L$ connects to every row in $F_{j'}$, so its degree is exactly
  $\deg_B((j', t)) = |F_{j'}| = D$.
On the other hand, every right row $\ell \in V_R$ has degree at most $\zeta$ in $G_0$, so it can belong to candidate sets $F_{j'}$ for at most $\zeta$ distinct virtual columns. Since each virtual column is cloned $L$ times on the left, the degree of any right row $\ell \in V_R$ is bounded by
\[
  \deg_B(\ell) \le \zeta L.
\]
We now verify Hall's marriage condition. Let $A \subseteq V_L$ be an arbitrary subset of left slots, and let $N_B(A) \subseteq V_R$ denote its neighborhood in $B$. We count the number of edges $e_B(A, N_B(A))$ between $A$ and $N_B(A)$ from both sides. From the left side, every slot in $A$ has degree $D$, so
\[
  e_B(A, N_B(A)) = |A| \cdot D.
\]
From the right side, every vertex in $N_B(A)$ has degree at most $\zeta L$, so
\[
  e_B(A, N_B(A)) \le |N_B(A)| \cdot (\zeta L).
\]
Combining these two relations gives
\[
  |A| \cdot D \le |N_B(A)| \cdot (\zeta L) \implies |N_B(A)| \ge \frac{D}{\zeta L} |A|.
\]
Since $D \ge \zeta L$ by Equation~\eqref{eq:D-hall-condition}, we have $|N_B(A)| \ge |A|$ for every subset $A \subseteq V_L$.

By Lemma~\ref{lem:hall-classical}, there exists a matching in $B$ that saturates all vertices of $V_L$. This matching assigns each slot $(j', t)$ to a distinct candidate row $\ell \in F_{j'}$. For each $j' \in J_{\mathrm{branch}}$, the set of $L$ rows matched to its slots forms the reservoir $R_{j'} \subseteq F_{j'}$. Because the matching pairs each slot with a distinct right vertex, the reservoirs $R_{j'}$ across all $j' \in J_{\mathrm{branch}}$ are pairwise disjoint, completing the proof.
\end{proof}

We now combine these combinatorial components to establish the existence of the full forest of witness trees.

\begin{theorem}
\label{thm:witness-forest-existence}
For a matrix $S \in \R^{n \times k}$ with row sparsity $\zeta$, set $q = \frac{4\zeta M}{\alpha}$. In the $C_4$-free subgraph $G_0$, there exist $T = \Theta(r / \zeta^5)$ mutually vertex-disjoint witness trees $T_1, \ldots, T_T$ such that:
\begin{enumerate}
  \item Each tree $T_s$ consists of a center row $i_s^*$ and at most $\zeta q$ leaves, with total size
  \begin{equation}\label{eq:B-tree-def}
    B_{\mathrm{tree}} = 1 + \zeta q = \Theta\left( \frac{\zeta^2}{\alpha} \right).
  \end{equation}
  \item The trees $T_1, \ldots, T_T$ are mutually vertex-disjoint: $V(T_s) \cap V(T_u) = \emptyset$ for all $s \ne u$.
  \item For each tree $T_s$, the cancellation vector $x^{(s)} \in \R^n$ supported on $V(T_s)$ satisfies
  \begin{equation}\label{eq:tree-cancellation-bound}
    \|x^{(s)}\|_2^2 \ge 1 \quad \text{and} \quad \|S^\top x^{(s)}\|_2^2 \le \frac{\alpha}{4}.
  \end{equation}
\end{enumerate}
\end{theorem}

\begin{proof}
By Lemma~\ref{lem:turan-centers}, we extract $T = \Theta(r / \zeta^5)$ center rows $i_1^*, \ldots, i_T^*$ whose virtual-column supports are pairwise disjoint and which do not belong to any candidate set $F_{j'}$. By Lemma~\ref{lem:hall-reservoirs}, since $D \ge \zeta L$, there exist mutually disjoint leaf reservoirs $R_{j'} \subseteq F_{j'}$ of size $L = q + q(\zeta - 1)^2$ for all active branch virtual columns.

For each center row $i_s^*$, its virtual columns correspond one-to-one with its branch columns $\supp(s_{i_s^*}) \subseteq [k]$, and we apply Lemma~\ref{lem:single-tree-disjointness} within $R_{j'}$ to select $q$ leaves per branch column. Even if two different center rows have virtual columns originating from the same underlying column in $[k]$, their assigned reservoirs $R_{j'}$ are mutually disjoint subsets of $V(G_0)$, and the center rows are distinct and do not belong to any candidate set, so the resulting trees $T_1, \ldots, T_T$ are mutually vertex-disjoint: $V(T_s) \cap V(T_u) = \emptyset$.

Finally, for each tree $T_s$, let $x^{(s)} \in \R^n$ be the cancellation vector defined in Lemma~\ref{lem:outer-energy-bound} on $V(T_s)$. Because $x_{i_s^*}^{(s)} = 1$, we have $\|x^{(s)}\|_2^2 = 1 + \sum_{a=1}^{d_s} \sum_{\ell \in L_a} (x_\ell^{(s)})^2 \ge 1$. Moreover, Lemma~\ref{lem:outer-energy-bound} guarantees that $(S^\top x^{(s)})_{j_a} = 0$ for every branch column $j_a \in \supp(s_{i_s^*})$, so the squared Euclidean norm of $S^\top x^{(s)}$ is supported entirely on the outer columns. Since every row in $G_0$ satisfies $\|s_{i_s^*}\|_2^2 \le M$, applying Equation~\eqref{eq:outer-energy-sum} and substituting $q = 4\zeta M / \alpha$ gives
\[
  \|S^\top x^{(s)}\|_2^2 = \sum_{c \notin \supp(s_{i_s^*})} |(S^\top x^{(s)})_c|^2 \le \frac{\zeta \|s_{i_s^*}\|_2^2}{q} \le \frac{\zeta M}{q} = \frac{\alpha}{4},
\]
which completes the proof.
\end{proof}

\section{Proof of the Lower Bound Theorem}
\label{sec:lower-bound-proof}

In this section, we complete the proof of the sparsity lower bound. We first formalize the reduction principle from Section~\ref{sec:overview}, which establishes that breaking injectivity on a random coordinate sample for a fixed matrix refutes OSI for any isotropic distribution.

\begin{lemma}
\label{lem:reduction-master}
For sufficiently large $r$ and $n \ge \Omega(r \log^3 r)$, let $\alpha \in (0, 1)$. Suppose that for every fixed matrix $S \in \R^{n \times k}$ with row sparsity $\zeta$ and $\|S\|_F^2 \le 10n$ satisfying the overlap condition, a uniformly random coordinate index set $R \sim \mathrm{Unif}\binom{[n]}{r}$ satisfies
\begin{equation}\label{eq:reduction-failure-target}
  \Prb_{R \sim \mathrm{Unif}\binom{[n]}{r}}\bracket{\smin(S_R^\top)^2 \ge \alpha} \le \frac{1}{4}.
\end{equation}
Then no isotropic distribution over row-$\zeta$-sparse matrices in $\R^{n \times k}$ satisfying the overlap condition with high probability can be an $(r, \alpha)$-OSI.
\end{lemma}

\begin{proof}
Suppose for contradiction that an isotropic distribution $\mathcal{D}$ over row-$\zeta$-sparse matrices in $\R^{n \times k}$ satisfies the overlap condition with probability at least $9/10$ and is an $(r, \alpha)$-OSI. By definition, every fixed orthonormal matrix $Q \in \R^{n \times r}$ satisfies
\[
  \Prb_{\Omega \sim \mathcal{D}}\bracket{\smin(\Omega^\top Q)^2 \ge \alpha} \ge \frac{19}{20}.
\]
In particular, this guarantee must hold for every coordinate subspace $Q_R = I_R$ where $R \subset [n]$ has size $r$.

Here, the minimum success probability over coordinate subspaces is bounded by the average failure probability:
\begin{align*}
  \min_{R^* \in \binom{[n]}{r}} \Prb_{\Omega \sim \mathcal{D}}\bracket{\smin(\Omega_{R^*}^\top)^2 \ge \alpha} 
  &\le \E_{R \sim \mathrm{Unif}\binom{[n]}{r}} \bracket{\Prb_{\Omega \sim \mathcal{D}}\bracket{\smin(\Omega_R^\top)^2 \ge \alpha}} \\
  &= \E_{\Omega \sim \mathcal{D}} \bracket{\Prb_{R \sim \mathrm{Unif}\binom{[n]}{r}}\bracket{\smin(\Omega_R^\top)^2 \ge \alpha}}.
\end{align*}
Next, by the isotropy condition $\E[\Omega\Omega^\top] = I_n$, each row satisfies $\E[\|s_i\|_2^2] = 1$, so $\E[\|\Omega\|_F^2] = n$. By Markov's inequality, $\Prb[\|\Omega\|_F^2 > 10n] \le 1/10$, and by hypothesis $\Omega$ violates the overlap condition with probability at most $1/10$. Define the good event $\mathcal{E}$ on which $\Omega$ satisfies both $\|\Omega\|_F^2 \le 10n$ and the overlap condition, so that $\Prb[\mathcal{E}^c] \le 1/5$. Whenever $\Omega$ satisfies $\mathcal{E}$, the hypothesis in Equation~\eqref{eq:reduction-failure-target} applies to $S = \Omega$, giving $\Prb_R[\smin(\Omega_R^\top)^2 \ge \alpha] \le 1/4$. Conditioning on $\mathcal{E}$ yields
\begin{align*}
  \min_{R^* \in \binom{[n]}{r}} \Prb_{\Omega \sim \mathcal{D}}\bracket{\smin(\Omega_{R^*}^\top)^2 \ge \alpha}
  &\le \E_{\Omega \sim \mathcal{D}}\bracket{\Prb_{R}\bracket{\smin(\Omega_R^\top)^2 \ge \alpha}} \\
  &\le \Prb[\mathcal{E}^c] + \E_{\Omega \sim \mathcal{D}}\bracket{\Prb_{R}\bracket{\smin(\Omega_R^\top)^2 \ge \alpha} \;\middle|\; \mathcal{E}} \\
  &\le \frac{1}{5} + \frac{1}{4} = \frac{9}{20}.
\end{align*}
Therefore, there exists a specific coordinate subset $R^* \subset [n]$ of size $r$ such that $\Prb_\Omega[\smin(\Omega_{R^*}^\top)^2 \ge \alpha] \le 9/20$. This strictly contradicts the requirement that $\Prb_\Omega[\smin(\Omega_{R^*}^\top)^2 \ge \alpha] \ge 19/20$, completing the proof.
\end{proof}

Therefore, it now remains only to establish Equation~\eqref{eq:reduction-failure-target}: showing that every sparse matrix satisfying the overlap condition has its injectivity broken on a random coordinate sample with high probability will complete the entire proof.

\thmlowerbound*

\begin{proof}
Suppose for contradiction that the row sparsity satisfies
\begin{equation}\label{eq:sparsity-contradiction-assumption}
  \zeta \le c \sqrt{\frac{\log r}{\log \log r}}
\end{equation}
for a sufficiently small constant $c > 0$. By Lemma~\ref{lem:reduction-master}, it suffices to show that for every fixed matrix $S \in \R^{n \times k}$ with row sparsity $\zeta$ and $\|S\|_F^2 \le 10n$ satisfying the \hyperref[def:condition-h]{overlap condition}, a random coordinate index set $R \sim \mathrm{Unif}\binom{[n]}{r}$ satisfies $\Prb_R[\smin(S_R^\top)^2 \ge \alpha] \le 1/4$.

If $|V_{\mathrm{clean}}| < 13n / 16$, then Lemma~\ref{lem:heavy-row-filter} and Lemma~\ref{lem:trivial-refutations} immediately imply $\Prb_R[\smin(S_R^\top)^2 \ge \alpha] \le 1/10 \le 1/4$. Otherwise, assume $|V_{\mathrm{clean}}| \ge 13n / 16$. We analyze the probability that a uniform random sample $R \sim \mathrm{Unif}\binom{[n]}{r}$ captures at least one witness tree. Under the two-stage sampling coupling, we first draw an intermediate row set $V_0 \subset [n]$ of size $n_0 = \Theta(r \zeta^6 / \alpha)$ uniformly at random from $[n]$, and then draw $R \subset V_0$ of size $r$ uniformly at random from $V_0$, whose marginal distribution is identical to $R \sim \mathrm{Unif}\binom{[n]}{r}$. By Lemma~\ref{lem:c4-free-construction} and Theorem~\ref{thm:witness-forest-existence}, with probability at least $7/8$ over the choice of $V_0$, the pruned $C_4$-free subgraph $G_0$ on $V(G_0) \subseteq V_0 \cap V_{\mathrm{clean}}$ satisfies $|V(G_0)| \ge 5n_0 / 8$ and contains $T = \Theta(r / \zeta^5)$ mutually vertex-disjoint witness trees $T_1, \ldots, T_T$, each of size $B_{\mathrm{tree}} = \Theta(\zeta^2 / \alpha)$. Conditioned on such a realization of $V_0$, since $V(T_s) \subseteq V(G_0) \subseteq V_0$, drawing $r$ rows uniformly without replacement from $V_0$ includes each vertex of $T_s$ with conditional probability at least $(r - B_{\mathrm{tree}}) / n_0 \ge r / (2n_0) = \Omega(1 / \zeta^6)$. For any fixed witness tree $T_s$, the probability that all $B_{\mathrm{tree}}$ of its vertices are captured by $R$ satisfies
\[
  p = \Prb[V(T_s) \subseteq R \mid V_0] \ge \left( \frac{r}{2n_0} \right)^{B_{\mathrm{tree}}} = \left( \Omega\left(\frac{1}{\zeta^6}\right) \right)^{\Theta(\zeta^2 / \alpha)} = \exp\left( -O\left( \frac{\zeta^2}{\alpha} \log \zeta \right) \right).
\]
Under the assumption in Equation~\eqref{eq:sparsity-contradiction-assumption}, choosing $c$ sufficiently small ensures that $O(\frac{\zeta^2}{\alpha} \log \zeta) \le \frac{1}{2} \log r$, which yields $p \ge 1 / \sqrt{r}$.

Because the $T$ trees are mutually vertex-disjoint, sampling without replacement exhibits negative correlation across the capture events $\{V(T_s) \subseteq R\}$. Consequently, the conditional probability that the random sample misses all $T$ trees satisfies
\[
  \Prb\bracket{\bigcap_{s=1}^T \{V(T_s) \not\subseteq R\} \;\middle|\; V_0} \le (1 - p)^T \le \exp(-Tp).
\]
Since $T = \Theta(r / \zeta^5)$, we have $Tp \ge \Theta(r^{1/2} / \zeta^5) \ge r^{1/4} \ge 3$ for sufficiently large $r$. Therefore, conditioned on $V_0$ satisfying the event of Lemma~\ref{lem:c4-free-construction}, with probability at least $1 - \exp(-3) \ge 9/10$ the sample $R$ completely captures at least one witness tree $T_{s^*} \subseteq R$.

On this capture event, all vertices of $V(T_{s^*})$ belong to $R$. Restricting the cancellation vector $x^{(s^*)} \in \R^n$ from Theorem~\ref{thm:witness-forest-existence} to the coordinates in $R$ produces a non-zero vector $z \in \R^r$ that breaks injectivity on $R$, satisfying $\|z\|_2 = \|x^{(s^*)}\|_2 \ge 1$ and
\[
  \smin(S_R^\top)^2 \le \frac{\|S_R^\top z\|_2^2}{\|z\|_2^2} \le \frac{\alpha}{4} < \alpha.
\]
Combining the failure probability at most $1/8$ of $V_0$ in Lemma~\ref{lem:c4-free-construction} and the conditional failure probability at most $1/10$ of tree capture gives $\Prb_R[\smin(S_R^\top)^2 \ge \alpha] \le 1/8 + 1/10 = 9/40 \le 1/4$. By Lemma~\ref{lem:reduction-master}, no isotropic distribution over row-$\zeta$-sparse matrices satisfying the overlap condition with high probability can be an $(r, \alpha)$-OSI.
\end{proof}

\section*{Acknowledgments}

MS's work is supported by the NSF under award 2402314.
RAM would like to acknowledge the DARPA DIAL and DARPA AIQ programs for providing partial support of this work.
Kijun Shin and David Woodruff would like to thank partial support from a Simons Investigator Award and Office of Naval Research (ONR) grant number N000142112647.

\bibliographystyle{alpha}
\bibliography{references}

@article{tikhomirov2020singularity,
  title={Singularity of random Bernoulli matrices},
  author={Tikhomirov, Konstantin},
  journal={Annals of Mathematics},
  volume={191},
  number={2},
  pages={593--634},
  year={2020},
  publisher={Department of Mathematics, Princeton University, Princeton, New Jersey, USA}
}

@article{jain2020sharp,
  title={Sharp invertibility of random Bernoulli matrices},
  author={Jain, Vishesh and Sah, Ashwin and Sawhney, Mehtaab},
  journal={arXiv preprint arXiv:2010.06553},
  year={2020}
}

@article{rudelson3956smallest,
  title={The smallest singular value of a random rectangular matrix, 2009},
  author={Rudelson, Mark and Vershynin, Roman},
  journal={URL https://arxiv. org/abs/0802.3956},
  volume={5},
  number={2}
}

@article{costello2006random,
  title={Random symmetric matrices are almost surely nonsingular},
  author={Costello, Kevin P and Tao, Terence and Vu, Van},
  year={2006}
}

@article{costello2010rank,
  title={On the rank of random sparse matrices},
  author={Costello, Kevin P and Vu, Van},
  journal={Combinatorics, Probability and Computing},
  volume={19},
  number={3},
  pages={321--342},
  year={2010},
  publisher={Cambridge University Press}
}

@article{demichele2024rank,
  title={On the rank, kernel, and core of sparse random graphs},
  author={DeMichele, Patrick and Glasgow, Margalit and Moreira, Alexander},
  journal={Random Structures \& Algorithms},
  volume={65},
  number={4},
  pages={719--793},
  year={2024},
  publisher={Wiley Online Library}
}

@article{glasgow2025exact,
  title={The exact rank of sparse random graphs},
  author={Glasgow, Margalit and Kwan, Matthew and Sah, Ashwin and Sawhney, Mehtaab},
  journal={arXiv preprint arXiv:2303.05435},
  volume={1},
  number={3},
  pages={4},
  year={2025}
}

@book{du1995minimax,
  title={Minimax and applications},
  author={Du, Ding-Zhu and Pardalos, Panos M},
  volume={4},
  year={1995},
  publisher={Springer Science \& Business Media}
}

@article{CEMT25,
  author    = {Chris Cama{\~n}o and Ethan N. Epperly and Raphael A. Meyer and Joel A. Tropp},
  title     = {Faster Linear Algebra Algorithms with Structured Random Matrices},
  journal   = {arXiv preprint arXiv:2508.21189},
  year      = {2025},
  url       = {https://arxiv.org/abs/2508.21189}
}

@article{Heidary26,
  author    = {Diar Heidary},
  title     = {{SparseStack} Is an Optimal Oblivious Subspace Embedding},
  journal   = {arXiv preprint arXiv:2609.02978v2},
  year      = {2026},
  url       = {https://arxiv.org/abs/2609.02978v2}
}

@article{Tropp15,
  author    = {Joel A. Tropp},
  title     = {An Introduction to Matrix Concentration Inequalities},
  journal   = {Foundations and Trends in Machine Learning},
  volume    = {8},
  number    = {1--2},
  pages     = {1--230},
  year      = {2015},
  url       = {https://arxiv.org/abs/1501.01571}
}

@article{Tropp25,
  author    = {Joel A. Tropp},
  title     = {Comparison Theorems for the Minimum Eigenvalue of a Random Positive-Semidefinite Matrix},
  journal   = {Communications of the American Mathematical Society},
  volume    = {6},
  pages     = {844--895},
  year      = {2026},
  note      = {arXiv:2501.16578v1},
  url       = {https://arxiv.org/abs/2501.16578v1}
}

@inproceedings{Sarlos06,
  author    = {Tam{\'a}s Sarl{\'o}s},
  title     = {Improved Approximation Algorithms for Large Matrices via Random Projections},
  booktitle = {Proceedings of the 47th Annual IEEE Symposium on Foundations of Computer Science (FOCS)},
  pages     = {143--152},
  year      = {2006}
}

@inproceedings{CW13,
  author    = {Kenneth L. Clarkson and David P. Woodruff},
  title     = {Low Rank Approximation and Regression in Input Sparsity Time},
  booktitle = {Proceedings of the 45th Annual ACM Symposium on Theory of Computing (STOC)},
  pages     = {81--90},
  year      = {2013}
}

@inproceedings{NN13,
  author    = {Jelani Nelson and Huy L. Nguy{\^e}n},
  title     = {{OSNAP}: Faster Numerical Linear Algebra Algorithms via Sparser Subspace Embeddings},
  booktitle = {Proceedings of the 54th Annual IEEE Symposium on Foundations of Computer Science (FOCS)},
  pages     = {117--126},
  year      = {2013}
}

@inproceedings{Cohen16,
  author    = {Michael B. Cohen},
  title     = {Nearly Tight Oblivious Subspace Embeddings by Trace Inequalities},
  booktitle = {Proceedings of the 27th Annual ACM-SIAM Symposium on Discrete Algorithms (SODA)},
  pages     = {278--287},
  year      = {2016}
}

@article{HMT11,
  author    = {Nathan Halko and Per-Gunnar Martinsson and Joel A. Tropp},
  title     = {Finding Structure with Randomness: Probabilistic Algorithms for Constructing Approximate Matrix Decompositions},
  journal   = {SIAM Review},
  volume    = {53},
  number    = {2},
  pages     = {217--288},
  year      = {2011}
}

@article{Woodruff14,
  author    = {David P. Woodruff},
  title     = {Sketching as a Tool for Numerical Linear Algebra},
  journal   = {Foundations and Trends in Theoretical Computer Science},
  volume    = {10},
  number    = {1--2},
  pages     = {1--157},
  year      = {2014}
}

@article{MT20,
  author    = {Per-Gunnar Martinsson and Joel A. Tropp},
  title     = {Randomized Numerical Linear Algebra: Foundations and Algorithms},
  journal   = {Acta Numerica},
  volume    = {29},
  pages     = {403--572},
  year      = {2020}
}

@incollection{McD89,
  author    = {Colin McDiarmid},
  title     = {On the Method of Bounded Differences},
  booktitle = {Surveys in Combinatorics, 1989},
  series    = {London Mathematical Society Lecture Note Series},
  volume    = {141},
  pages     = {148--188},
  publisher = {Cambridge University Press},
  year      = {1989}
}

@article{HRT26,
  author    = {Han Huang and Mark Rudelson and Konstantin Tikhomirov},
  title     = {Well-Invertible Column Subsets of Sparse Matrices Are Rare},
  journal   = {arXiv preprint arXiv:2607.05384},
  year      = {2026},
  url       = {https://arxiv.org/abs/2607.05384}
}

@techreport{Car79,
  author      = {Yair Caro},
  title       = {New Results on the Independence Number},
  institution = {Tel-Aviv University},
  year        = {1979}
}

@article{Hal35,
  author    = {Philip Hall},
  title     = {On Representatives of Subsets},
  journal   = {Journal of the London Mathematical Society},
  volume    = {s1-10},
  number    = {1},
  pages     = {26--30},
  year      = {1935}
}

@article{KN14,
  author    = {Daniel M. Kane and Jelani Nelson},
  title     = {Sparser {Johnson}--{Lindenstrauss} Transforms},
  journal   = {Journal of the ACM},
  volume    = {61},
  number    = {1},
  pages     = {4:1--4:23},
  year      = {2014}
}

@article{Tur41,
  author    = {Paul Tur{\'a}n},
  title     = {Eine {Extremalaufgabe} aus der {Graphentheorie}},
  journal   = {Matematikai {\'e}s Fizikai Lapok},
  volume    = {48},
  pages     = {436--452},
  year      = {1941}
}

@techreport{Wei81,
  author      = {Victor K. Wei},
  title       = {A Lower Bound on the Stability Number of a Graph},
  institution = {Bell Laboratories Technical Memorandum 81-11217-9},
  year        = {1981}
}

@inproceedings{CDDR24,
  title={Optimal embedding dimension for sparse subspace embeddings},
  author={Chenakkod, Shabarish and Derezi{\'n}ski, Micha{\l} and Dong, Xiaoyu and Rudelson, Mark},
  booktitle={Proceedings of the 56th Annual ACM Symposium on Theory of Computing},
  pages={1106--1117},
  year={2024}
}

@article{MDMEM23,
  title={Randomized numerical linear algebra: A perspective on the field with an eye to software},
  author={Murray, Riley and Demmel, James and Mahoney, Michael W and Erichson, N Benjamin and Melnichenko, Maksim and Malik, Osman Asif and Grigori, Laura and Luszczek, Piotr and Derezi{\'n}ski, Micha{\l} and Lopes, Miles E and others},
  journal={arXiv preprint arXiv:2302.11474},
  year={2023}
}

@inproceedings{CNRSP26,
  title={GPU-parallelizable randomized sketch-and-precondition for linear regression using sparse sign sketches},
  author={Chen, Tyler and Niroula, Pradeep and Ray, Archan and Subrahmanya, Pragna and Pistoia, Marco and Kumar, Niraj},
  booktitle={2026 IEEE International Parallel and Distributed Processing Symposium Workshops (IPDPSW)},
  pages={411--420},
  year={2026},
  organization={IEEE}
}

@inproceedings{MW17,
  title={Sublinear time low-rank approximation of positive semidefinite matrices},
  author={Musco, Cameron and Woodruff, David P},
  booktitle={2017 IEEE 58th Annual Symposium on Foundations of Computer Science (FOCS)},
  pages={672--683},
  year={2017},
  organization={IEEE}
}

@inproceedings{MM13,
  title={Low-distortion subspace embeddings in input-sparsity time and applications to robust linear regression},
  author={Meng, Xiangrui and Mahoney, Michael W},
  booktitle={Proceedings of the forty-fifth annual ACM symposium on Theory of computing},
  pages={91--100},
  year={2013}
}

@article{CDMD24,
  title={Optimal oblivious subspace embeddings with near-optimal sparsity},
  author={Chenakkod, Shabarish and Derezi{\'n}ski, Micha{\l} and Dong, Xiaoyu},
  journal={arXiv preprint arXiv:2411.08773},
  year={2024}
}

@inproceedings{CDMD26,
  title={Optimal subspace embeddings: Resolving Nelson-Nguyen conjecture up to sub-polylogarithmic factors},
  author={Chenakkod, Shabarish and Derezi{\'n}ski, Micha{\l} and Dong, Xiaoyu},
  booktitle={Proceedings of the 2026 Annual ACM-SIAM Symposium on Discrete Algorithms (SODA)},
  pages={4501--4559},
  year={2026},
  organization={SIAM}
}

@article{DMMS11,
  title={Faster least squares approximation},
  author={Drineas, Petros and Mahoney, Michael W and Muthukrishnan, Shan and Sarl{\'o}s, Tam{\'a}s},
  journal={Numerische mathematik},
  volume={117},
  number={2},
  pages={219--249},
  year={2011},
  publisher={Springer}
}

@inproceedings{CCF02,
  title={Finding frequent items in data streams},
  author={Charikar, Moses and Chen, Kevin and Farach-Colton, Martin},
  booktitle={International colloquium on automata, languages, and programming},
  pages={693--703},
  year={2002},
  organization={Springer}
}

@inproceedings{TZ04,
  title={Tabulation based 4-universal hashing with applications to second moment estimation.},
  author={Thorup, Mikkel and Zhang, Yin},
  booktitle={SODA},
  volume={4},
  pages={615--624},
  year={2004}
}

@inproceedings{WDLSA09,
  title={Feature hashing for large scale multitask learning},
  author={Weinberger, Kilian and Dasgupta, Anirban and Langford, John and Smola, Alex and Attenberg, Josh},
  booktitle={Proceedings of the 26th annual international conference on machine learning},
  pages={1113--1120},
  year={2009}
}

@inproceedings{PP13,
  title={Fast and scalable polynomial kernels via explicit feature maps},
  author={Pham, Ninh and Pagh, Rasmus},
  booktitle={Proceedings of the 19th ACM SIGKDD international conference on Knowledge discovery and data mining},
  pages={239--247},
  year={2013}
}

@article{RT08,
  title={A fast randomized algorithm for overdetermined linear least-squares regression},
  author={Rokhlin, Vladimir and Tygert, Mark},
  journal={Proceedings of the National Academy of Sciences},
  volume={105},
  number={36},
  pages={13212--13217},
  year={2008},
  publisher={National Academy of Sciences}
}

@misc{LWDMZ26,
      title={Stellar Colosseum: A Many-Agent Harness for Long-Horizon Research in Mathematics and Theoretical Computer Science}, 
      author={Honghao Lin and David P. Woodruff and Yuan Deng and Jieming Mao and Song Zuo and Vahab Mirrokni},
      year={2026},
      eprint={2609.15983},
      archivePrefix={arXiv},
      primaryClass={cs.AI},
      url={https://arxiv.org/abs/2609.15983}, 
}

\end{document}